\documentclass{aa} 
\usepackage{graphicx}
\usepackage{txfonts}
\usepackage{multirow}
\usepackage{caption, threeparttable}
\usepackage{xcolor}
\usepackage{float}
\begin{document}

\newcommand{\teff}{$T_\mathrm{eff}$}
\newcommand{\logg}{$\log g$}
\newcommand{\feh}{[Fe/H]}
\newcommand{\microturb}{$\xi_\mathrm{micro}$}
\newcommand{\sife}{[Si/Fe]}
\newcommand{\mgfe}{[Mg/Fe]}
\newcommand{\afe}{[$\alpha$/Fe]}
\newcommand{\co}{CO($\nu =2-0$)\,}

\newcommand{\water}{H$_2$O}
\newcommand{\invcm}{cm$^{-1}$}
\newcommand{\kms}{km\,s$^{-1}$}
\newcommand{\mic}{$\mu \mathrm m$}

\defcitealias{Nandakumar:2023}{Paper\,I}
\defcitealias{Nandakumar:2023b}{Paper\,II}

\titlerunning{NIR Abundance Trends for 21 Elements} 
\authorrunning{Nandakumar et al.}

   \title{M Giants with IGRINS
}
   \subtitle{III. Abundance Trends for 21 Elements in the Solar Neighborhood from High-Resolution, Near-Infrared Spectra}

   \author{G. Nandakumar
            \inst{1}
            \and
          N. Ryde
          \inst{1}
          \and
         R. Forsberg
          \inst{1}            
        \and
       M. Montelius
         \inst{2}
                \and
        G. Mace
        \inst{3}
  \and
        H. J\"onsson
        \inst{4}
    \and
        B. Thorsbro
        \inst{5,1}
          }

   \institute{Division of Astrophysics, Department of Physics, Lund University, Box 43, SE-221 00 Lund, Sweden\\
              \email{govind.nandakumar@fysik.lu.se}
                 \and
    Kapteyn Astronomical Institute, University of Groningen, Landleven 12, 9747 AD Groningen, The Netherlands
              \and
    Department of Astronomy and McDonald Observatory, The University of Texas, Austin, TX 78712, USA
     \and
     Materials Science and Applied Mathematics, Malmö University, SE-205 06 Malmö, Sweden
     \and
     Observatoire de la C\^ote d'Azur, CNRS UMR 7293, BP4229, Laboratoire Lagrange, F-06304 Nice Cedex 4, France
     }

   \date{Received ; accepted }

 
  \abstract
   {In order to be able to investigate the chemical history of the entire Milky Way, it is imperative to also study the dust-obscured regions, where most of the mass lies,  in detail. The Galactic Center is an example of such a region of interest to study. Due to the intervening dust along the line-of-sight, near-infrared spectroscopic investigations are necessary.}
   {The aim is to demonstrate that M giants observed at high spectral resolution in the H and K bands (1.5-2.4\,\mic) can yield useful abundance-ratio trends versus metallicity for 21 elements. These elements can therefore be studied also for heavily dust-obscured regions of the Galaxy, such as the Galactic Center. The abundance-ratio trends will be important for the further investigation of the Galactic chemical evolution in these regions.}  
   {We have observed near-infrared spectra of 50 M giants in the solar neighbourhood at high signal-to-noise and at a high spectral resolution with the IGRINS spectrometer on the GEMINI South telescope. The full H and K bands are recorded simultaneously at $R=45,000$. We adopted the fundamental stellar parameters for these stars from \cite{Nandakumar:2023}, with \teff\ ranging from 3400 to 3800 K. With a manual spectral synthesis method, we have derived stellar abundances for 21 atomic elements, namely F, Mg, Si, S, Ca, Na, Al, K, Sc, Ti, V, Cr, Mn, Co, Ni, Cu, Zn, Y, Ce, Nd, and Yb. We have systematically studied useful spectral lines of all these elements in the H and K bands.}
   {We demonstrate what elements can be analysed from H- and K-band high-resolution spectra, and we show which spectral lines can be used for an abundance analysis, showing them line by line. We discuss the 21 abundance-ratio trends and compared them with those determined from APOGEE and from the optical GILD sample. Especially, the trends of the heavy elements Cu, Zn, Y, Ce, Nd, and Yb are possible to retrieve from high-resolution H- and K-band spectra. This opens up these nucleosynthetic channels, including both the s- and the r-process, in dust-obscured populations. The [Mn/Fe] versus [Fe/H] trend is shown to be more or less flat at low metallicities, implying that existing NLTE correction are relevant. }
    {With high-resolution, near-infrared spectra it is possible to determine reliable abundance-ratio trends versus metallicity for 21 elements, including elements formed in several different nucleosynthetic channels. The important neutron-capture elements, both s- and r-dominated elements, are doable. This opens up the possibility to study the chemical evolution in detail also in dust-obscured region of the Milky Way, such as the Galactic Center. M giants are useful bright probes for these regions and for future studies of extra-galactic stellar populations. A careful analysis of high quality spectra is needed to retrieve all these elements, often from weak and blended lines. A spectral resolution of $R\gtrsim 40,000$ is a further quality which helps in deriving precise abundances for this range of elements. In comparison to APOGEE, we can readily obtain the abundances for Cu, Ce, Nd and Yb from the H-band, which demonstrates an advantage of analysing high-resolution spectra.} 


   \maketitle
%

\section{Introduction}
\label{sec:intro}
\vspace{-5pt}
In order to get a full picture of the formation and evolution of the Milky Way, the chemistry of the dust-obscured parts along the Galactic plane, where most of the mass lies, needs to be thoroughly investigated. The dust prevents optical studies toward the inner Galactic plane, especially the Milky Way center. The emerging possibility to investigate the chemistry through near-infrared observations are now maturing, opening up these areas for study \citep[e.g. ][]{frogel:99,carr:00,ramirez:00,cunha:07,rich:07,rich:12,ryde:15,ryde:2016_metalpoor,ryde:2016_bp2,Nandakumar:18,Guerco:2022,Nieuwmunster:2023}. The main developments are toward more sensitive instruments and the possibility to efficiently record larger parts of the near-infrared spectrum, such as one or more full telluric transmission bands\footnote{Y-band at 0.95-1.1\,\mic; J-band at 1.1-1.4\,\mic; H-band at 1.5-1.8\,\mic; and K-band at 2.0-2.4\,\mic.} at once.

The stellar probes used for the chemical investigation of the inner regions of the Milky Way need to be bright. When K giants are not bright enough, e.g. in the Nuclear Star Cluster, methods to analyse the ubiquitous, brighter, but cooler M giants in the near-infrared are needed. While the optical spectra of M giants are swamped by molecules, their near-infrared spectra (0.95-2.4\,\mic) can be analyzed \citep[see, for example, the analysis in][]{Hayes:2022}. We, therefore, need methods to retrieve stellar parameters and accurate and precise abundances from near-infrared spectra of not only K giants but also M giants. 

Abundance determination of local-disk stars have been done based on optical high-resolution spectra for over 50 years, providing a large set of methods and spectral lines with accurate data from which one can obtain global stellar parameters and chemical abundances of stars. However, the near-infrared is less ventured than the optical, with often less accurate atomic data for spectral lines of interest, if identified at all. 
As such, to determine adequate stellar parameters and chemical abundances, identifying usable lines and developing a careful analysis of high-resolution spectra is required. In this series of papers, we are trying to do just that.  

In order to determine a range of nucleosynthetic channels and, therefore, a range of different elements, high spectral resolution ($R\gtrsim40,000$) is needed. Several instruments with such capabilities, to certain extents, exist already, such as the GIANO \citep{Origlia:2014}, WINERED \citep{winered}, CRIRES+ \citep{crires} spectrographs and the IGRINS spectrometer \citep[IGRINS;][]{Yuk:2010}. In this study we use the latter, IGRINS, that allows for observing spectra covering both the H- and K-bands at high-resolution at once. IGRINS has been extensively used in a broad range of near infrared spectroscopic studies such as in the investigation of environments surrounding young stars \citep{Kaplan:2017,Kaplan:2021}, young stellar objects \citep{Lee:2016,YSOsi:2023,YSOsii:2023}, planetary nebulae \citep{Sterling:2016,Madonna:2018}, very metal poor stars \citep{Afsar:2016}, carbon stars \citep{carbonstar:2023}, and evolved field stars as well as those in open and globular clusters \citep[e.g.][]{afsar:2018,afsar:2019,afsar:2020,montelius:22,afsar:2023,Brady:2023,Holanda:2024}

The urgent need to develop methods for deriving reliable stellar parameters and stellar abundances from high-resolution, near-infrared spectra of M giants, is not least relevant because the
immense possibilities of observing more distant M giants that will be expanded on with upcoming instruments \citep[such as ESO's Multi-Object Optical and Near-infrared Spectrograph, MOONS,][]{moons:12} and future extremely large telescopes such as ESO's Extremely Large Telescope \citep[ELT,][]{ELT2008,ELT2014} and the Thirty Meter Telescope \citep[TMT,][]{TMT} and the high-resolution, near-infrared spectrometers projected for them \citep{hires:13,hires:16,tmt_nir:19}.   







Previous publications in this series are \citet{Nandakumar:2023}, from now on \citetalias{Nandakumar:2023}, and \citet{Nandakumar:2023b}, from now on \citetalias{Nandakumar:2023b}. In these we demonstrate a method for obtaining stellar parameters, $\alpha$ and fluorine abundances for a set of 50 cool M-giant stars 
in the solar neighborhood. In this paper, we extend the set of chemical abundances that are derived to include odd-Z, iron-peak, and neutron-capture elements, showing in total 21 elements apart from iron. This range of elements is important and opens the opportunity of obtaining the full chemical view of the Galaxy and its components. Especially the neutron-capture elements introduces new nucleosynthetic channels with their own evolutionary timescales \citep[see, e.g.][]{manea:23}.  
Based on the method demonstrated in this study, it will now readily be possible to investigate also the massive, dust obscured parts along the Galactic plane. Future infrared surveys and studies of the Galactic central parts performed at high spectral resolution ($R>40,000$) would therefore be rewarding. 
 
The details of the observations and data reduction are provided in the section~\ref{sec:obs} followed by the analysis in the section~\ref{sec:analysis}. In section~\ref{sec:results}, we show the elemental-abundance trends for the $\alpha$-, odd-z, iron-peak, and neutron-capture elements for the 50 solar-neighborhood stars analysed in this series. Further comparison with the trends from a well-studied optical giant-star sample (GILD) and from APOGEE are made in section~\ref{sec:discussion}.

\section{Observations and Data reduction}
\label{sec:obs}

We will determine the abundances of 21 elements for 50 
M giants (\teff$< 4000$~K) from high-resolution spectra observed with  the Immersion GRating INfrared Spectrograph \citep[IGRINS;][]{Yuk:2010,Wang:2010,Gully:2012,Moon:2012,Park:2014,Jeong:2014}.
These cool giants were selected from the APOGEE datarelease DR16 \citep{ApogeeDR16}.

We have  observed the entire H- and K-bands (1.5 - 1.75 and 2.05 - 2.3 \mic, respectively) at a spectral-resolution of $R\sim45,000$,
thus giving access to a wealth of spectral lines enabling a detailed study of a range of elements. Earlier abundance studies have shown the strength of high-resolution spectra recorded with IGRINS \citep[see][]{afsar:2018,afsar:2019,afsar:2020,montelius:22,afsar:2023}. For instance, \citet{afsar:2018} determined abundances for 21 elements for three horizontal-branch stars of $5000<$\teff$< 5300$\,K showing a range of suitable spectral lines that are internally self-consistent.
Some elements are better measured in the wavelength range of IGRINS than in the optical range.

All of the observed stars  presented here lie in the solar neighbourhood, and serve as a good reference sample. 
A subset with 44 of these were observed with IGRINS mounted on the Gemini South telescope \citep{Mace:2018} within the programs GS-2020B-Q-305 and GS-2021A-Q302 from January to April 2021. The other six nearby M giants were available in the IGRINS spectral library \citep{park:18,rrisa}, and were observed at McDonald Observatory \citep{Mcdonald}. Details of all these observations are provided in \citetalias{Nandakumar:2023}, where the stellar parameters and the [$\alpha$/Fe] (Mg, Si, and Ca) and Ti are presented. 

The spectral reductions were done with the IGRINS PipeLine Package \citep[IGRINS PLP;][]{Lee:2017} to optimally extract the telluric corrected, wavelength calibrated spectra after flat-field correction \citep{Han:2012,Oh:2014}. The spectra were then resampled and normalized in {\tt iraf} \citep{IRAF} but to take care of any residual modulations in the continuum levels, we put considerable focus on defining specific local continua around every line that elemental abundances were determined from. Finally, the spectra are shifted to laboratory wavelengths in air after a stellar radial velocity correction. The average signal-to-noise ratios (SNR)\footnote{SNR is provided by RRISA \citep[The Raw $\&$ Reduced IGRINS Spectral Archive;][]{rrisa} and is the average SNR for H and K-band, respectively, and is per resolution element. It varies over the orders and it is lowest at the ends of the orders} of  the spectra  generally have SNR$>100$.


\section{Analysis}
\label{sec:analysis}


The spectroscopic analysis in this work is carried out using the spectral synthesis method wherein the stellar parameters and elemental abundances of a star are determined by fitting its observed spectrum with a synthetic spectrum. The synthetic spectrum is generated using the spectroscopy made easy \citep[SME;][]{sme,sme_code} tool by calculating the spherical radiative transfer through a relevant stellar atmosphere model defined by its fundamental stellar parameters. We choose the stellar atmosphere model by interpolating in a grid of one-dimensional (1D) Model Atmospheres in a Radiative and Convective Scheme (MARCS) stellar atmosphere models \citep{marcs:08}.

\subsection{Stellar parameters}
\label{sec:stellarparameters}

Fundamental stellar parameters, namely the effective temperature (\teff), surface gravity (\logg), metallicity (\feh) and microturbulence ($\xi_\mathrm{micro}$), are crucial in spectroscopic analysis. These parameters form the basis for a reliable determination of elemental abundances from respective absorption lines in the observed spectrum. Thus, it is important to determine accurate fundamental stellar parameters.
\citetalias{Nandakumar:2023} devised a novel method to determine these for M giants (\teff\,$<$ 4000 K) by a spectral synthesis method also employing SME and MARCS stellar atmosphere models. In this study, we use the same stars in \citetalias{Nandakumar:2023} and the determined stellar parameters from that paper. In the method, a set of \teff-sensitive OH lines, a few chosen CO and Fe lines are synthesized and fitted, while setting \teff, \feh, $\xi_\mathrm{micro}$, and the C and N abundance values as free parameters. The best-fit synthetic spectrum provided the values for these free parameters. The surface gravity, \logg, is constrained based on the \teff\, and \feh\, values from 3-10 Gyr Yonsei-Yale isochrones \citep{Demarque:2004}. The main assumption in this method is the value of oxygen abundance that is taken from a simple functional form of the [O/Fe] versus [Fe/H] trend in \cite{Amarsi:2019} for thin and thick-disk stars \citepalias[see also Figure~1 in][]{Nandakumar:2023}.


The method was validated in \citetalias{Nandakumar:2023} by determining the stellar parameters for the stars also analysed here. The six nearby, well-studied M-giants, with well-determined stellar parameters, were used to benchmark the method. Furthermore, they determined $\alpha$-element trends versus metallicity for the stars showcasing the accuracy and precision of the derived stellar parameters. These are given in Table 2 in \citealt{Nandakumar:2023b}. The derived stellar parameters are in the range of 3350 to 3800\,K in \teff, 0.3 to 1.15\,dex in \logg, and $-0.9$ to 0.25\,dex in \feh. The separation  into the high-$\alpha$ and low-$\alpha$ populations was originally based on the Mg abundances determined by APOGEE. Our magneisum abundance estimates in \citetalias{Nandakumar:2023} (also see Section~\ref{sec:alpha} and Figure~\ref{fig:mg_trend} in this work) further confirms this classification.




\subsection{Linelist}
\label{sec:Linelist}

For the spectral synthesis we use a line list containing the spectral lines' wavelengths, lines strengths in the form of
$\log gf$-values, excitation energies, broadening parameters, and hyperfine splitting values, if necessary. As a starting point, we use an updated version of the VALD line list \citep{vald,vald4,vald5}. For the many lines that lack reliable experimental transition probabilities, we estimated $\log gf$-values by determining them astrophysically. The $\log gf$-values for the H-band lines have been determined from a high SNR reflected solar spectrum of Ceres, measured with IGRINS \citep{montelius:21}. The values have been validated by computing abundances of some 34 K-giants and comparing them to high resolution optical abundances from the Giants in the Local Disk sample (GILD; J\"onsson et al. in prep.). For the K-band lines, we determined $\log gf$-values using the high-resolution infrared solar flux spectrum of \cite{Wallace:2003} and testing them on the high-resolution ($R \sim$ 100,000), infrared spectrum of the bright giant Arcturus\footnote{\citet{Ramirez:2011} determined the fundamental parameters of Arcturus (or $\alpha$ Boo) as \teff$=4290$ K, \logg$=1.7$, and \feh$=-0.5$ dex.} from the Arcturus atlas \citep{Hinkle:1995}. For lines with hyperfine structure splitting, the $\log gf$-values have been assigned according to their relative linestrength within the LS multiplet \citep{Cowan:1981,montelius:21}.  

For many lines, we adopted the broadening parameters from the ABO theory \citep{anstee_investigation_1991,Anstee1995,Barklem1997a,Barklem1998b} or from the spectral synthesis code BSYN based on routines from MARCS \citep{marcs:08}. The line data for the CO, CN, and OH lines were adopted from the line lists of \citet{li:2015}, \citet{brooke:2016}, and \citet{sneden:2014}, respectively. The molecular lines are important since they often act as blending lines in atomic spectral lines.

 The details of the lines for the elements Mg, Si, Ca (except for the six new lines in this work), Ti, F, and Yb are provided in \citetalias{Nandakumar:2023}, \citetalias{Nandakumar:2023b}, and \cite{montelius:22}. The details about the H-band lines used in this work except for cerium and neodymium can be found in \citet{montelius:21}. We provide the details about cerium and neodymium H-band lines and K-band lines of other elements below \footnote{We refrain from providing a full line list with excitation energies, $\log gf$ and broadening parameters since they are updated astrophysically and hence model-dependent.}.


{\bf Sulphur} For the K-band sulphur line (S{\sc i}), we adjusted the $\log gf$-value from theoretical calculations in \cite{BQZ} to fit the line in the high resolution solar spectrum. The broadening value for the line is adopted from the ABO theory. 

{\bf Sodium} For the three K-band sodium lines (Na{\sc i}), hyperfine transition information and $\log gf$-values are available in the VALD linelist originally determined 
experimentally from \cite{arqueros:1988}. Broadening parameters were determined from the ABO theory for all three lines. The wavelengths of the first two lines had to be shifted forward by $\sim$ 0.1 \AA\, to fit them in the high resolution spectra of the Sun and Arcturus with synthetic spectra. 


{\bf Aluminium} Among the three K-band aluminium lines (Al{\sc i}), VALD provides hyperfine transition lines from \cite{Chang:1990} for the first two lines, and no line information for the third line. \cite{Nordlander:2017} provides line data for all three lines but only one wavelength per line for the first two lines, i.e., no hyperfine transition information. We could also fit these lines in the high resolution spectra of the Sun and Arcturus with the line data from \cite{Nordlander:2017}. The abundances determined for the stars in this work from the first two lines with line data from VALD were found to be lower by $\sim$0.2 dex compared to those determined with line data from \cite{Nordlander:2017}. Hence, we adopted the line data from \cite{Nordlander:2017} for all three K-band aluminium lines to determine abundances.

{\bf Calcium} We have included six K-band calcium lines (Ca{\sc i}) lines in addition to the five lines in H- and K-band used in \citetalias{Nandakumar:2023}. We updated the $\log gf$-values of all six lines astrophysically using the high resolution solar spectrum, and tested them using high resolution Arcturus spectrum. 

{\bf Scandium}  The line list extracted from VALD provided the hyperfine structure information for the three scandium lines (Sc{\sc ii}) in the K-band from \cite{EH} and \cite{ACHBG}. The $\log gf$-values have been adopted from \cite{Pehlivan:2015} determined based on laboratory measurements.  
In addition, we shifted the wavelengths of the second and third line by 0.13 \AA\, and 0.05 \AA\, respectively to fit them in the high resolution spectra of the Sun and Arcturus.





{\bf Nickel}  In addition to the H-band lines, five K-band nickel lines (Ni{\sc i}) have been used to determine nickel abundances in this work. The $\log gf$-values for these lines from the VALD database are from \cite{K08}. Only two lines in the high resolution spectra of the Sun and Arcturus are fit by synthetic spectra using the VALD $\log gf$-values. We estimated astrophysical $\log gf$-values for the remaining three lines. 


 
{\bf Yttrium} The two ionized yttrium-lines (Y{\sc ii}) in the K-band used in this work are weak in the high resolution spectra of the Sun and Arcturus, thus making it difficult to reliably estimate astrophysical $\log gf$-values. In addition, there are no information on hyperfine transitions for these lines. Hence, we adopt the line data provided by VALD from \cite{K06}. The only way we could determine the reliability of these lines and their $\log gf$-values is by inspecting the synthetic spectra fit to the observed lines that are stronger in cool M-giants, and then to compare the resulting abundance trends with that of the trends derived from well studied lines in optical spectra (see the discussion in Sections \ref{sec:neutroncapture} and \ref{sec:discussion}).

{\bf Cerium} \cite{Cunha:2017} determined astrophysical $\log gf$-values for ionized cerium lines (Ce{\sc ii}) in the H-band, using a combination of a high-resolution Fourier Transform Spectrometer (FTS) spectrum of Arcturus and an APOGEE spectrum of a metal-poor, but s-process enriched, red-giant star. We adopted these $\log gf$-values for the four cerium lines used in this work to determine cerium abundances.

{\bf Neodymium} We use the $\log gf$-values of the three ionized neodymium lines (Nd{\sc ii}) in the H-band, determined astrophysicaly in \cite{Hasselquist:2016}. The metal-poor, K-giant 2MASSJ16011638-1201525 was used in \cite{Hasselquist:2016} to determine the $\log gf$-values for these lines.

\subsection{Non-local thermodynamic equilibrium (NLTE) grids}
\label{sec:nlte}
We have included non-local thermodynamic equilibrium (NLTE) grids for the elements C, N, O, Na, Mg, Al, Si, K, Ca, Mn and Fe (\citealt{NLTE}, \citealt{lind17} and \citealt{amarsi16} with subsequent updates; Amarsi priv. comm.), with the departure coefficients computed using the MPI-parallelized NLTE radiative transfer code \texttt{Balder} \citep{Amarsi:2018}.

\begin{figure*}
  \includegraphics[width=\textwidth]{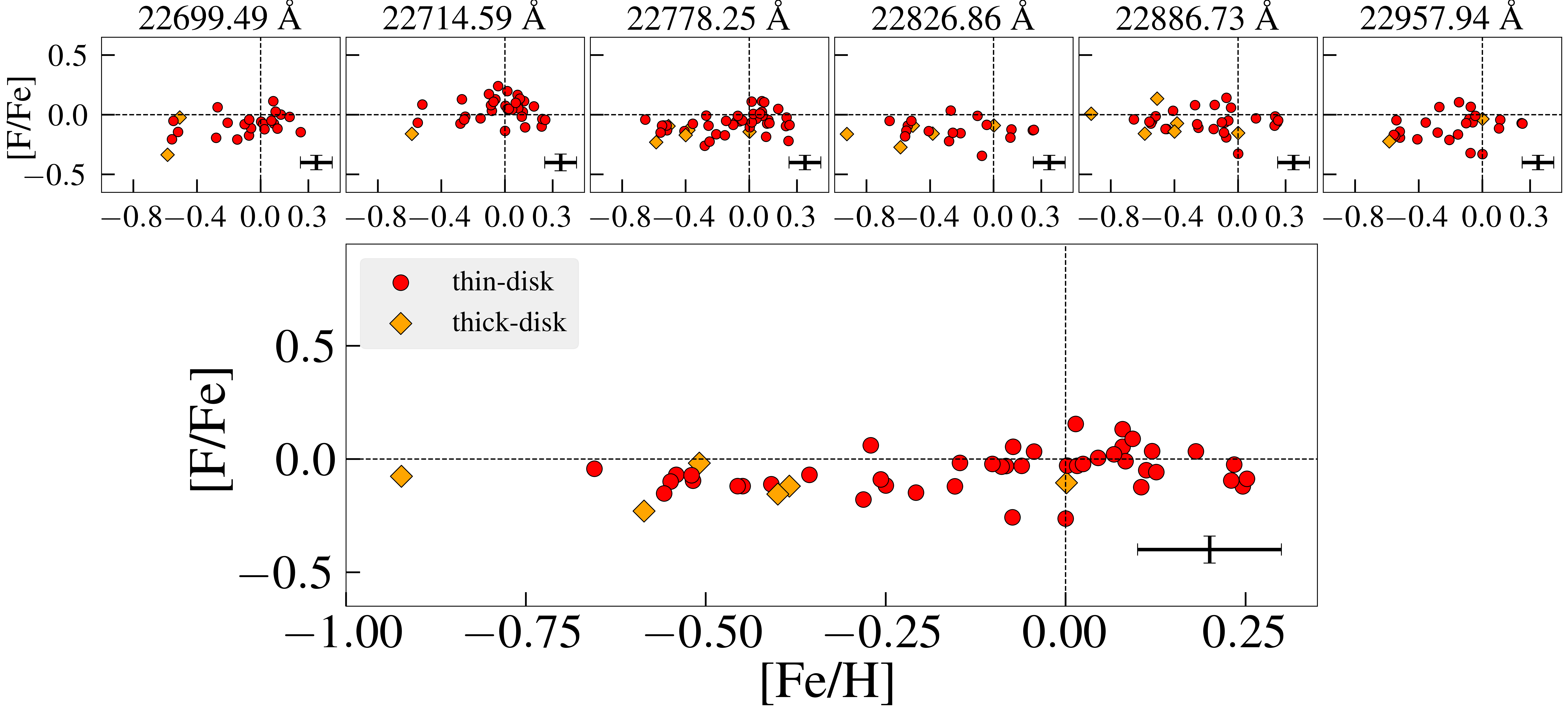}
  \caption{ [F/Fe] versus [Fe/H] trends for 50 M giants in our sample. Figures in the upper panel show the trends determined from five individual HF lines. The mean [F/Fe] versus [Fe/H] trend from all six lines is shown in the bottom panel. Red filled circles and orange diamonds denote the thin- and thick-disk stars, respectively. The error bar in the bottom right part in the top panels and the bottom panel indicate the uncertainties in abundances determined from each line and the mean uncertainty determined as the standard error of mean (see text in the Section~\ref{sec:results} for more details).  }
  \label{fig:f_trend}%
\end{figure*}

 \begin{figure*}
  \includegraphics[width=\textwidth]{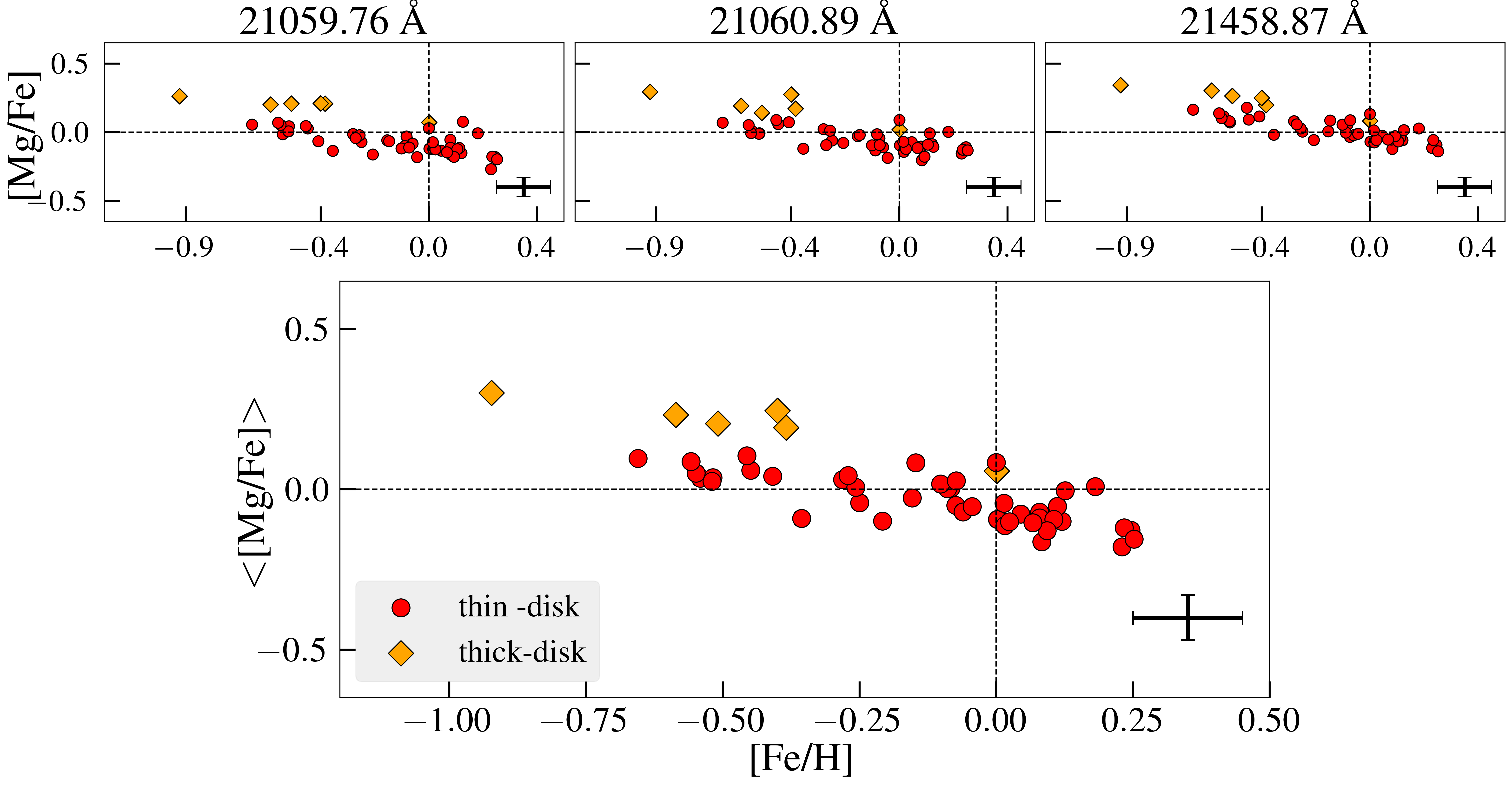}
  \caption{ [Mg/Fe] versus [Fe/H] for 50 M giants in our sample. Arrangement of figures and markers are similar to Figure~\ref{fig:f_trend}.  }
  \label{fig:mg_trend}%
\end{figure*}

\begin{figure*}
  \includegraphics[width=\textwidth]{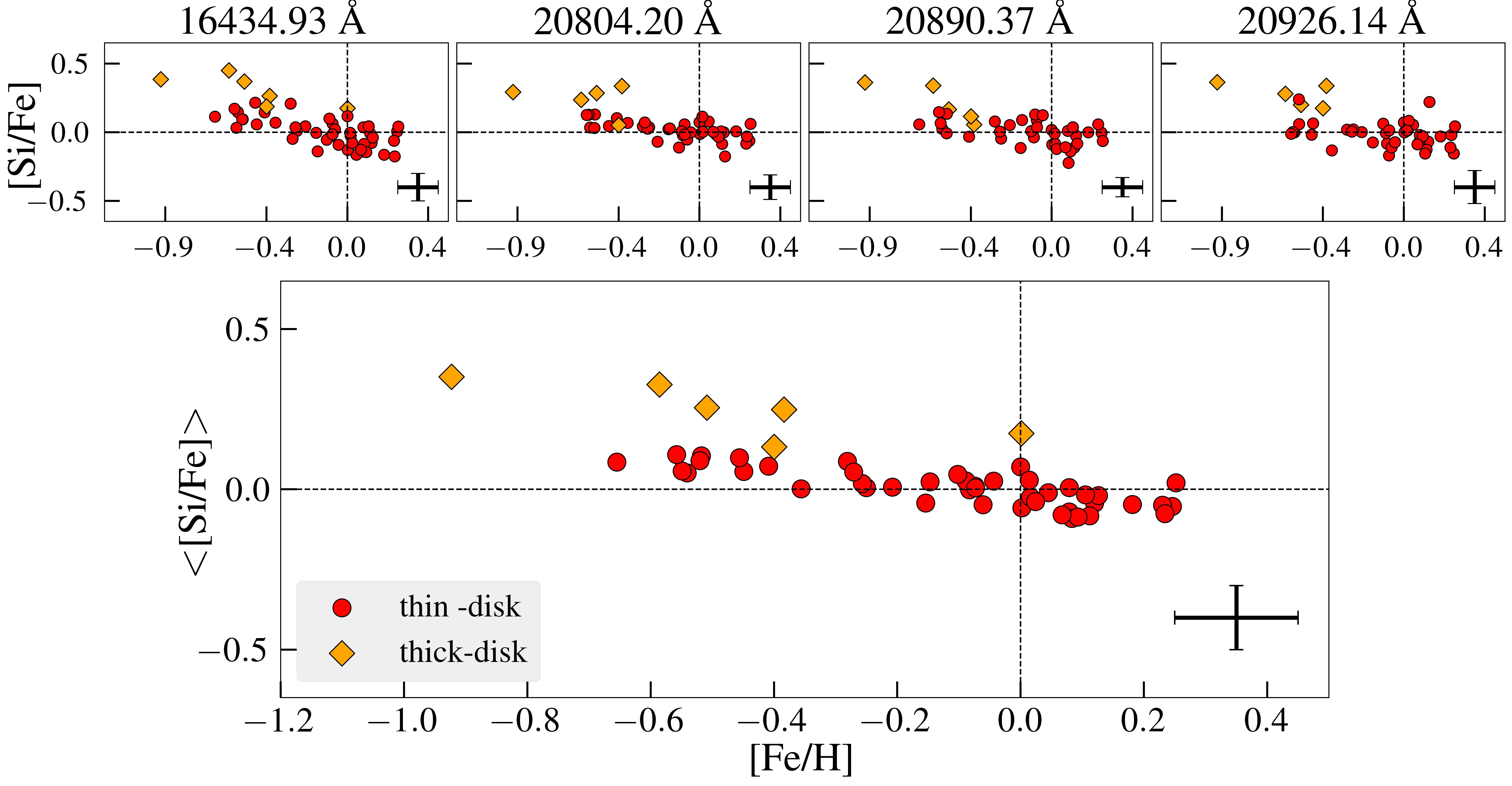}
  \caption{ [Si/Fe] versus [Fe/H] for 50 M giants in our sample. Arrangement of figures and markers are similar to Figure~\ref{fig:f_trend}. }
  \label{fig:si_trend}%
\end{figure*}

\begin{figure}
  \includegraphics[width=\columnwidth]{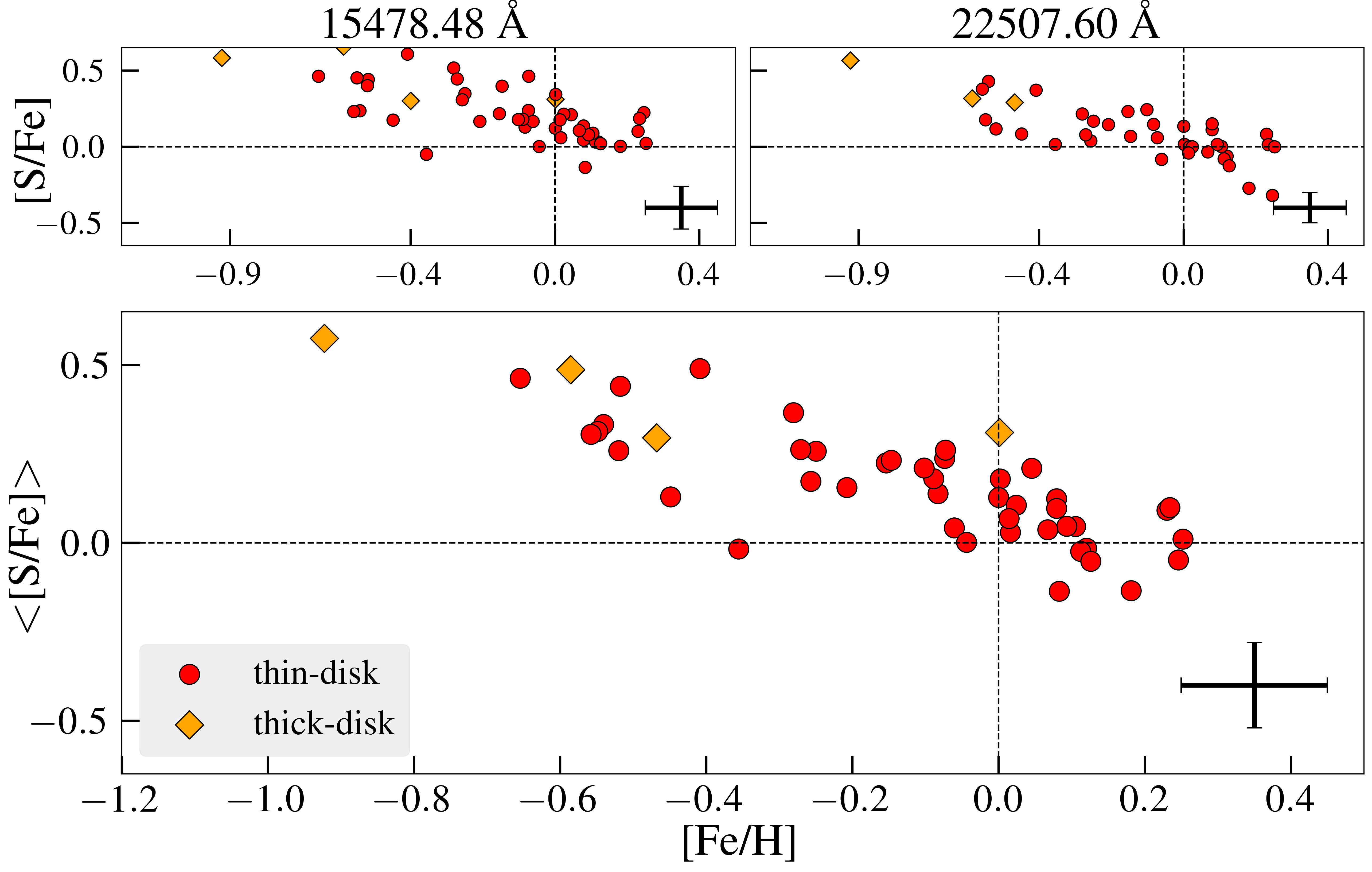}
  \caption{ [S/Fe] versus [Fe/H] for 48 M giants in our sample. Arrangement of figures and markers are similar to Figure~\ref{fig:f_trend}. }
  \label{fig:s_trend}%
\end{figure}

\begin{figure*}
  \includegraphics[width=\textwidth]{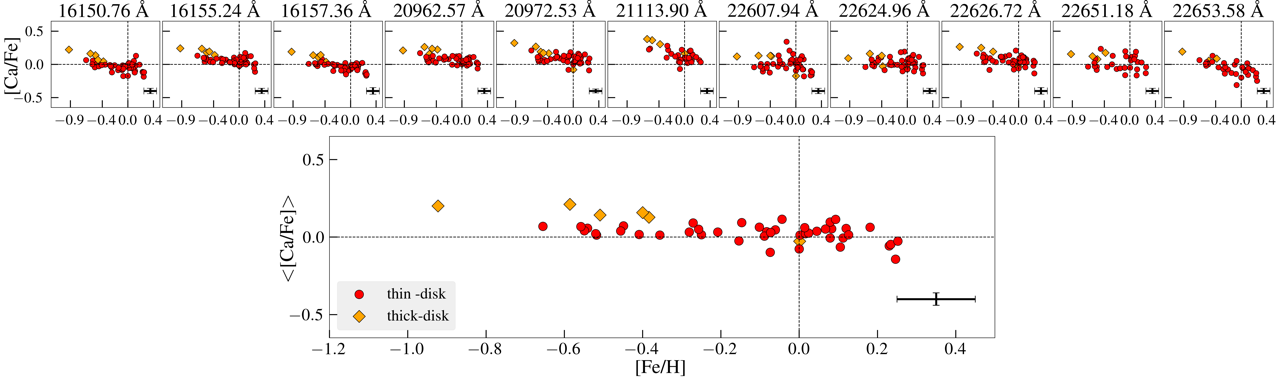}
  \caption{ [Ca/Fe] versus [Fe/H] for 50 M giants in our sample. Arrangement of figures and markers are similar to Figure~\ref{fig:f_trend}. }
  \label{fig:ca_trend}%
\end{figure*}

\section{Results}
\label{sec:results}
We determine the metallicities and chemical abundances of 21 elements for the sample of 50 M-giants in the solar neighbourhood. In this section, we show the abundance trends for these elements determined from each elemental lines as well as the mean estimated abundances as a function of metallicity, after the careful removal of abundances determined from bad noisy lines and those affected by strong telluric contamination. We group them as $\alpha-$, odd-Z, iron-peak, and neutron-capture elements. Fluorine is discussed separately. The tables with line-by-line and mean abundances of all elements for the 50 stars will be made available online. 

We choose one of the metal-rich star, 2M14261117-6240220, in order to determine uncertainties in estimated abundances for all 21 elements resulting from the typical uncertainties in the stellar parameters. We follow the method described in \citetalias{Nandakumar:2023} where we redetermined 50 values of the abundances from all selected lines of each element by setting the stellar parameters to values randomly chosen from normal distributions with the actual stellar parameter value as the mean and the typical uncertainties ($\pm$100\,K in \teff, $\pm$0.2\,dex in \logg, $\pm$0.1\,dex in \feh, and $\pm$0.1\,km\,s$^{-1}$ in $\xi_\mathrm{micro}$) as the standard deviation. The uncertainty in the abundance determined from each line is then taken to be the mid-value of the difference between the 84$^{th}$ and 16$^{th}$ percentiles of the abundance distribution. The final abundance uncertainty is the standard error of mean of uncertainties from all chosen lines of each element. The mean uncertainties range from 0.04 to 0.12\,dex with the maximum uncertainty of 0.12\,dex estimated for sulfur.

For all elements except fluorine and cerium, we have scaled the abundances with respect to solar abundance values from \cite{solar:sme}. We used A$_{\odot}$(F) = 4.43 from \cite{Lodders:2003} and A$_{\odot}$(Ce) =  1.58 from \cite{Grevesse:2015} for fluorine and cerium respectively.

\begin{figure*}
  \includegraphics[width=\textwidth]{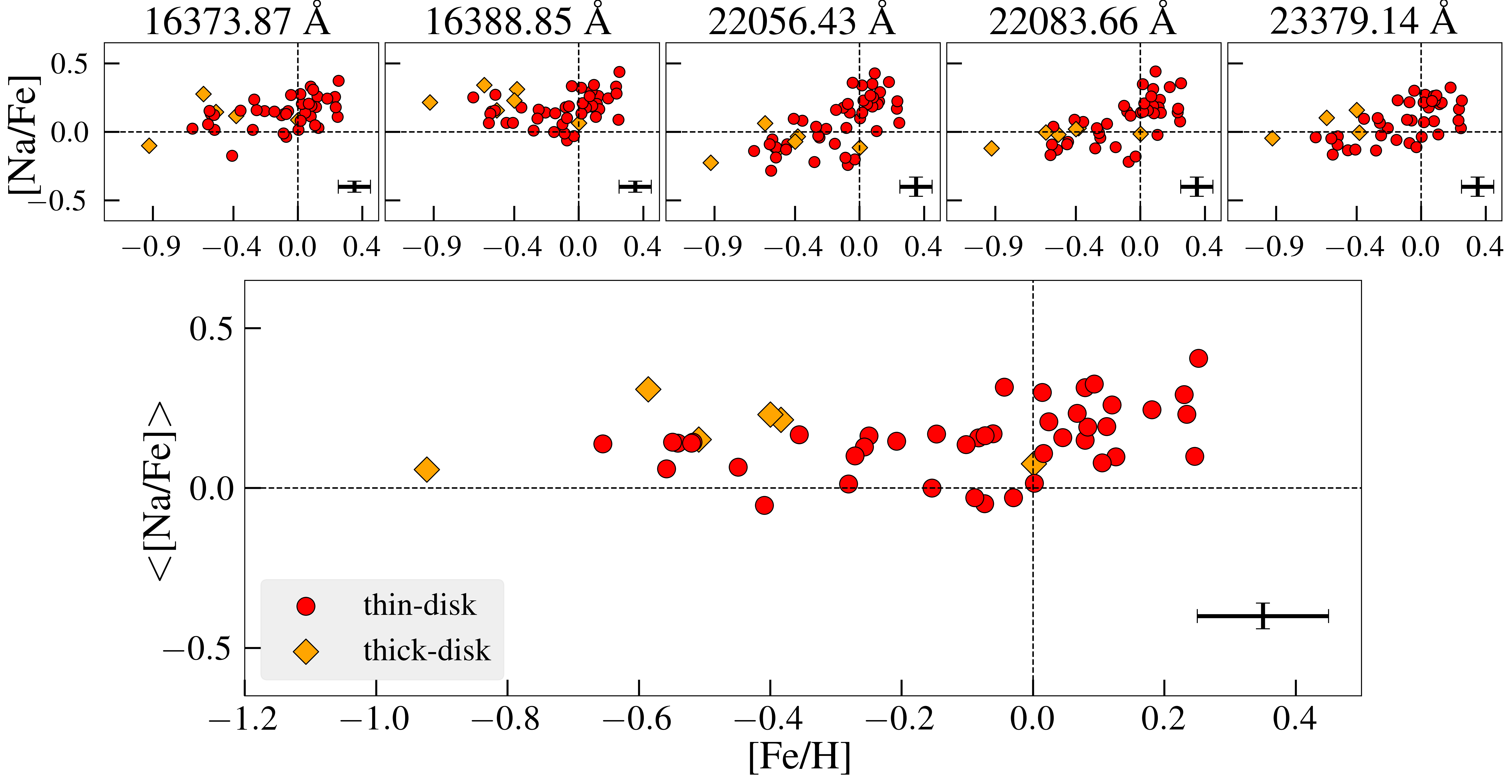}
  \caption{ [Na/Fe] versus [Fe/H] for 50 M giants in our sample. Arrangement of figures and markers are similar to Figure~\ref{fig:f_trend}. We exclude [Na/Fe] abundances from  the three K-band lines to estimate the mean [Na/Fe] abundance since the abundances from these lines are sub-solar for the sub-solar metallicity stars. This is inconsistent with the abundances from H-band lines, as well as in APOGEE and other optical studies (see also Figures~\ref{fig:apogee_trend},~\ref{fig:all_bawlas_trend}, and ~\ref{fig:all_gild_trend}).}
  \label{fig:na_trend}%
\end{figure*}


\begin{figure*}
  \includegraphics[width=\textwidth]{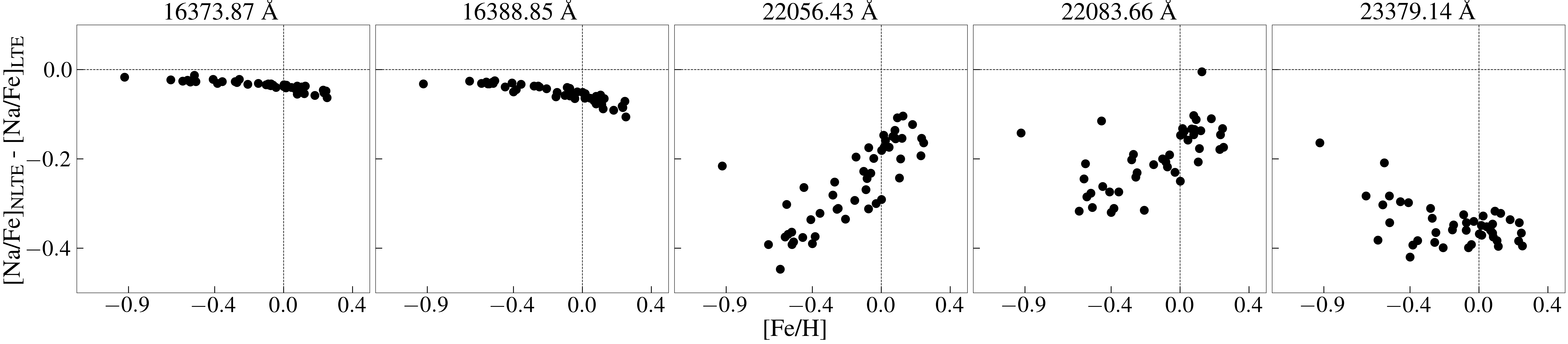}
  \caption{ NLTE - LTE differences versus [Fe/H] for sodium (Na) abundances from the two H-band and three K-band lines for 50 M giants in our sample. NLTE corrections are higher and ranges from $\sim -0.4$\,dex to $\sim -0.1$\,dex for the K-band lines. }
  \label{fig:na_nlte_lte}%
\end{figure*}

\begin{figure*}
  \includegraphics[width=\textwidth]{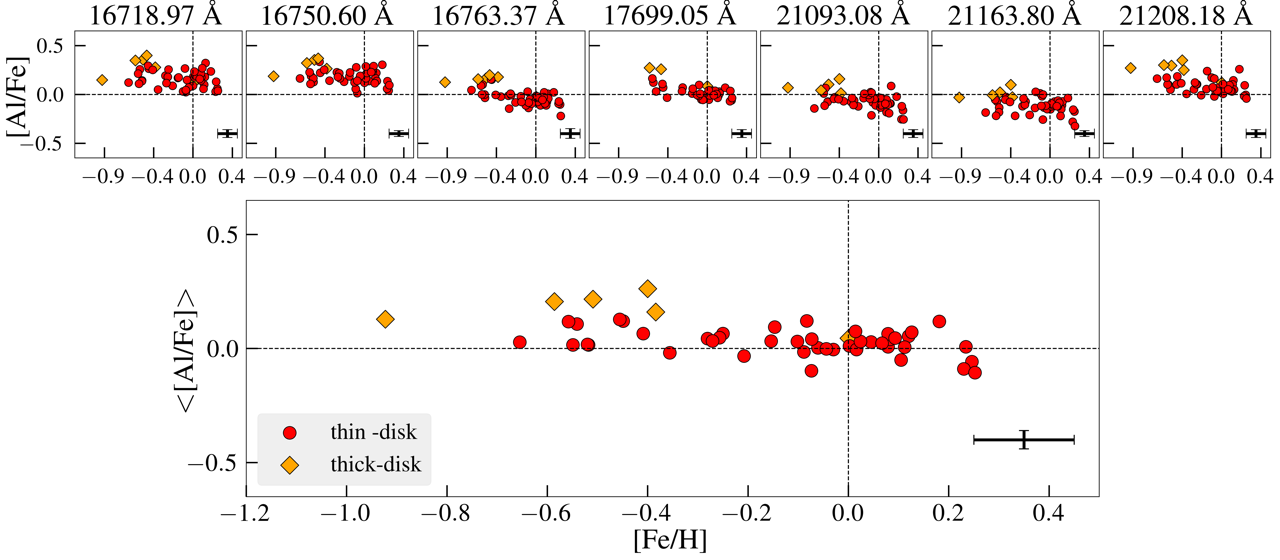}
  \caption{ [Al/Fe] versus [Fe/H] for 50 M giants in our sample. Arrangement of figures and markers are similar to Figure~\ref{fig:f_trend}. }
  \label{fig:al_trend}%
\end{figure*}

\begin{figure}
  \includegraphics[width=\columnwidth]{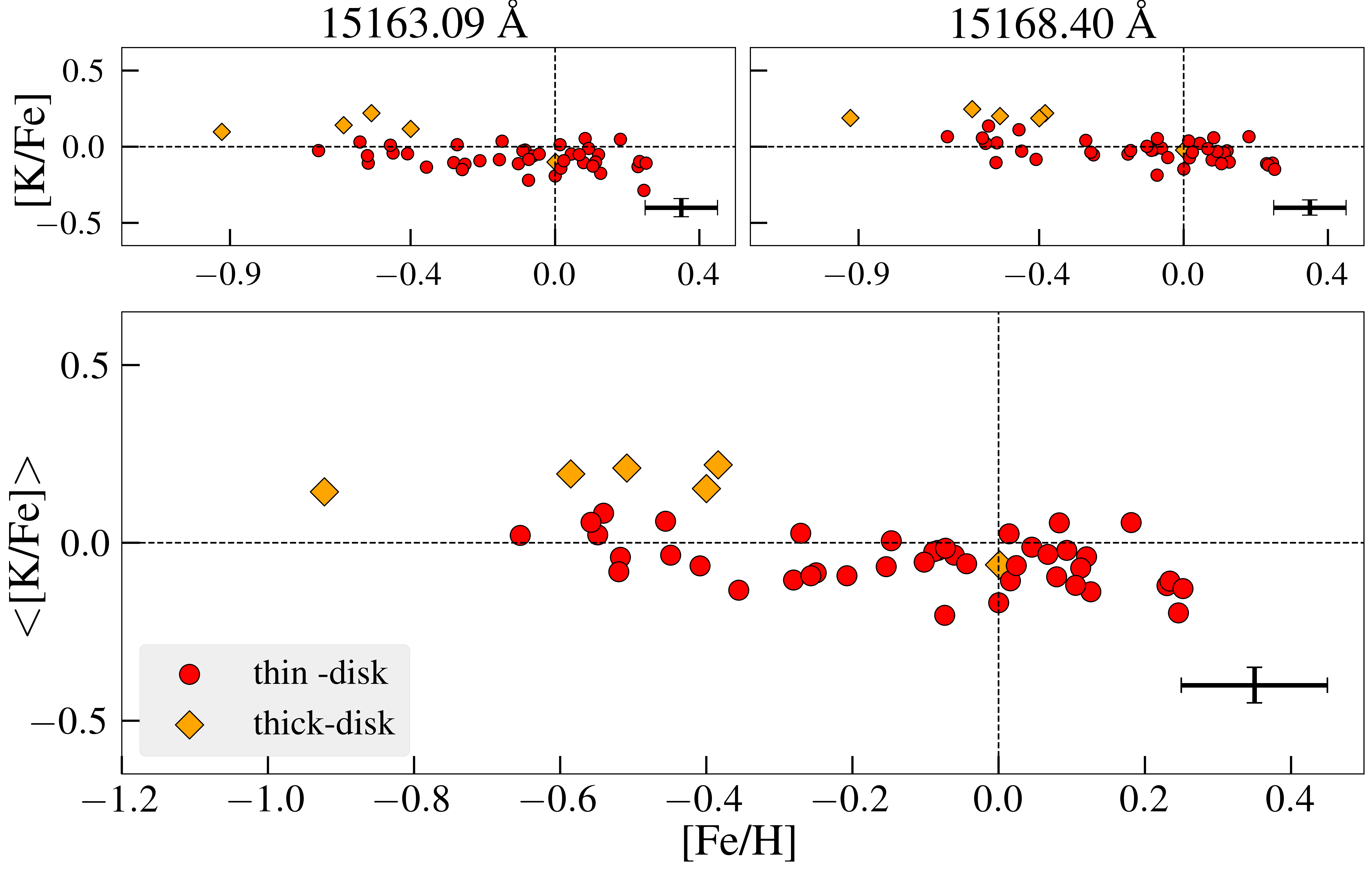}
  \caption{ [K/Fe] versus [Fe/H] for 48 M giants in our sample. Arrangement of figures and markers are similar to Figure~\ref{fig:f_trend}. }
  \label{fig:k_trend}%
\end{figure}

\subsection{Fluorine (F)}
\label{sec:fluorine}
{\bf Fluorine} The fluorine abundances are determined from HF molecular lines in the K-band. \citetalias{Nandakumar:2023b} carried out a detailed analysis of 10 vibrational-rotational, R-branch lines in IGRINS spectra of the 50 solar-neighborhood stars used in this work. This was done to find the best set of HF lines from which to determine the fluorine abundances for this type of stars, and avoiding saturated lines. They suggested to use the three bluest HF lines for cool (\teff\,$<$ 3500 K) and metal-rich (\feh\,$>$ 0 dex) M giants, and three additional ones for M giants at sub-solar metallicities. All HF lines are intrinsically very temperature sensitive, which means that we should expect a larger spread in the derived fluorine abundances.  We find a flat [F/Fe] versus [Fe/H] trend, but with indications of a slightly higher level of [F/Fe] for stars with \feh\,$>$ 0 dex, in the solar neighbourhood sample, see Figure~\ref{fig:f_trend} \citepalias[also presented in][]{Nandakumar:2023b}.


\begin{figure*}
  \includegraphics[width=\textwidth]{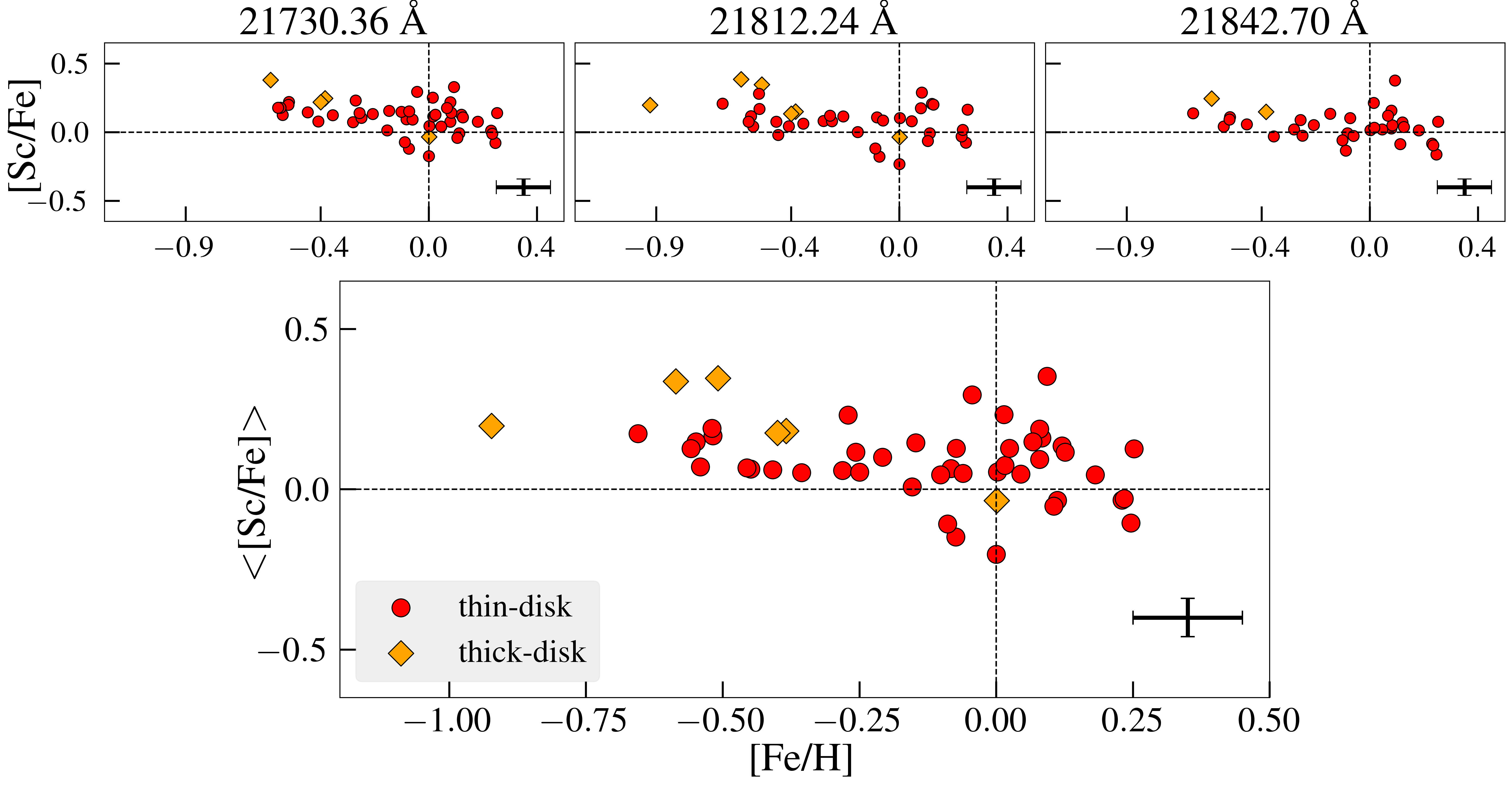}
  \caption{ [Sc/Fe] versus [Fe/H] for 50 M giants in our sample. Arrangement of figures and markers are similar to Figure~\ref{fig:f_trend} }
  \label{fig:sc_trend}%
\end{figure*}

\begin{figure}
  \includegraphics[width=\columnwidth]{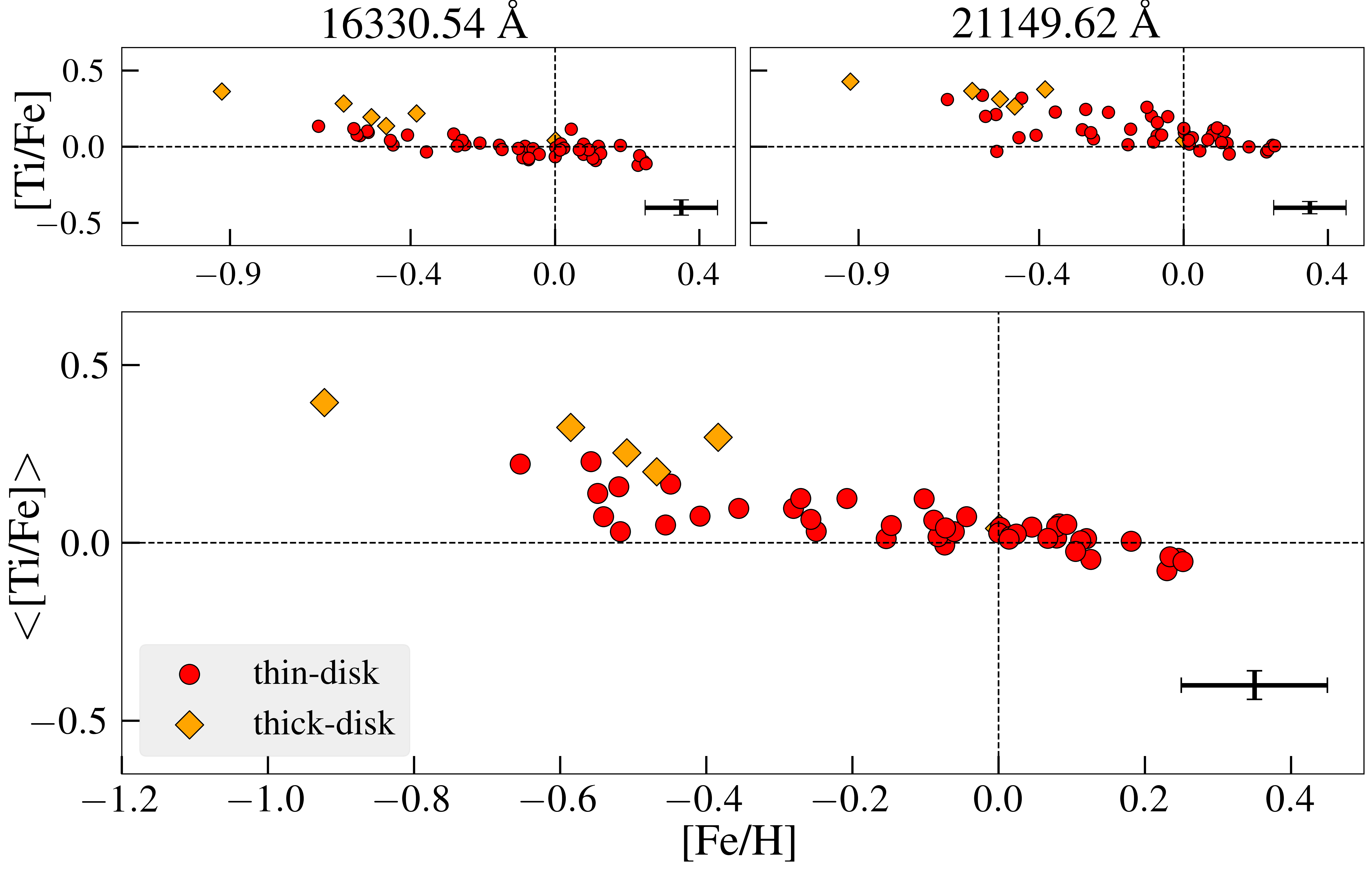}
  \caption{ [Ti/Fe] versus [Fe/H] for 50 M giants in our sample. Arrangement of figures and markers are similar to Figure~\ref{fig:f_trend} }
  \label{fig:ti_trend}%
\end{figure}

\begin{figure}
  \includegraphics[width=\columnwidth]{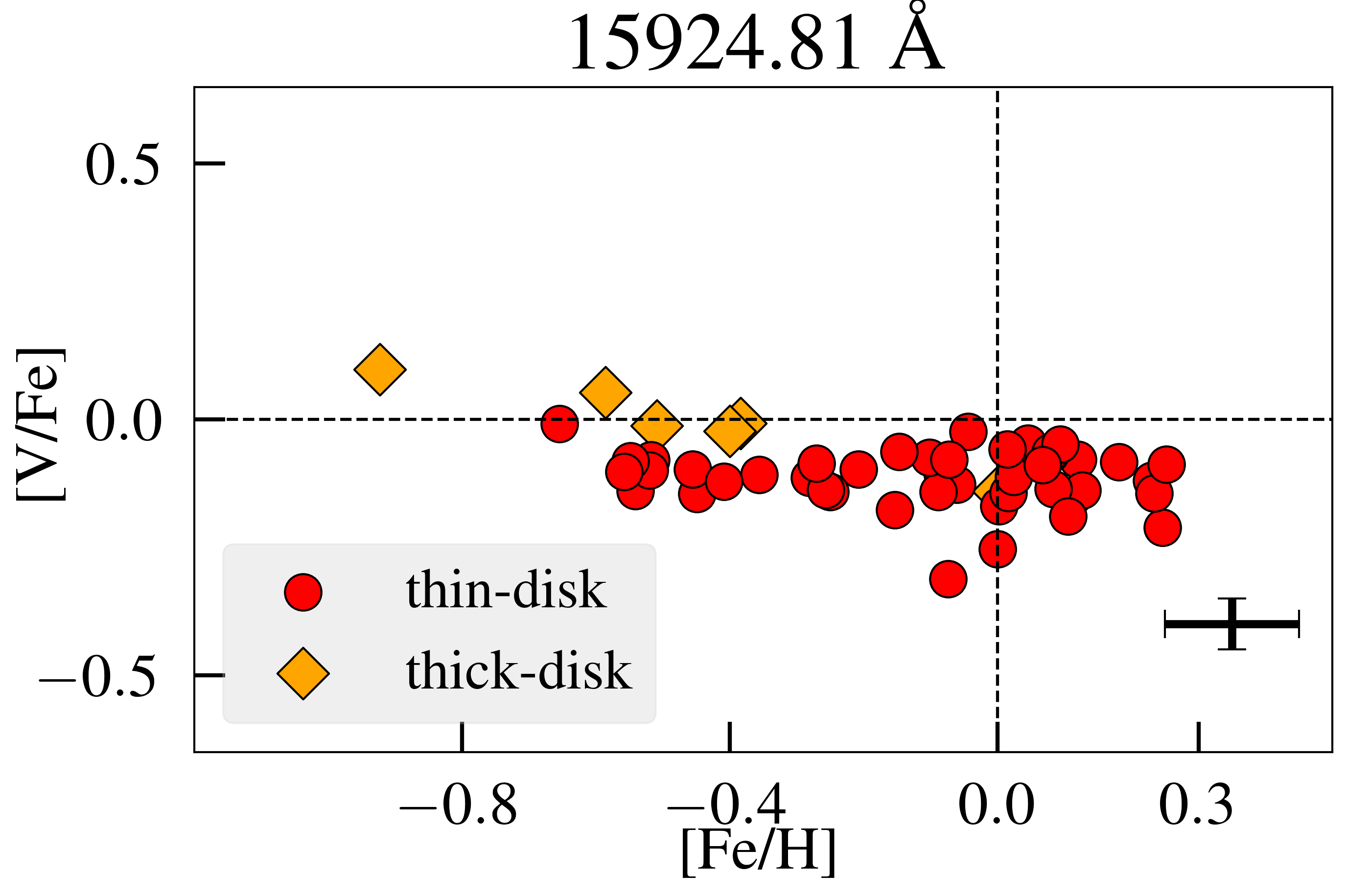}
  \caption{ [V/Fe] versus [Fe/H] for 50 M giants in our sample. Arrangement of figures and markers are similar to Figure~\ref{fig:f_trend}. }
  \label{fig:v_trend}%
\end{figure}

\begin{figure*}
  \includegraphics[width=\textwidth]{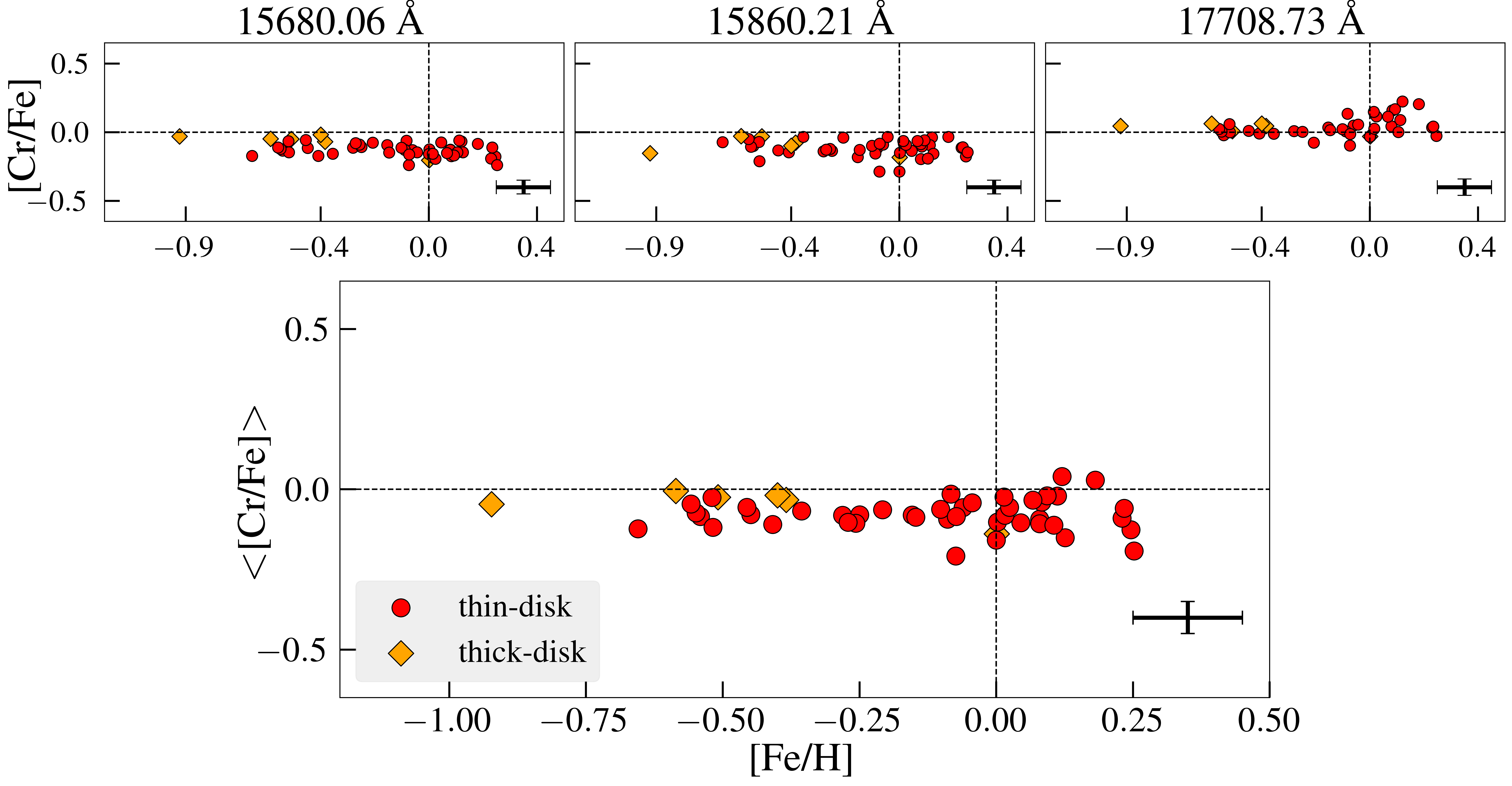}
  \caption{ [Cr/Fe] versus [Fe/H] for 50 M giants in our sample. Arrangement of figures and markers are similar to Figure~\ref{fig:f_trend}. }
  \label{fig:cr_trend}%
\end{figure*}

\begin{figure}
  \includegraphics[width=\columnwidth]{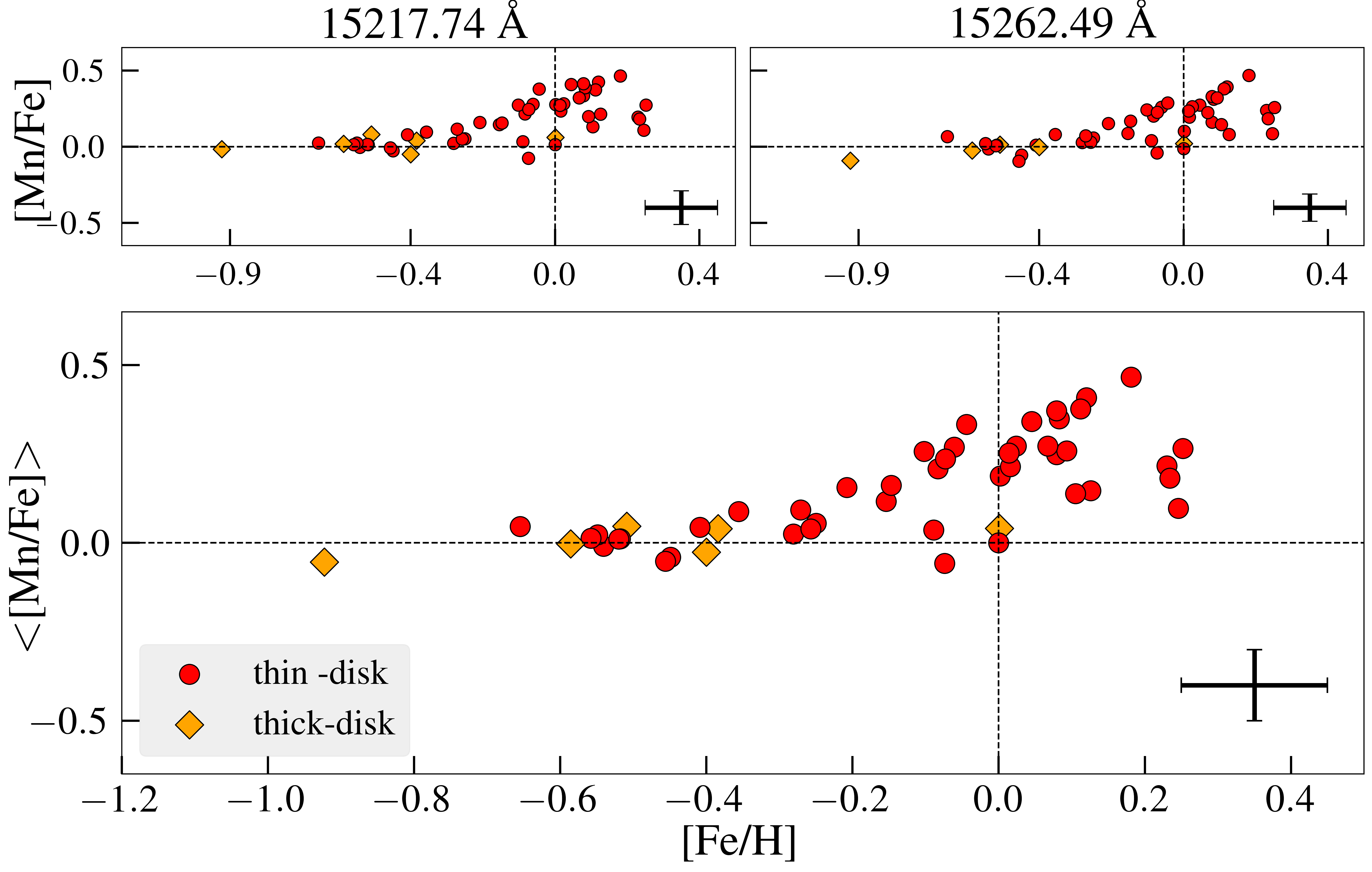}
  \caption{ [Mn/Fe] versus [Fe/H] for 50 M giants in our sample. Arrangement of figures and markers are similar to Figure~\ref{fig:f_trend}. }
  \label{fig:mn_trend}%
\end{figure}

\begin{figure}
  \includegraphics[width=\columnwidth]{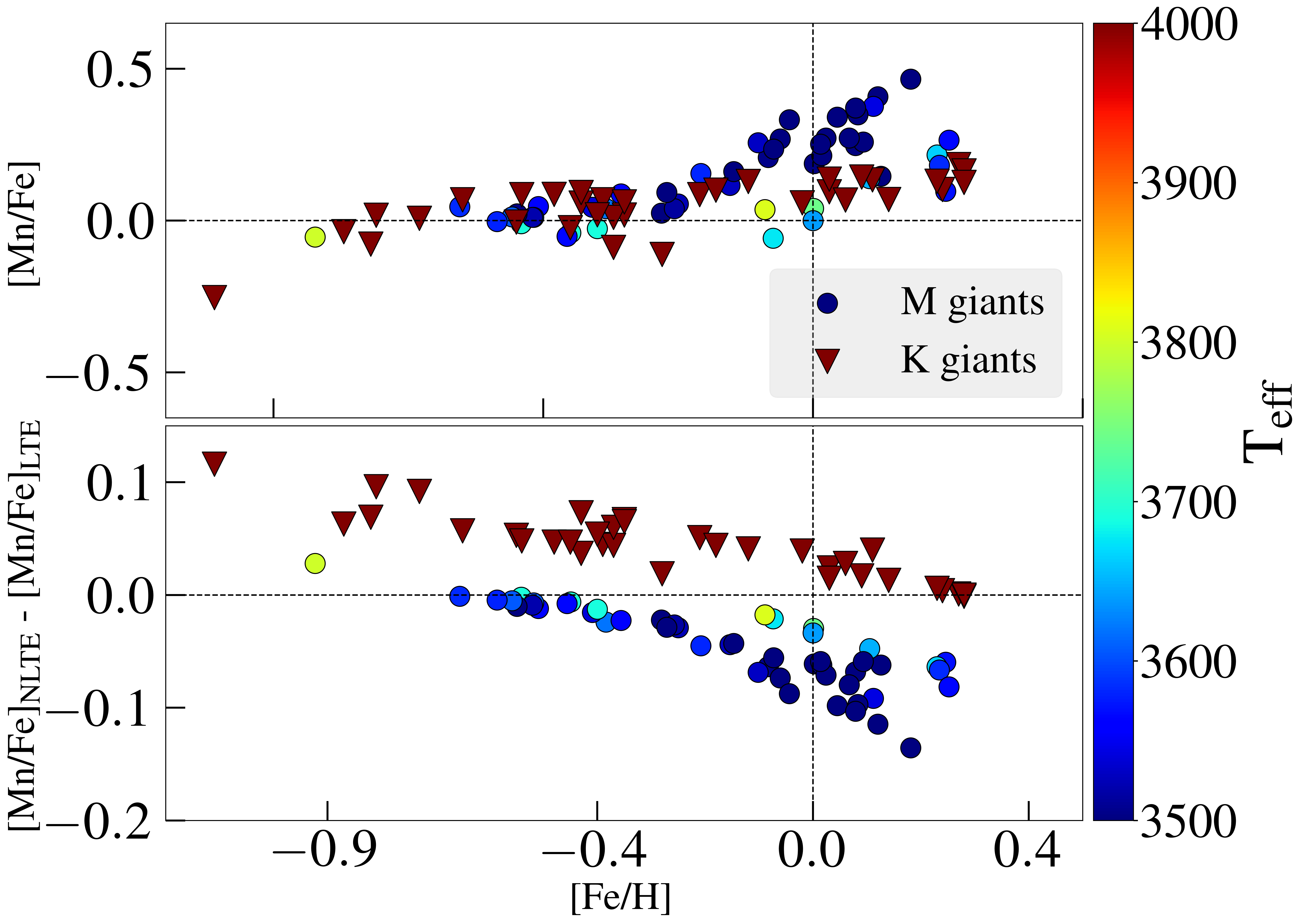}
  \caption{ Upper panel: [Mn/Fe] versus [Fe/H] for solar neighbourhood M-giants in this work (filled circles) and for the warmer K-giants (\teff$>4000$\,K) analysed in \citet{montelius:21} (inverted triangles), color coded in \teff\, including the NLTE corrections and determined from the same H-band lines. Bottom panel: NLTE corrections (NLTE-LTE) to manganese abundances based on the calculations by \cite{NLTE} for M-giants in this work and K-giants analysed in \citet{montelius:21} indicated with the same markers as in upper panel. The NLTE corrections for the abundance derived from these lines depend on the effective temperatures of the stars.}
  \label{fig:mn_trend_teff}%
\end{figure}

\begin{figure}
  \includegraphics[width=\columnwidth]{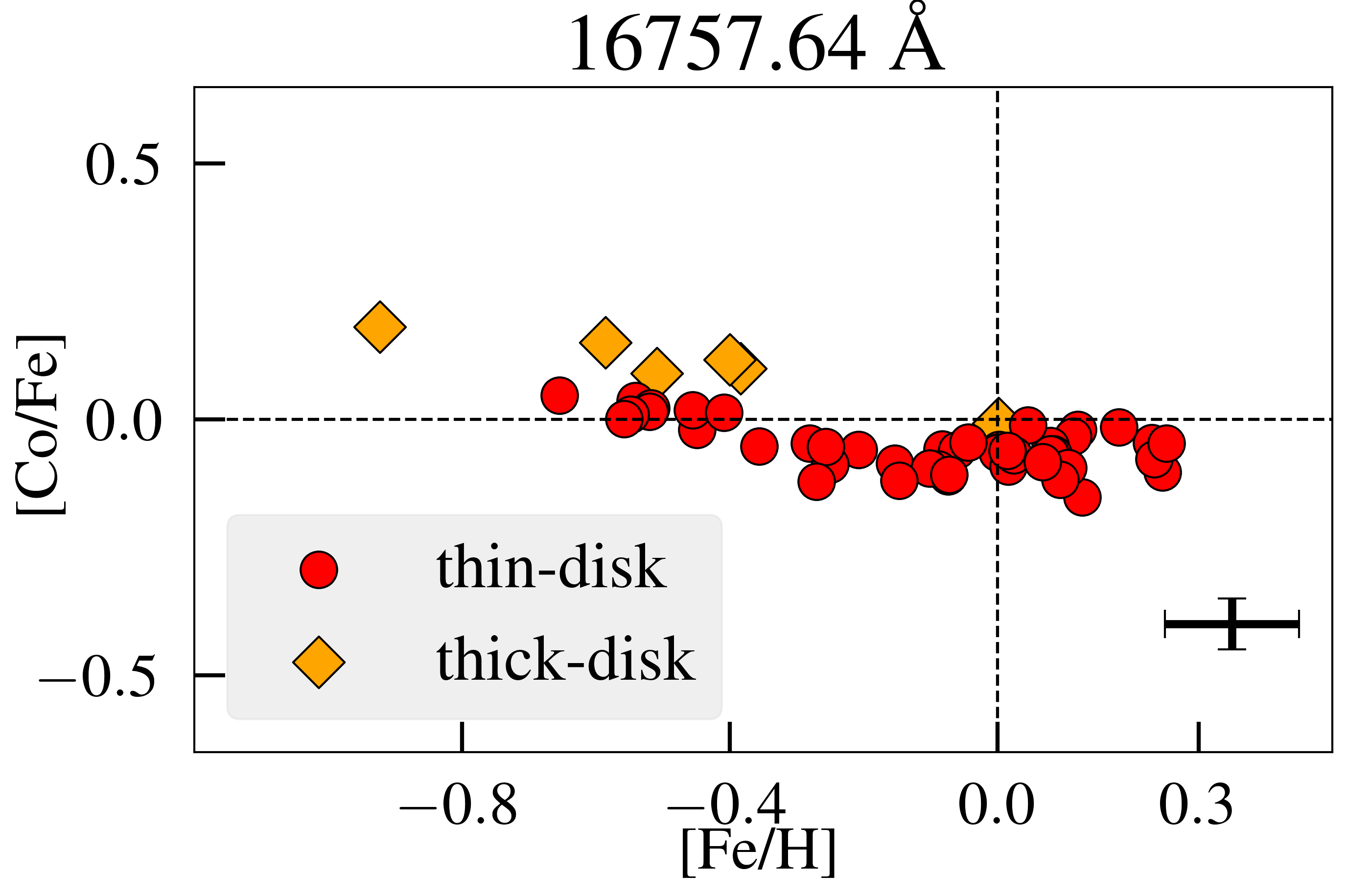}
  \caption{ [Co/Fe] versus [Fe/H] for 50 M giants in our sample. Arrangement of figures and markers are similar to Figure~\ref{fig:f_trend}. }
  \label{fig:co_trend}%
\end{figure}

\begin{figure*}
  \includegraphics[width=\textwidth]{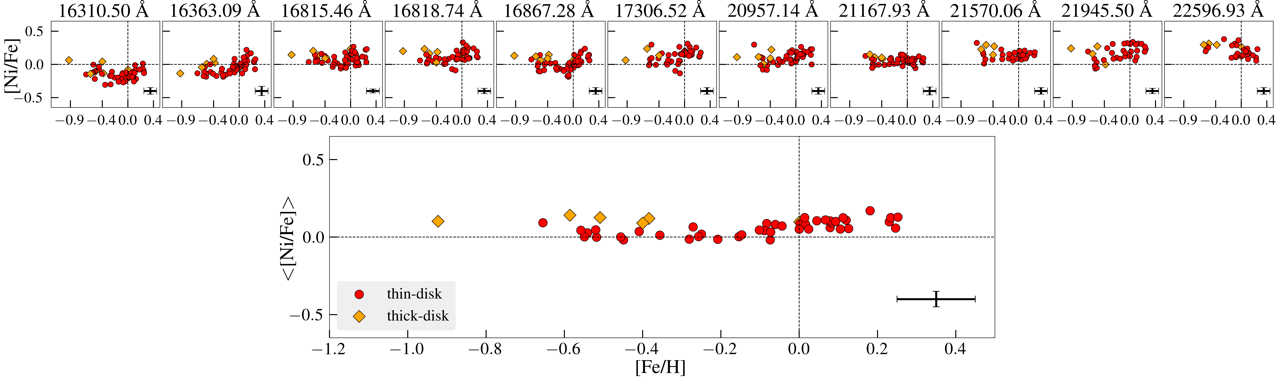}
  \caption{ [Ni/Fe] versus [Fe/H] for 50 M giants in our sample. Arrangement of figures and markers are similar to Figure~\ref{fig:f_trend}. }
  \label{fig:ni_trend}%
\end{figure*}

\begin{figure}
  \includegraphics[width=\columnwidth]{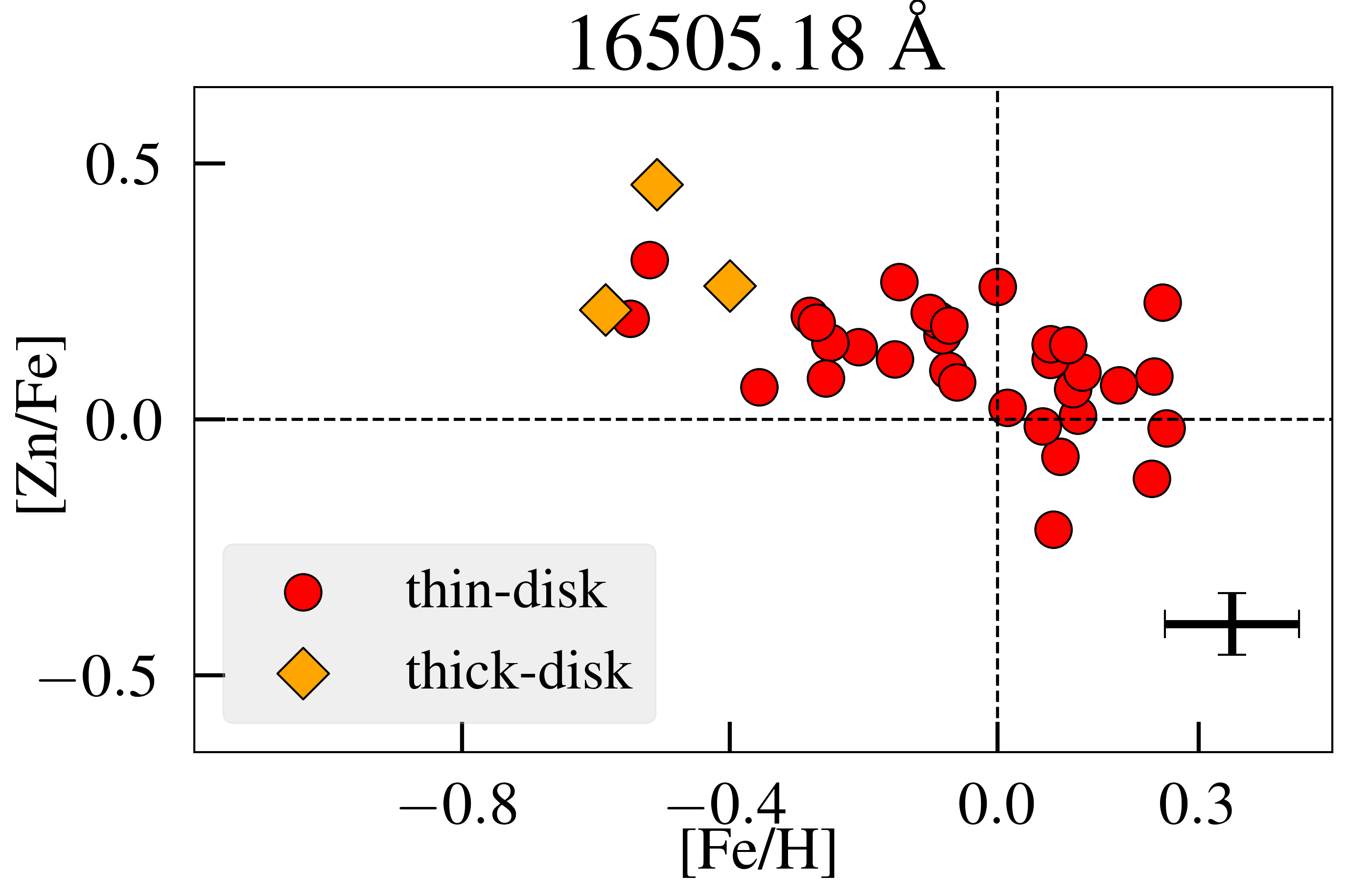}
  \caption{ [Zn/Fe] versus [Fe/H] for 35 M giants in our sample. Arrangement of figures and markers are similar to Figure~\ref{fig:f_trend}. }
  \label{fig:zn_trend}%
\end{figure}

\begin{figure}
  \includegraphics[width=\columnwidth]{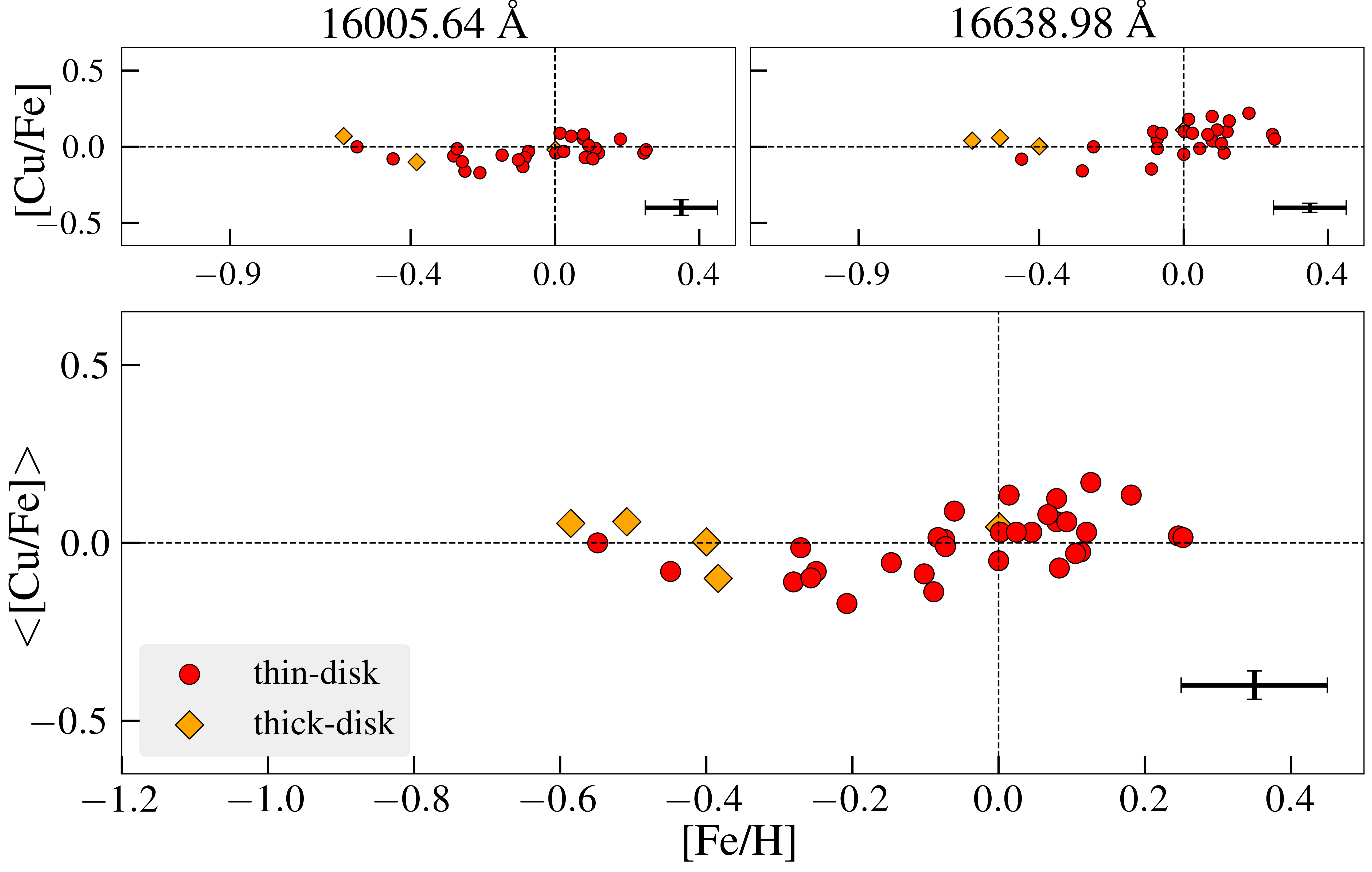}
  \caption{ [Cu/Fe] versus [Fe/H] for 36 M giants in our sample. Arrangement of figures and markers are similar to Figure~\ref{fig:f_trend}. }
  \label{fig:cu_trend}%
\end{figure}

\subsection{$\alpha$ elements (Mg, Si, S, and Ca)}
\label{sec:alpha}

{\bf Magnesium, silicon, sulphur, and calcium} The $\alpha$ elements are the most commonly studied group of elements with very well-investigated abundance trends for different Milky Way stellar populations. We have determined the abundances of magnesium (Mg), silicon (Si), sulphur (S) and calcium (Ca) from multiple lines in the H- and K-bands, see Figure~\ref{fig:alphaspectra}. For Mg and Si we determined abundances from the same lines as used in \citetalias{Nandakumar:2023}. We also used NLTE corrections for Mg, Si, and Ca lines.

For sulphur we used one H- and K-band line each to determine its abundances and for calcium we used six more lines than used in \citetalias{Nandakumar:2023} in the K-band, resulting in a total of 11 Ca-lines. The abundance trends for Mg, Si, S, and Ca are shown in the Figures~\ref{fig:mg_trend}, \ref{fig:si_trend}, \ref{fig:s_trend}, and \ref{fig:ca_trend}, respectively, with the trends from each individual line in the top panels and the mean abundance estimated from all lines in the bottom panel. The solar neighbourhood thin- and thick-disk stars in each figure are shown by red filled circles and orange diamonds, respectively. Except in the case of Mg, for which only K-band lines are used, both H- and K-band lines have been used to determine the abundances for rest of the $\alpha$ elements.

There is a clear enhancement in the mean abundances of all $\alpha$ elements (not as clear for the  sulphur abundances though) for the solar neighbourhood thick-disk stars leading to a separation between the thin and thick-disk populations. This is also seen in optical analyses for Mg, Si, and Ca \citep[see, for instance,][]{jonsson:17}, but less so for S \citep[see, e.g.,][]{costa:20,Perdigon:2021}. This separation is more evident in the metal-poor regime (\feh\,$< -0.4$\,dex) where five out of the six thick-disk stars lie. Except for Si, there is no clear enhancement seen for the thick-disk star with solar metallicity. As mentioned in \citetalias{Nandakumar:2023}, this is possibly due the fact that the Si-abundance was only determined from one good line, 16434.93 \AA. 

We have chosen to list Ti among the iron-peak elements \citep[see also][]{sneden:16}, and not as an $\alpha$ element, since the nucleosynthetic formation-channel of the main isotope, $^{48}\mathrm{Ti}$, is through the decay of the radioactive $^{48}\mathrm{Cr}$ isotope in Si-burning zones in core-collapse supernova \citep{curtis:2019}. It can, therefore, be considered primary as an iron-peak element. Cr is created in SNe Ia and SNe II in comparable amounts \citep{clayton:2003,lomaeva:19}.

\subsection{Odd-Z elements (Na, Al, and K)}
\label{sec:oddz}
Among the odd-Z elements, we determined abundances of sodium (Na), aluminium (Al), and potassium (K). For the determination of the Na and Al abundances, we used lines from both the H- and K-bands, while only two H-band lines are used to determine K abundances, see Figure~\ref{fig:oddzspectra}. 

{\bf Sodium} The sodium abundances from each line show an increasing [Na/Fe] trend with increase in metallicity, see Figure~\ref{fig:na_trend}. For sodium, we use NLTE corrections. At sub-solar metallicities, the Na abundances determined from the two H-band lines have super-solar values, whereas the lines in the K-band have sub-solar values resulting in larger positive slopes. The trend found for local dwarfs by \citet{bensby:2014} and for local giants by GILD (see Figure~\ref{fig:all_gild_trend}), both without NLTE corrections, are consistent with the trend determined using the two H-band lines. 

This difference between the H- and K-band lines abundances trends can be attributed to the NLTE corrections to the three K-band line abundances from \cite{NLTE}. Figure~\ref{fig:na_nlte_lte} shows the NLTE-LTE differences in the [Na/Fe] abundances determined from each line for all stars. The NLTE corrections to the H-band line abundances are very low ($\lesssim - 0.1$\,dex) for all [Fe/H]. Meanwhile, there is a large variation in the NLTE corrections to the K-band lines that range from $\sim-0.4$\,dex to $\sim -0.1 $\,dex. Hence, one plausible explanation for the sub-solar [Na/Fe] values determined from the K-band lines at sub-solar metallicities is that the NLTE corrections are too large for these metallicities. Abundances determined from the different lines should ideally yield the same abundance trends. In their work involving the characterisation of two low-latitude star cluster candidates towards the Galactic bulge with IGRINS, \cite{Lim:2022} found significant and systematic discrepancies between some abundance ratios measured from H- and K-band spectra, particularly for Al, Ca, and Ti. They identified NLTE effects as one of the possible reasons (in addition to the effects of atmosphere parameters and atomic parameters) for this, and also found that abundance differences are reduced by NLTE corrections but not eliminated completely.

Thus, we exclude Na abundances from the three K-band lines while determining the mean [Na/Fe] abundances in the  bottom panel of Figure~\ref{fig:na_trend}. The final, mean-[Na/Fe] trend resembles the trend found for local dwarfs by \citet{bensby:2014} with a stretched out N-shape. The trend for the metal-poor, thick-disk stars shows a hint of a separate trend compared to that of the thin-disk stars, following the upper envelope of the thin-disk trend. Furthermore, the thick-disk trend seems to turn down for decreasing metallicities, which is also expected from chemical-evolution models \citep[see, for example,][]{kobayashi:20}.



{\bf Aluminum} We have determined Al abundances from seven lines, four of them in the H-band and three in the K-band, all with NLTE corrections. There are differences in the abundance trends from each line in terms of the determined values and the magnitude of scatter, see Figure~\ref{fig:al_trend}. The mean [Al/Fe] values (bottom panel) determined from all the lines, though, show a clear difference between the thin- and thick-disk stars. The mean [Al/Fe] of the solar-neighbourhood, thin-disk stars are generally near solar (in the super-solar regime), a bit lower compared to the clear increasing slope for decreasing metallicities found in local dwarfs \citep{bensby:2014}. But we see a clear slope in the  trends from individual Al lines. The solar-neighbourhood, thick-disk stars are enhanced by $\sim 0.2$\,dex, and decreases as metallicity decreases, which is also expected from chemical-evolution models \citep[see, for example,][]{kobayashi:20}. 


{\bf Potassium} To determine the [K/Fe] abundances, two neighboring H-band lines separated by $\sim$ 5\,\AA\, were used, see Figure~\ref{fig:k_trend}. We use NLTE corrections for potassium. For the metal-poor, thick-disk stars there is a clear enhancement in [K/Fe] determined from both the lines, which is not found in the study by \citet{takeda:20}, but is in APOGEE for K giants \citep[see, e.g.,][]{friden:23}. The mean [K/Fe] values for the thin-disk stars are slightly sub-solar and show a slightly declining trend, as expected form optical observations and models \citep{takeda:02,prantzos:18,takeda:20}. There are not many studies on the Galactic evolution of K in the Milky Way disks. 

\subsection{Iron-peak elements (Sc, Ti, V, Cr, Mn, Co, and Ni)}
\label{sec:ironpeak}
We have determined abundances of all seven iron-peak elements: scandium (Sc), titanium (Ti), vanadium (V), chromium (Cr), manganese (Mn), cobalt (Co), and nickel (Ni). While both H- and K-band lines have been used to determine the titanium and nickel abundances, only H-band lines have been used to determine abundances for the other iron-peak elements, apart from Sc whose abundance is determined from lines only in the K-band (see Figure~\ref{fig:ironpeakspectra}). 

{\bf Scandium} All the three Sc lines show a similar trend, see Figure~\ref{fig:sc_trend}. The mean [Sc/Fe] trend shows a hint of a generally decreasing slope with metallicity, with the thick-disk stars showing higher Sc abundances than those of the thin-disk. This trend is similar to the trend for disk dwarfs and giants in \citet{battistini:2015} and \citet{lomaeva:19}, respectively.
We note that the scatter in our data for super-solar metallicities are larger than in the optical studies. One plausible explanation for this scatter is that the K-band Sc{\sc i} lines are found to be very \teff-sensitive, especially at \teff\,$<$ 4000 K , \citep[see, e.g.,][]{thorsbro:2018}, which means that we should expect a larger scatter, just like for fluorine. 


{\bf Titanium} For Ti, we determined abundances from the same two lines used in \citetalias{Nandakumar:2023}, one in the H and one in the K-band, with the former showing the tightest trend, see Figure~\ref{fig:ti_trend}. The thick-disk is clearly separated from the thin-disk 
resembling the dwarf-star trends in, for instance, \citet{bensby:2014} or the giant-star trends in, for instance, \citet{jonsson:17,lomaeva:19}.


{\bf Vanadium} We determined the [V/Fe] abundances from one H-band line. Thin-disk stars have sub-solar values and metal-poor thick-disk stars have enhanced super-solar values, see Figure~\ref{fig:v_trend}. Both the dwarf-star trend in \citet{battistini:2015} and giant-star trend in \citet{lomaeva:19} from optical analyses show similar enhancement for the thick-disk stars in their sample, although their trend is higher by 0.1-0.2 dex. 

{\bf Chromium} The [Cr/Fe] trend is flat. The Cr abundances determined from the two lines at bluer wavelengths in the H-band are sub-solar for both thin- and thick-disk stars and show a tight trend, see Figure~\ref{fig:cr_trend}. The [Cr/Fe] abundances determined from the reddest wavelength line is more uncertain since this wavelength region has some significant telluric lines affecting the quality of the spectra. The trend derived from this line is largely solar for metal-poor stars and super-solar at high metallicities with an increased scatter, see Figure~\ref{fig:cr_trend}. The mean [Cr/Fe] values are, however, mainly sub-solar at all metallicities with the slightly larger scatter for metal-rich stars mainly caused by this reddest line. Overall, this shows an expected flat trend for Cr. The abundance differences between the thin- and thick-disk stars is small, although the metal-poor, thick-disk stars show a consistent enhancement in [Cr/Fe] compared to the thin-disk stars from all three lines. This is very similar to the finding from the optical analysis for giants in e.g. \citet{lomaeva:19}.


{\bf Manganese} Manganese is of special interest since its abundance trends depend sensitively on the properties and type of the Type-Ia-supernovae progenitors \citep{reyes:20}, and there are indications of metallicity-dependent SNe II yields \citep{Woosley:1995}. However, its chemical-evolution trends are debated; A range of optical and near-infrared trends as well as Galactic chemical evolution models show an increasing trend with metallicity \citep{battistini:2015,Zasowski:2019,lomaeva:19,kobayashi:11,vasini:2023}. On the other hand, \citet{bergemann:08} found that NLTE abundances are larger, especially for the most metal-poor stars, with correction up to 0.5\,dex. This would lead to an essentially flat trend with metallicity \citep[see also][]{battistini:2015,lomaeva:19}, with the yields being less metallicity dependent. The question is whether the chemical evolution models or the NLTE corrections need modifications. 

In the analysis of K giants (\teff > 4000\,K) by \citet{montelius:21}, Mn-abundances was determined from high-resolution IGRINS spectra using the two Mn-lines at 15218 and 15262\,\AA, including the important hyperfine-structure components. He found a considerably flatter trend, especially at subsolar metallicities, than that found earlier. He included NLTE corrections based on the calculations by \cite{NLTE}, which are shown by the inverted dark-red triangles in the bottom panel of Figure~\ref{fig:mn_trend_teff}. They are positive for all the stars with \teff$>4000$\,K. In the earlier optical studies, the NLTE corrections are also positive, flattening the positive slope at low metallicites \citep{bergemann:19}. \citet{friden:23} confirms the flat trend when comparing a set of stars analysed both by APOGEE and IGRINS; he does not find the increasing Mn abundances for increasing metallicity as provided by APOGEE. 

Here, we analyse cooler stars (\teff$<4000$\,K) observed with the same high spectral-resolution IGRINS spectrometer and determined manganese abundances from the same two lines including the NLTE corrections from \cite{NLTE}. In Figure~\ref{fig:mn_trend}, our mean [Mn/Fe] trend confirms the flat trend at \feh\,$<-0.2$, and here the NLTE corrections are negligible for the coolest stars (see filled circles in the bottom panel of Figure~\ref{fig:mn_trend_teff}), and the lines can, therefore, be assumed to have formed in LTE conditions. For higher metallicities we see a large scatter and an increasing trend. In the upper panel of Figure~\ref{fig:mn_trend_teff}, we plot the [Mn/Fe] abundances determined here and for the warmer stars (\teff$>4000$\,K) analysed in \citet{montelius:21}, color coded in \teff\, with circles and triangles, respectively, including the NLTE corrections. The trend is flat for all stars when including these corrections, except for the coolest, metal-rich stars. 

For the warmer stars of \citet{montelius:21}, and for the warmer of our cool stars (\teff$>3550-4000$\,K), there is no temperature dependence of the derived abundances, and these show a flat trend also at high metallicities. This is expected since the abundance should not depend on the temperature of the stars. However, for the coolest stars, there is a clear and unwanted \teff-dependence of the overabundance of Mn for super-solar metallicities. This shows that we can not trust the Mn abundances for the coolest stars for super-solar metallicities. This could be caused by the temperature-dependence of the NLTE corrections being too small. The NLTE corrections are negligible for the metal-poor stars, but for the more metal-rich stars the corrections are larger and lead to lowering the derived abundances, see the lower panel in Figure~\ref{fig:mn_trend_teff}. This would mean that the NLTE correction would need to be larger in order to lower the [Mn/Fe] abundances down to the trend shown by the warmer stars. Another reason could be general difficulties in the analysis of the coolest stars. For these cool, metal-rich stars, the Mn lines get increasingly stronger with the risk of getting saturated, which means that the uncertain microturbulence will affect the abundance determination more than wanted. The wavelength region is, however, quite free from telluric contamination.  

Thus, the abundance analyses of the Mn trends from high-resolution, near-infrared stellar spectra of cool metal-poor stars, based on lines that are formed, more or less, in LTE conditions, indicate that the flat [Mn/Fe] versus [Fe/H]-trend presumably is correct and that the nucleosynthesis of Mn, therefore, is empirically more like a normal iron-peak element.
The flat trend is strengthened by the other studies based on optical and near-infrared lines, however, requiring NLTE corrections \citep[see, e.g.,][]{bensby:2014}. Since the metallicity-dependent NLTE corrections are different for different lines (optical and near-infrared) and very different for different \teff\, of the stars, see Figure~\ref{fig:mn_trend_teff}, this indicates that the NLTE corrections are correct for optical and near-infrared spectral lines (but not for the coolest and most metal-rich stars) and that the cosmic nucleosynthesis of Mn is, thus, still uncertain. 

{\bf Cobalt} Only one line has been used to determine the [Co/Fe] abundances, from which we obtain a very tight trend with very low scatter, see Figure~\ref{fig:co_trend}. The trend for the thin-disk stars starts with super-solar values at lowest metallicities that transitions to sub-solar values at [Fe/H]$\sim -0.4$\,dex and continues at sub-solar values as the metallicity increases. All five metal-poor, thick-disk stars have enhanced [Co/Fe] values, and the solar metallicity thick-disk star has a solar [Co/Fe] value that is closer to the thin-disk stars.  The thin- and thick-disk trends are clearly separated. These trends follow the dwarf-star trends found in \citet{battistini:2015}, although our trends are systematically lower by $\sim0.1$~dex. The trend is also very similar to the giant trend from the optical analysis by \citet{lomaeva:19}, although their trend is also higher by $0.1-0.2$\,dex.

{\bf Nickel} To determine the [Ni/Fe] abundances, we used six lines in the H-band and five lines in the K-band, and there are significant differences in the measured abundances and dispersion or scatter from each line, see Figure~\ref{fig:ni_trend}. The mean [Ni/Fe] values for the thin-disk stars, though, show a tight trend that increases slightly with increasing metallicity. This is very similar to the dwarf-star trend in \citet{bensby:2014} and giant trend in \citet{lomaeva:19}. The thick-disk stars show enhancements with super-solar values. 

\begin{figure}
  \includegraphics[width=\columnwidth]{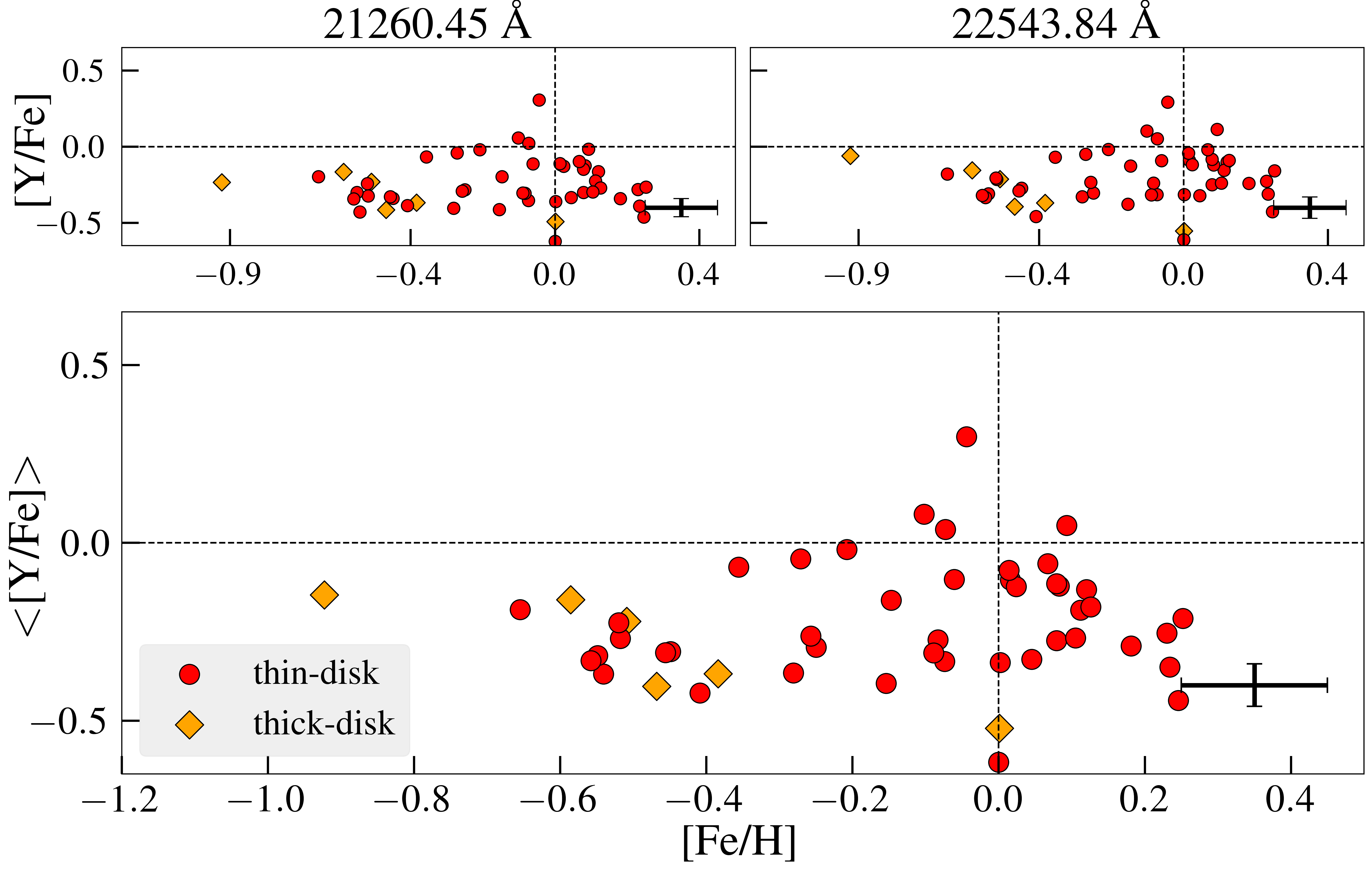}
  \caption{ [Y/Fe] versus [Fe/H] for 50 M giants in our sample. Arrangement of figures and markers are similar to Figure~\ref{fig:f_trend}. }
  \label{fig:y_trend}%
\end{figure}

\begin{figure*}
  \includegraphics[width=\textwidth]{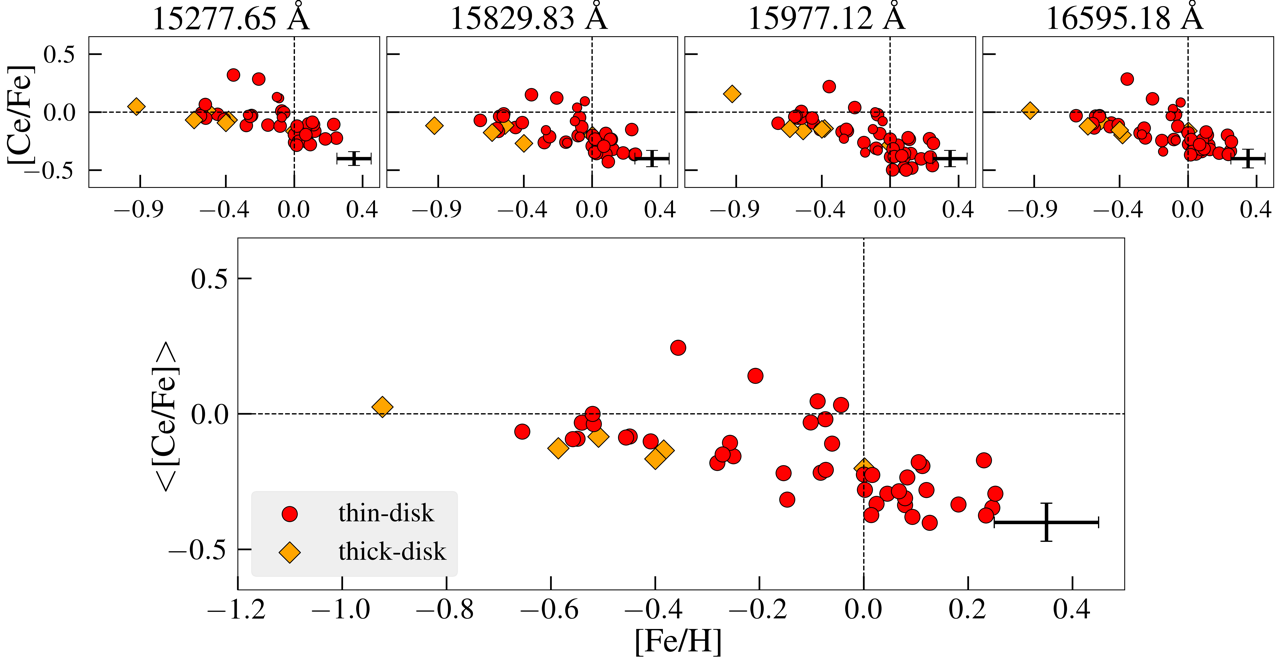}
  \caption{ [Ce/Fe] versus [Fe/H] for 50 M giants in our sample. Arrangement of figures and markers are similar to Figure~\ref{fig:f_trend}. }
  \label{fig:ce_trend}%
\end{figure*}

\begin{figure*}
  \includegraphics[width=\textwidth]{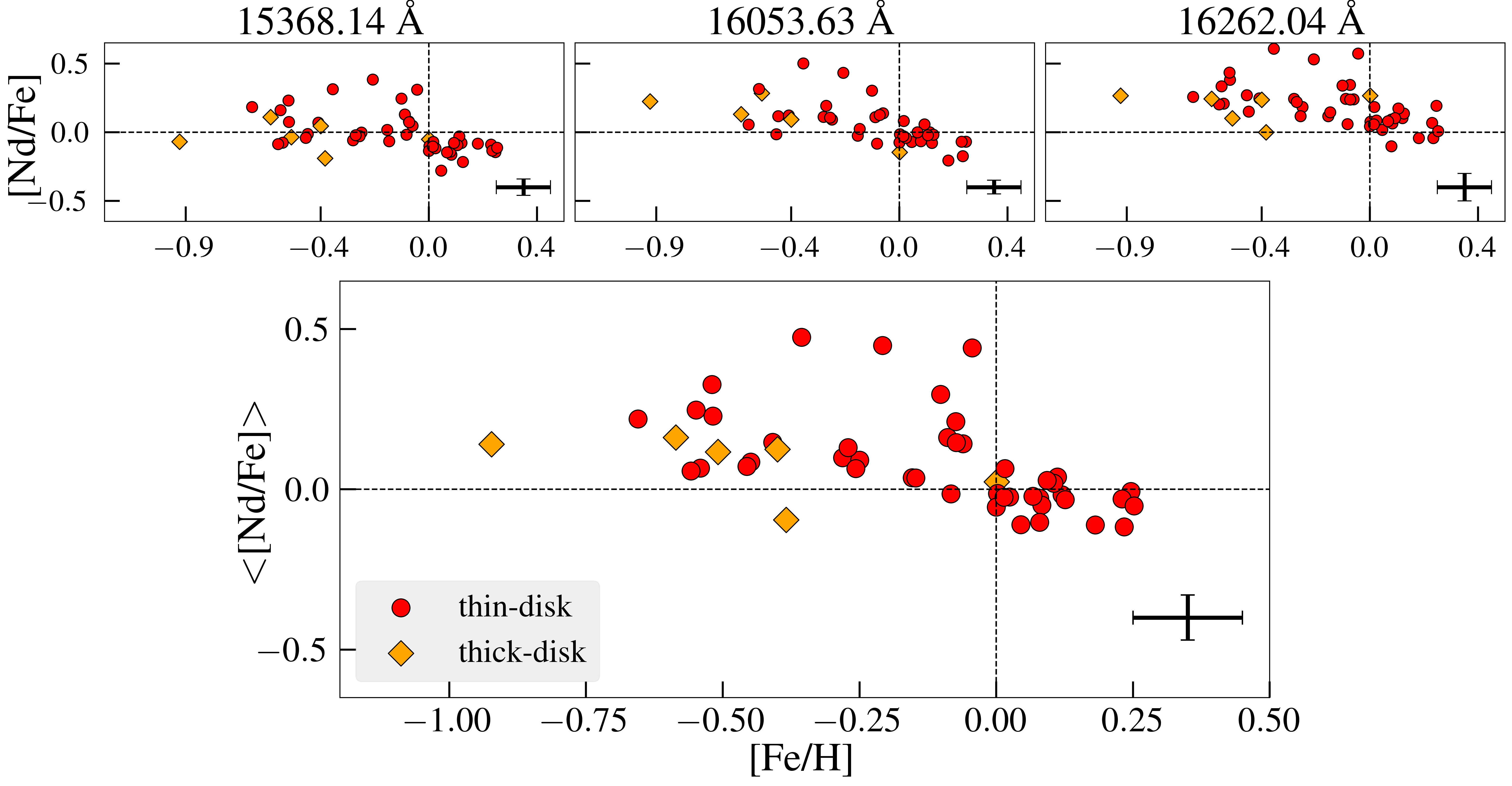}
  \caption{ [Nd/Fe] versus [Fe/H] for 50 M giants in our sample. Arrangement of figures and markers are similar to Figure~\ref{fig:f_trend}. }
  \label{fig:nd_trend}%
\end{figure*}

\begin{figure}
  \includegraphics[width=\columnwidth]{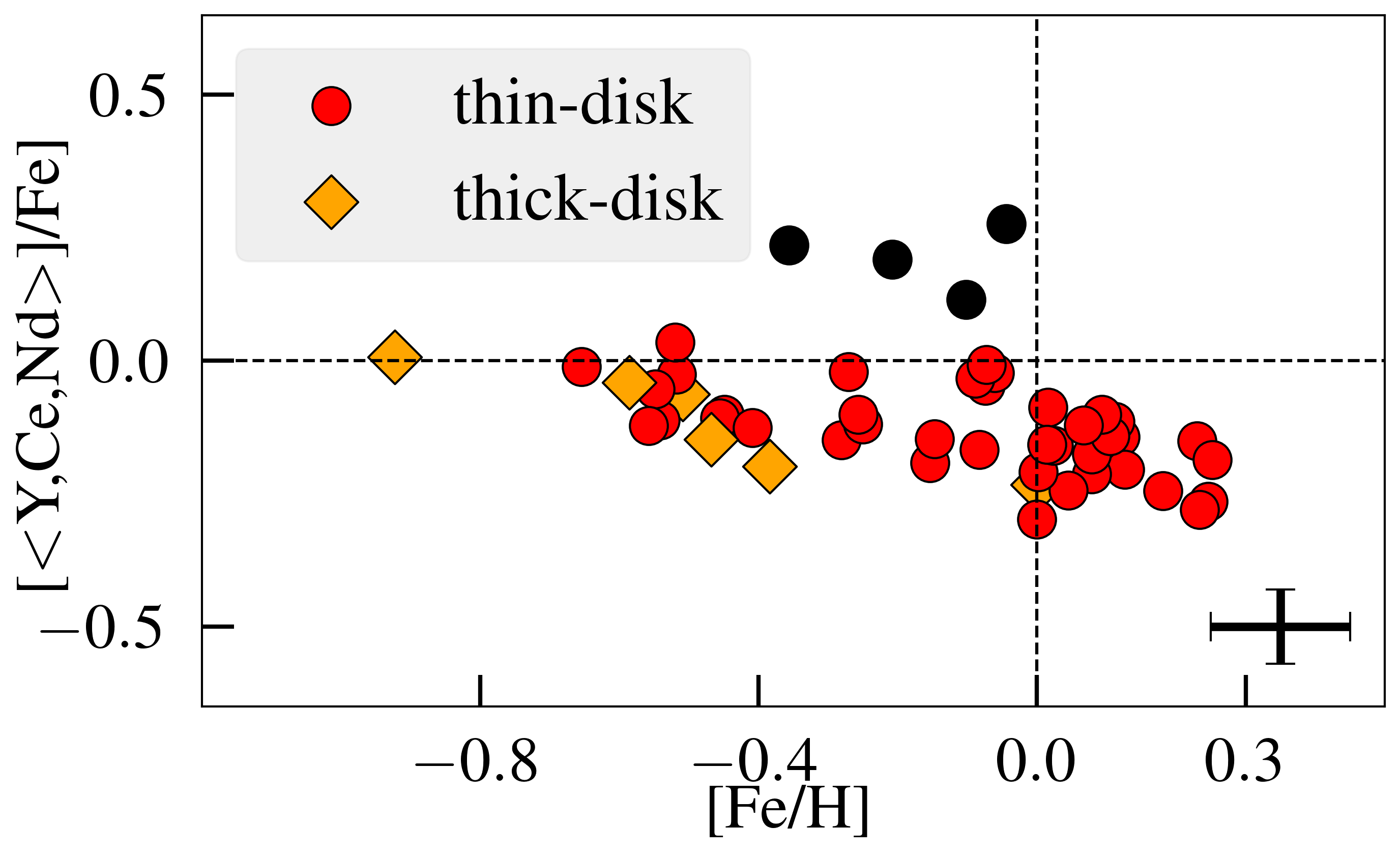}
  \caption{ Mean abundances of the three s-process elements, Y, Ce, and Nd, versus [Fe/H] for 50 M giants in our sample. Black filled circles represent the stars that show high abundances of all these s-process elements at the same time. Remaining markers are similar to Figure~\ref{fig:f_trend}.  }
  \label{fig:sproc_mean_trend}%
\end{figure}

\begin{figure}
  \includegraphics[width=\columnwidth]{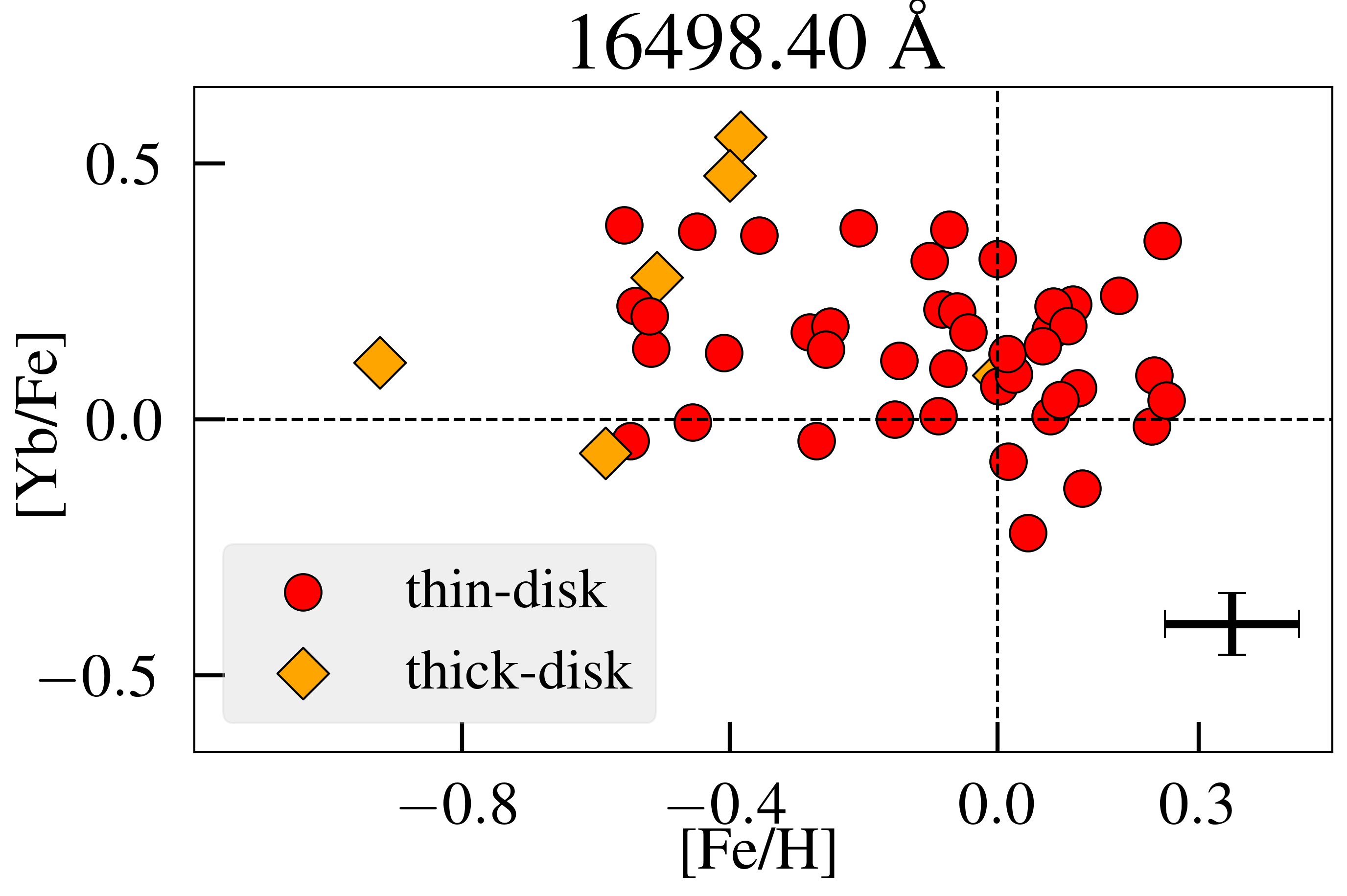}
  \caption{ [Yb/Fe] versus [Fe/H] for 49 M giants in our sample. Arrangement of figures and markers are similar to Figure~\ref{fig:f_trend}. }
  \label{fig:yb_trend}%
\end{figure}

\begin{figure*}
  \includegraphics[width=\textwidth]{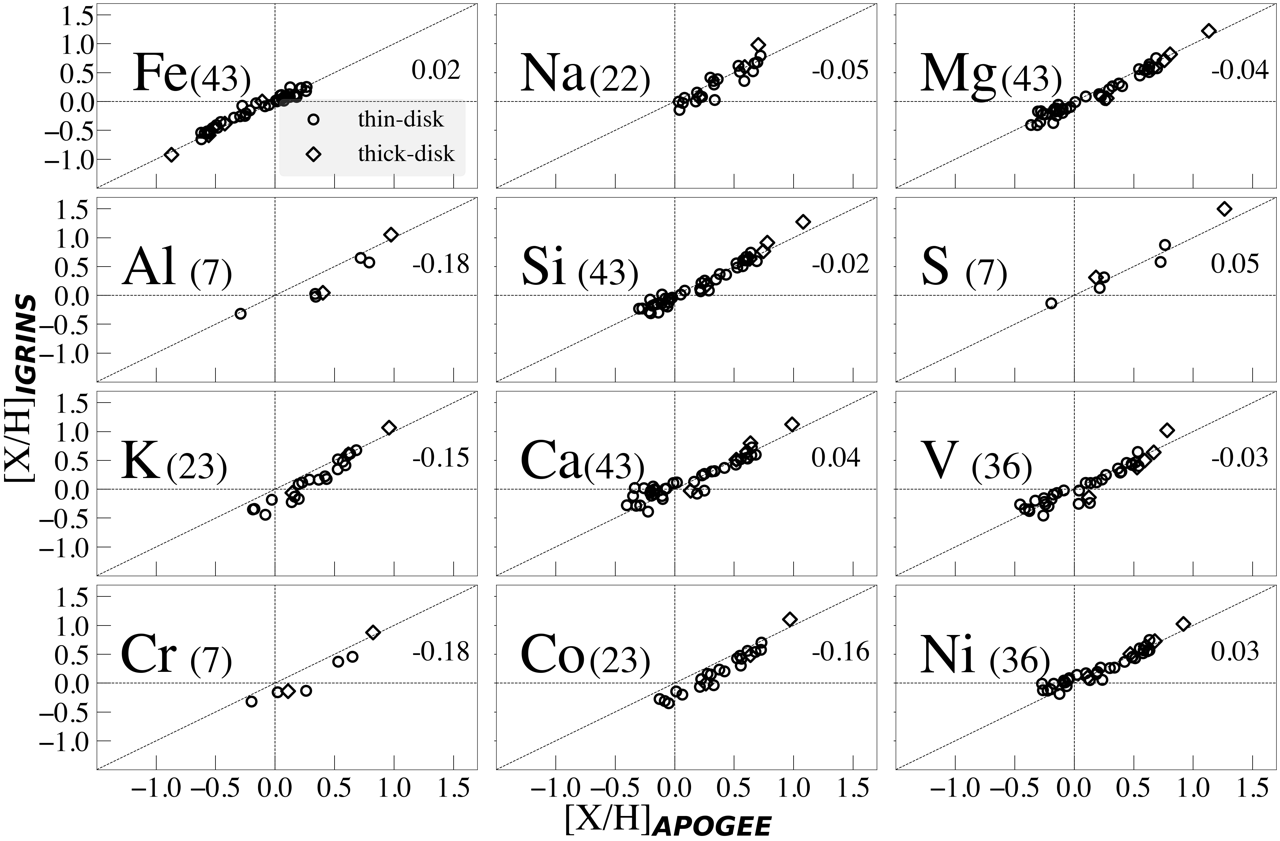}
  \caption{ Elemental abundances with respect to hydrogen ([X/H]) including metallicity and 11 elements determined from IGRINS spectra versus those in APOGEE DR17 catalog for stars in common between APOGEE and this work. The black open circles and black open diamonds represent stars chemically classified as thin- and thick-disk respectively, in our sample. Number of stars with valid APOGEE abundance estimates for each element are listed within the brackets in each panel. The linear one-to-one relation is shown by the black dashed line and the mean difference values in metallicity and elemental abundances between this work and APOGEE are listed in the top right part in each panel.  }
  \label{fig:apogee_trend}%
\end{figure*}

\subsection{Transition elements between iron-peak and neutron-capture elements (Zn)}

The elements after Ni transition to the weak-s dominated elements \citep[nuclides with mass numbers up to A = 90;][]{prantzos:18}. There are strong observational indications that Cu has a mainly weak-s origin \citep[e.g.][]{delgado:17,heitor:20}. The origin of Zn is, however, complex and uncertain. According to \citet{clayton:2003} it is, indeed, synthesized by the weak s-process, but also through radioactive decay from products synthesised in explosive nucleosynthesis in core-collapse supernovae through alpha-rich freeze out. 

{\bf Zinc} We measure the zinc abundance from only one H-band line, which makes it less certain. The abundance trend resembles an $\alpha$ element, decreasing with metallicity, see Figure \ref{fig:zn_trend}. The thick-disk trend  is marginally higher than that of the thin disk. It is obvious that the Zn trend is not similar to the trend of Cu, which is, as mentioned, mainly synthesised by weak-s process, see Section \ref{sec:weak}. This is interesting since the weak-s process also contribute to the origin of Zn to a certain degree. Although similar to the trend of \citet{daSilveira:18}, our trend shows a larger negative slope as the metallicity grows than other optical studies like \citet{mishenina:02,bensby:2014,delgado:17}. All these studies, however, show a flatter trend than the metallicity-dependent, secondary behaviour (weak-s), trend that Cu shows. For example, \citet{mishenina:02} clearly demonstrate in a [Cu/Zn] trend, that the [Cu/Fe] trend is more metallicity dependent than that of [Zn/Fe]. This shows a clear empirical difference between the cosmic origin of Cu and Zn.

\subsection{Neutron-capture elements (Cu, Y, Ce, Nd, and Yb)}
We have determined the abundances of copper (Cu), yttrium (Y), cerium (Ce), neodymium (Nd), and ytterbium (Yb), which are all mainly synthesised by neutron-capture processes. Cu is synthesised primarily in the weak-s process in massive stars as a secondary element \citep[e.g.][]{pagin:10,heitor:20,Cu_ges:21}.  However, to some extent, it is also synthesised in explosive nucleosynthesis in the inner shells of core-collapse supernovae through alpha-rich freeze out, as a primary element 
\citep[e.g.][]{pagin:10}. 

Most neutron-capture elements are produced by a combination of the s- and r-processes. For Y and Ce, the main s-process dominates with s/r=70/30 [in \%] and 85/15, respectively, in the solar isotopic mixture \citep{bisterzo:14,prantzos:20}. Neodymium has a s/r=60/40 ratio and can therefore still be considered as a predominantly s-process element. On the other hand, ytterbium has a ratio close to 50/50 \citep{bisterzo:14,prantzos:20,kobayashi:20}, which makes it the element with the highest contribution from the r-process in its origin among the elements presented in this study. The more the r-process contributes the more the abundance trends will resemble and reveal an r-process contribution.

While only K-band lines were used in the case of yttrium, only H-band lines were used for the copper, cerium, neodymium, and ytterbium abundance determinations. Figure~\ref{fig:neutroncapturespectra} shows the H- and K-band lines of neutron-capture elements used in this study.

\subsubsection{Weak s-elements (Cu)}
\label{sec:weak}

{\bf Copper} Both the lines used to determine the copper abundance are weak, and the long-wavelength wing of the bluest line is somewhat blended with an Fe I line-wing, see Figure~\ref{fig:neutroncapturespectra}. As a result, reliable Cu abundances could be determined for fewer stars, see Figure~\ref{fig:cu_trend}. 
The mean [Cu/Fe]-abundances show a wave-like trend, starting off at slightly super-solar values at the lowest metallicities, to then decrease to sub-solar values at sub-solar metallicities, but then again to increase at super-solar metallicities. The thick-disk stars tend to lie above that of the thin-disk. These solar-neighbourhood trends nicely follow the expected trends as seen in optical studies both from dwarfs and giants, see, for e.g., \citet{delgado:17,Rebecca_phd}. 





\subsubsection{s-process dominated elements (Y, Ce, and Nd)}
\label{sec:neutroncapture}

{\bf Yttrium} The derived yttrium abundances from the two K-band lines show similar sub-solar values and show a similar scatter and trends as a function of metallicity, see Figure \ref{fig:y_trend}. The thick-disk stars have [Y/Fe] values that follow the lower envelope of the thin-disk trend. The scatter, or the broad distribution of [Y/Fe] abundances for a given metallicity for the thin-disk stars, as well as the fact that the thick-disk trend lies below the thin-disk 'cloud' is also seen in [Y/Fe] versus [Fe/H] trends from optical studies \citep[see a clear example in][]{taut:21,Rebecca_phd}. This is also expected from theoretical Galactic chemical evolution models of, e.g., Zr, which has a similar behaviour as Y \citep{Grisoni:2020}. The scatter in the thin-disk distribution is most likely real and therefore of cosmic origin and  not from measurement uncertainties. 

The normalization, in which the solar-neighbourhood trend for the thin-disk is expected to go through the solar abundance ratio at [Fe/H]=0, is comparable to the optical determined trend (see Figure~\ref{fig:all_gild_trend} in Section~\ref{sec:GILDcompare}). This indicates that the $\log gf$-values obtained from VALD might not be too bad for these yttrium lines and can be utilized in obtaining reliable abundances, see Section \ref{sec:Linelist}. 


{\bf Cerium} The [Ce/Fe] versus [Fe/H] trends from all four lines are similar, with a downward trend, i.e., the super-solar metallicity stars are found to have the lowest [Ce/Fe] values, see Figure \ref{fig:ce_trend}. Just like in the case of the trend for yttrium, the thick-disk stars follow the lower envelope of the thin-disk trend. The similar scatter or broad distribution in [Ce/Fe] abundances for a given metallicity, is also clearly seen in studies of optical lines for solar-neighbourhood giants \cite[see e.g.][]{battistini:2016,delgado:17,forsberg:19,taut:21}, and from near-infrared lines in the APOGEE Open Cluster Chemical Abundances and Mapping survey \citep{sales:22}. The normalization, in which the solar-neighbourhood trend for the thin-disk is expected to go through the solar-abundance ratio at [Fe/H]=0, is, however, too low for our near-infrared determination compared to the optical and point to that the $\log gf$-values might be too strong, leading to too low derived abundances of Ce from these lines.


{\bf Neodymium} The [Nd/Fe] abundances are determined from three H-band lines which show similar trends, see Figure \ref{fig:nd_trend}. The [Nd/Fe] determinations for the thick-disk stars follow the lower envelope of the thin-disk. These trends fit well in the overall picture of neutron-capture elements in the disk components found in \citet{taut:21,Rebecca_phd}, with the thick-disk following the lower envelope of the thin-disk 'cloud'. These features are also found in the cases of the s-process dominated elements Y and Ce, mentioned earlier. This points to a high contribution from the s-process in the cosmic origin of neodymium. The mean [Nd/Fe]-trend is, in general, similar to the mean [Ce/Fe] trend, but shifted along the y-axis. This might be a sign of a normalisation uncertainty, but we also note that the Nd trend is indeed generally higher compared to the Ce trend also in the optical GILD study, see Figure \ref{fig:all_gild_trend}.  


{\bf Mean s-process} In Figure \ref{fig:sproc_mean_trend}, we plot the mean abundances of the three s-process dominated elements, Y, Ce, and Nd, versus the metallicity. We can see that the three elements show star-by-star similarities. For example, the four stars at $-0.4<$\feh$<0.0$ marked in black in the figure, all show high abundances of these s-process elements at the same time (part of the broader distribution of s-processes abundances for a given metallicity, referred as the 'cloud' earlier). 

We show that these stars exhibit the same pattern, i.e. having relatively higher [s/Fe]-abundances for all of these three elements, obtained from different lines and different atomic data, to point at  the precision of the determined s-process abundances. Indeed, the thick-disk stars all follow the lower envelope of the thin disk for all these elements in the same way too. It is worthwhile to note that these four stars with relatively higher [s/Fe]-abundances are not abnormally high, as seen in Sect.~\ref{sec:GILDcompare} and Fig.~\ref{fig:all_gild_trend}, where the four stars are found within the s-process/cerium scatter "cloud". The point here is rather that they exhibit a similar pattern in [Y,Ce,Nd/Fe], as explained above.

\subsubsection{Element with $50/50$ s- and r-process contributions (Yb)} 
 
{\bf Ytterbium} The ytterbium abundance is measured from the singly-ionized ytterbium-line (Yb\,{\sc ii}) in the H-band at 16498.40 \AA. This line is severely blended with a CO molecular absorption line. In cool M-giants spectra, the molecular absorption lines get stronger, and a careful analysis of the Yb line has to be carried out in order to retrieve a reliable elemental abundances from such a blended line \citep[see discussion in][]{montelius:22}. We make sure that the close-by CO-lines from the same vibrational band are well fitted by the synthetic spectrum, while determining the Yb abundances (see for e.g. Figure~\ref{fig:neutroncapturespectra}). This gives us confidence in the [Yb/Fe] abundance we have determined from this difficult line. This is the first time that [Yb/Fe] has been determined from cool M-giants in the $-1.0$\,dex $<$ \feh\, $<  0.3$\,dex metallicity range. 

There is a large scatter ($\sim0.5$\,dex) in the [Yb/Fe] values. However, this agrees well with the trend and scatter derived by \citet{montelius:22} from warmer stars, which in general show weaker CO blending with the Yb line. Although there is a large scatter, the general picture that the thick-disk and the thin-disk [Yb/Fe]-values more or less overlap, is very similar to what \citet{Rebecca_phd} find for another 50/50 s/r-process element, namely praseodymium, Pr. \citet{Rebecca_phd} shows clearly in her study of 10 s- and r-process elements, with differing contributions of the two processes, that the thick-disk trend goes from lying below the thin-disk trend for a  'pure' s-process element (like Ce) to the thick-disk trend lying above the thin-disk trend for a 'pure' r-process element (like Eu). The thick-disk trends for the elements with more equal mixtures of the two origins, tend to lie between these extremes, which we observe here for Yb, reassuring us of the quality of the abundances.


\section{Discussion}
\label{sec:discussion}

\subsection{Comparison with other studies }
In the previous section, we presented detailed abundance trends of the 21 elements.  
Next, we need to compare and validate the abundance trends of these elements with the abundance trends from other studies. First we compare the abundance trends of the 44 of our stars that were also observed in APOGEE with the abundances provided in the APOGEE DR17 catalog \citep{ApogeeDr17}. We also compare our abundances trends for all 50 stars with the abundance trends of 6 elements with weaker and blended spectral features for $\sim$19 000 cool stars (\teff\,$<  4000$\,K) from APOGEE determined using the BACCHUS code in \cite{Hayes:2022}. Finally, we also compare with the abundance trends of $\sim$500 stars from the Giants in the Local Disk sample (GILD; J\"onsson et al. in prep., which builds upon and improves the analysis described in \citet{jonsson:17}) for which the stellar parameters and abundances have been determined from optical FIES spectra. For ytterbium, we compared our trend with the ytterbium abundances determined from the same line in the IGRINS H-band spectra of K giants in \cite{montelius:22}.


\begin{figure*}
  \includegraphics[width=\textwidth]{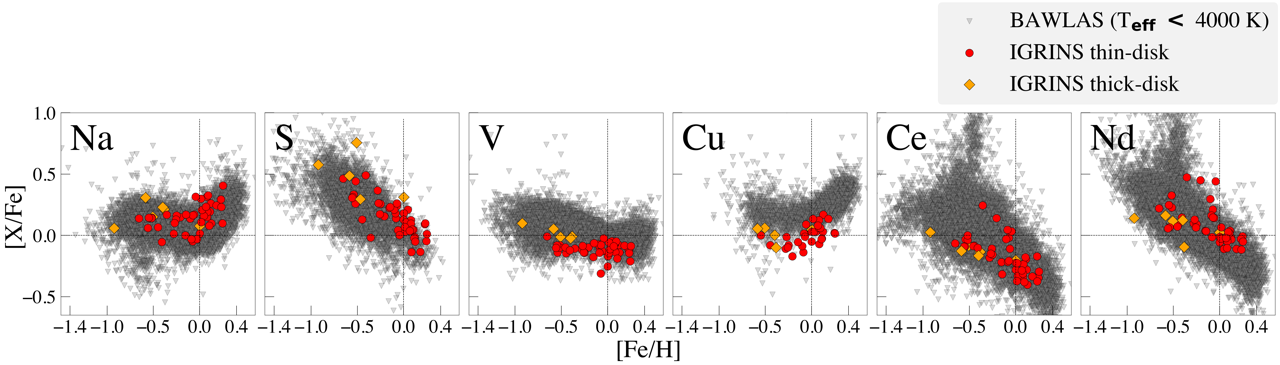}
  \caption{ Abundances trend versus metallicity of 6 elements determined for M giants in our sample (red filled circles and orange diamonds) and for $\sim$ 19 000 cool  (\teff\, $<$ 4000 K) giants (grey diamonds) from APOGEE DR17 reanalysed using the BACCHUS code \citep{Hayes:2022}.   }
  \label{fig:all_bawlas_trend}%
\end{figure*}

\subsubsection{Comparison with APOGEE solar neighborhood stars}
\label{sec:APOGEEcompare}


\begin{figure*}
  \includegraphics[width=\textwidth]{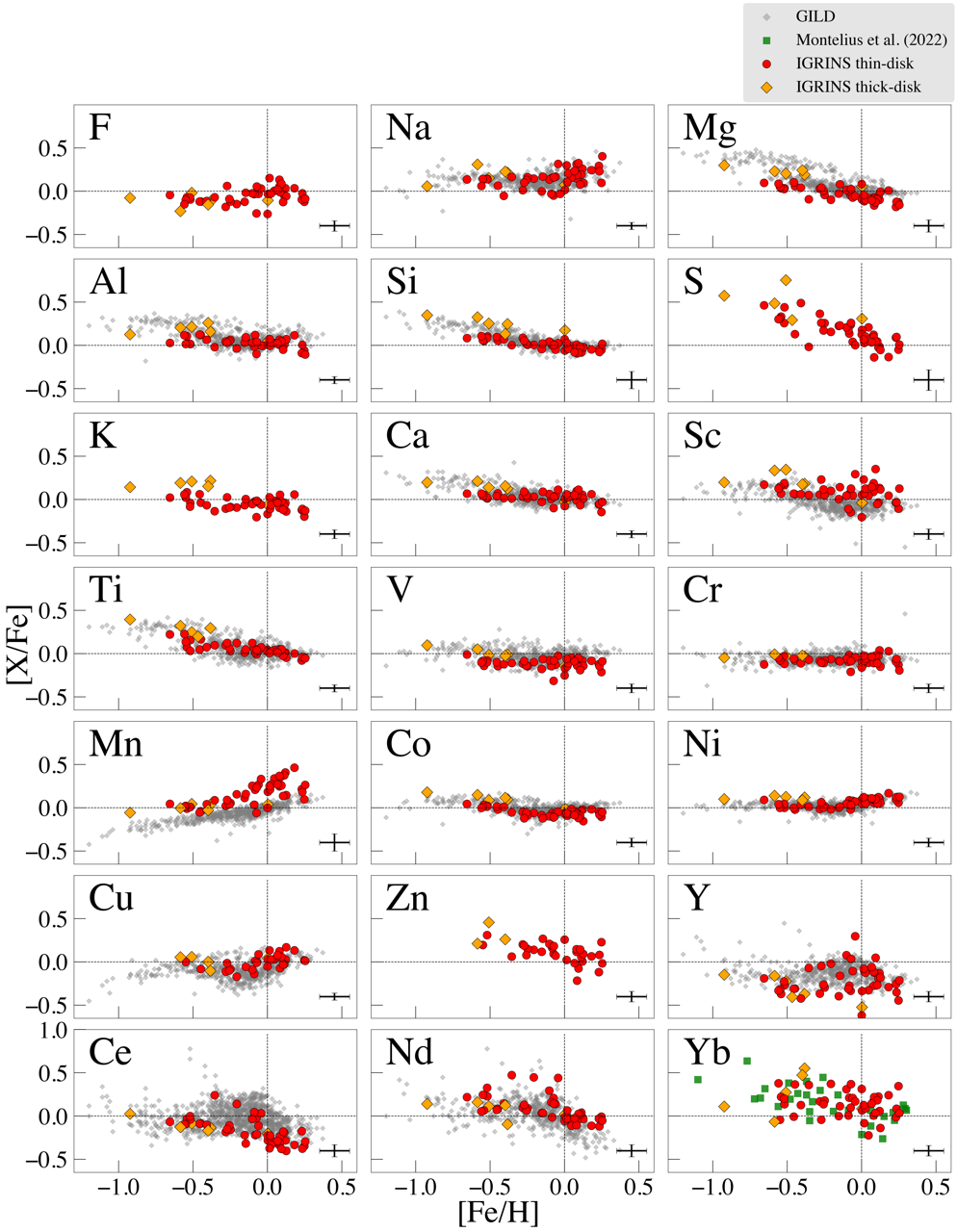}
  \caption{Abundance trends versus metallicity of 21 elements determined for the 50 M giants in this work (red filled circles and orange diamonds). Gray diamonds represent the abundances determined for $\sim$ 500 K giants in the solar neighbourhood from optical FIES spectra, for all elements except F, S, K, Zn, and Yb \citep[the latter is compared to abundances from][]{montelius:22}, that are provided in the giants in the local disk (GILD) catalog.}
  \label{fig:all_gild_trend}%
\end{figure*}

APOGEE has determined stellar parameters and elemental abundances for the 44 stars in our sample that we have in common. However, abundances of only 11 elements is provided in the DR17 catalog for these cool stars. None of the s-process elements are provided for these cool stars. 


We use APOGEE's spectroscopic values for the elemental abundances and not the calibrated ones, since the spectroscopic values are determined directly from the spectra which we also do in this work, making the comparison more relevant. We plot the one-to-one elemental abundances with respect to hydrogen ([X/H]) including metallicity and 11 elements determined from IGRINS spectra versus those in APOGEE DR17 catalog for stars in common between APOGEE and this work in Figure~\ref{fig:apogee_trend}. The mean differences determined for each elements are also listed in corresponding panels in the figure. We note that even though these stars observed by APOGEE and IGRINS, respectively, are the same, the stellar parameters are different since we have determined our own parameters and APOGEE provides their \citepalias[for a comparison see][]{Nandakumar:2023}. Another obvious difference is that APOGEE's spectral resolution is $R=22,500$ whereas the IGRINS resolution is twice that.

As indicated by the numbers listed within the brackets in each panel of the Figure~\ref{fig:apogee_trend}, APOGEE does not provide valid abundances for all 44 stars in our sample. It is also clear from the Figure~\ref{fig:apogee_trend} and the mean difference values that our abundance estimates are consistent with APOGEE for all elements except aluminium, potassium, chromium, and cobalt. For these elements, we find consistently lower values compared to APOGEE.

The offset in abundances in this work with respect to APOGEE may be attributed to a combination of different factors like differences in spectral resolution, line selection, line lists, pipelines etc between APOGEE and this work. Further, in his comparative study of analysing APOGEE and the higher-resolution IGRINS spectra, \citet{friden:23} shows that, in general, the higher spectral resolution helps in the analysis of stars at the metal-poor ends (for Ti, V, and Co) and metal-rich ends of the trends. The higher resolution also yield tighter trends for the neutron-capture elements, Ce, which follow the optical trends closer.

\subsubsection{Comparison with BAWLAS}
\label{sec:BAWLAScompare}

\citet{Hayes:2022} reanalyse 126,000 APOGEE DR17 spectra of giants with high signal-to-noise  with the BACCHUS code providing abundances in the {\it BACCHUS Analysis of Weak Lines in APOGEE Spectra} (BAWLAS) catalogue. The goal is to be able to use weaker and blended spectral features, which the automated and rigid APOGEE pipeline cannot handle as well. Since they take special care of and treat blends and upper limits, they are able to determine more precise abundances for Na, S, V, and Ce as a complement to the abundances given from the APOGEE pipeline. Furthermore, they also measure elements that APOGEE fails, such as P, Cu, and Nd. Similar to APOGEE DR17, chemical abundances derived for the above-mentioned elements have also been calibrated to the respective solar zero-point derived from the solar neighborhood samples in the catalog. Thus, \citet{Hayes:2022} show that more information can be extracted from surveys when a well-chosen sub-sample is reanalysed, capitalising on very high quality data. 

In Figure \ref{fig:all_bawlas_trend} we show the trends for six elements (Na, S, V, Cu, Ce, and Nd) determined in the BAWLAS study \citep{Hayes:2022} for $\sim$19 000 stars with \teff\, less than 4000 K in grey inverted triangles and our stars in red filled circles and orange diamonds. 

 We use the same lines employed by \cite{Hayes:2022} to derive abundances for Na and Nd, though we have astrophysically updated $\log gf$-values for Na lines and adopted $\log gf$-values from \cite{Hasselquist:2016} for Nd. For both Na and Nd, \cite{Hayes:2022} have calibrated their abundances with average zero-point offsets (with respect to the solar abundances in \citet{solar:sme}) of $\sim$0.2\,dex and $\sim$0.3\,dex respectively. Yet, our abundance trends are consistent with the BAWLAS abundances trends for Na and Nd. 
 
 Our abundance trends for the  rest of the four elements lie either on the upper envelope (for S) or lower envelope (for V, Cu, and Ce) of the BAWLAS abundances trends. One plausible explanation could be that for these four elements, we have used different lines or only a subset of lines used in \cite{Hayes:2022}. In addition, we have updated the $\log gf$-values astrophysically for certain elements (see Section~\ref{sec:Linelist}) while zero-point offsets have been made to abundances as listed in Table 2 in \cite{Hayes:2022}.

\subsubsection{Comparison with GILD}
\label{sec:GILDcompare}

We plot the abundances determined of the 21 elements for the 50 solar-neighborhood stars (red circles and orange diamonds) determined from IGRINS spectra and the GILD sample (grey diamonds) in Figure~\ref{fig:all_gild_trend}. While in the GILD project, high-resolution optical spectra of K giants are analysed, in the IGRINS project we have analysed high-resolution H- and K-band spectra of {\it M giants}, which are cooler. This means that there are no GILD stars that are in common with our IGRINS sample. GILD provides abundances for all elements in this study except fluorine, sulphur, and potassium. We compare the abundance trends of each group of elements ($\alpha$, odd-Z, iron-peak, and neutron-capture) in the following paragraphs.
 
A comparison of the $\alpha$-element abundance-trends for our IGRINS sample with the GILD sample has already been carried out in \citetalias{Nandakumar:2023}. They found the trends to be consistent with respect to each individual elemental-abundance trends. While the Si and Ca trends align well, the Mg trend in GILD \citep[which is similar to the optical dwarf trend of][]{bensby:2014} is significantly higher, both when compared to our IGRINS trend and that of APOGEE. Why the near-infrared Mg trends are lower than the optical one needs further investigations, but one reason could be inaccurate NLTE corrections. The thin-disk/thick-disk dichotomy in the metal-poor regime is also well evident for all three $\alpha$-elements and Ti in both the GILD and IGRINS samples. 
 
The abundance trends of the odd-Z elements, Na and Al, for the IGRINS sample are consistent with the GILD-sample abundance trends. An increase in [Na/Fe] at super-solar metallicities and the plateau at sub-solar metallicities are evident for both samples. The IGRINS trend, though, shows a hint of a downward trend for the lowest metallicity star, resulting in a N shape trend. This N shape is also seen in other optical trends, as mentioned earlier, like that of \citet{bensby:2014}. Similarly, the [Al/Fe] trends from both samples agree, except at the lowest metallicities (at [Fe/H] $< -0.5$\,dex), where the thick-disk stars in the GILD sample reach a plateau, while the IGRINS sample decreases for decreasing metallicities.



Among the eight iron-peak elements, all elements except chromium and manganese show significant differences between the thin- and thick-disk stars in both the GILD and IGRINS samples. The near-infrared [Sc/Fe] trend is more scattered than the optical one, but still with a clear thin/thick disk separation. The IGRINS [V/Fe] trend is tighter but lower than that of GILD. This might point at a normalisation problem. The thin-disk [Co/Fe] trends in both the samples also overlap. The flat [Cr/Fe] trend in the GILD sample is similar to the IGRINS sample with a majority of the GILD stars having slightly sub-solar values as is the case with the IGRINS sample. The [Mn/Fe] trends are different in slope and magnitude, but the optical GILD abundances are not corrected for departures from LTE. The [Ni/Fe] trends in both samples exhibit tight trends that increases slightly at super-solar metallicities, which could point at metallicity-dependent yields, perhaps coming in from production in weak-s, similarly to the trend seen in copper. 

When it comes to the s-elements, the [Cu/Fe] trend in the GILD sample is more scattered with the lowest values for stars at $-0.5 <$ \feh\,$< 0.0$\,dex and increases at super-solar metallicities. The IGRINS sample also exhibit a similar N-shaped trend, although the IGRINS trends is less-scattered.

The Y and Ce trends in IGRINS and GILD are remarkably similar. This further indicated that our intrinsic line strengths for Y are relevant, see Section \ref{sec:Linelist}. The thick disk follow the lower envelope of the thin disk and there is a cosmic scatter at $-0.5 <$ \feh\,$< 0.0$\,dex. The [Yb/Fe] trend is consistent with the [Yb/Fe] trend determined from the same H-band line in IGRINS spectra of {\it K giants} instead by \cite{montelius:22}, that also show large scatter. This scatter could, however, in some part be cosmic since the thick disk stars lie as expected compared to the pure s-elements. The [Nd/Fe] trend is, in general, higher than the optical [Nd/Fe] trend, and also have some stars with higher [Nd/Fe] similar to the few outliers in the GILD sample. One reason for this is a systematic offset due to uncertain intrinsic line-strengths ($\log gf$-values). 

Line strengths from the VALD database (in the case of yttrium) or astrophysical $\log gf$ values have been adopted from previous studies for neutron-capture lines (see section~\ref{sec:neutroncapture}). Meanwhile, multiple absorption lines in optical spectra for Y, Ce, and Nd have very reliable line strengths determined from experimental methods and are commonly used in the large-scale, optical spectroscopic survey such as GALAH. This could also be the reason for 
the slightly lower values of [Ce/Fe] in the IGRINS sample mainly following the lower values of the optical [Ce/Fe] trend. 
 

This points out the necessity of better linelist in H- and K-bands with reliable line strengths (preferably measured in the laboratory), accurate wavelengths, hyperfine splitting information, and broadening parameters to determine accurate abundances of the rare heavy elements in cool stars from high-resolution, near-infrared spectra. 

\subsection{Over all trends and precision }
\label{sec:precision}
From Figure \ref{fig:all_gild_trend} we can see how the thick-disk abundance trends clearly lie above those of the thin disk for the $\alpha$ elements, as well as for the elements Al, K, Sc, Ti, V, Co, and Ni. On the contrary, for all the s-process dominated elements (Y, Ce, and Nd), the thick-disk stars lie along the lower envelop of the thin-disk 'cloud'. This is indeed expected from Galactic chemical evolution models \citep[see, e.g.][]{Grisoni:2020}, as opposed to elements that are synthesised on a fast timescale, like the $\alpha$ elements, the weak s-process elements, like Cu, and the r-process dominated elements, like Eu. In our case Yb lies in between these cases with a 50/50 contribution from the s- and r-processes.

The precision of the abundances that are determined here are discussed in \citetalias{Nandakumar:2023} and \citetalias{Nandakumar:2023b}. They estimate abundance uncertainties in the range of $\sim$0.04 - 0.11\,dex resulting from typical uncertainties in stellar parameters ($\pm$100 K in \teff\,, $\pm$0.2 dex in \logg\,, $\pm$0.1 dex in \feh\,, and $\pm$0.1 km/s in $\xi_\mathrm{micro}$). However, it is interesting to see the general scatter in the abundance trends that are well determined. The [Si/Fe] abundance trend is, for example, very tight and are clearly derived with a very high precision, less than 0.05\,dex. Also the fact that the thick-disk trends in general align well is an indication of the high precision. In Figure \ref{fig:sproc_mean_trend}, we also see, as one more example, that the four stars that show high s-process contribution are high in all these stars at the same time, in spite of them having different stellar parameters and that the abundances are derived from totally different spectral lines, of course. This gives us confidence in the precision of our results and also that we believe that the scatter in the s-process thin-disk trends at $-0.5<$\feh$<0.0$ (the abundance 'cloud') are real and therefore of cosmic origin. This is also clearly seen in the GILD abundances that are derived from optical spectra for warmer stars \citep[see][]{Rebecca_phd}. 






\section{Conclusions}
\label{sec:conclusion}

We have demonstrated that it is possible to retrieve 21 reliable abundance trends versus metallicity, for F, Mg, Si, S, Ca, Na, Al, K, Sc, Ti, V, Cr, Mn, Co, Ni, Cu, Zn, Y, Ce, Nd, and Yb, from high-resolution H- and K-band spectra, here shown with IGRINS spectra of 50 solar-neighbourhood stars. Our trends can be used for a differential comparison with other stellar populations, in order to avoid any systematic uncertainties. 

The fact that we are able to investigate all these elements in near-infrared spectra is important now when near-infrared spectrometers are getting available and these can observe the obscure parts of the Milky Way. We have put an emphasis on M giants since these stars are the brightest giants and are needed in order to chemical characterize the heavily obscured parts of the Milky Way, for example the Galactic center region, including the nuclear stellar disk (NSD) and the nuclear star cluster (NSC). In order to be able to investigate the chemical history of the entire Milky Way, also the dust-obscured regions are needed to be studied in detail. 

Among the 21 elements there are several different nucleosynthetic channels that provide empirical data for the Galactic chemical evolution with different time-scales. The extended dimensions, that these different channels provide, will also provide decisive means to  determine differences between populations \citep[see, e.g.][]{manea:23}. Now, with the trends determined from infrared spectra, the interesting and important dust-obscured stellar populations, e.g. in the Galactic Center, can be investigated. The important neutron-capture elements, both s- and r-dominated elements, are thus possible to measure reliably. The notion that we can obtain Cu, Ce, Nd and Yb abundances from the H-band, points at the necessity of doing a dedicated, careful, and meticulous analysis of high quality spectra, in a similar way as done in \citet{Hayes:2022}. In our case we also advance by observing at higher resolution, $R \geq 45,000$, in order to even better disentangle blends and capture weak spectral features.  Several K band lines are identified for F, Mg, Si, S, Ca, Na, Al, Sc, Ti, and Ni. The s-element Y is only possible to get from the K band. While 21 elements are readily retrieved from IGRINS spectra, only 9 elements for these cool stars are provided in APOGEE. Note that the careful analysis of the BAWLAS study retrieve 3 extra elements. 

Our trends agree well with other larger sets of data done at optical wavelengths with proven methods. In our range of element trends, we see clear similarities between groups of elements (like the $\alpha$ elements) and differences between others. The precision is high enough to also clearly separate the thick and thin-disk trends for a wide variety of elements. We also confirm the increased cosmic scatter in the thin-disk, s-process element-trends for a given metallicity in the metallicity range of $-0.5<$\feh$<0.0$ (the abundance 'cloud'). To our knowledge, there is no theoretical explanation for this increased abundance distributions for given metallicities. We also clearly see the 'evolution' of the trends for the different neutron-capture elements from Cu to Yb, via Y, Ce, and Nd revealing the increasing r-process component of the elements' origins. We also find a flat [Mn/Fe] versus [Fe/H] trend at low metallicities, implying that the NLTE grids used for optical and other type of stars might be relevant and correct.  

Based on our findings, for future studies using  near-infrared, high-resolution spectroscopy, it is obvious that consistent NLTE corrections for the full parameter space as well as for more elements are needed, and the inconsistencies in abundance trends from different lines of the same element need to be further investigated. We also need more reliable line strengths and broadening parameters for near-infrared lines from experimental measurements and theoretical calculations.

Methods for deriving reliable stellar parameters and stellar abundances from high-resolution, near-infrared spectra of M giants, is also possible with other instruments than IGRINS. The GIANO and CRIRES+ spectrographs are available already, and in the near future the MOONS spectrometer will be a work horse although at APOGEE-type resolution. Future high-resolution, near-infrared spectrometers projected for the extremely large telescopes, such as the ELT and TMT, will open up immense possibilities of observing more distant M giants, also in other galaxies. 









\begin{acknowledgements}
We thank the anonymous referee for the constructive comments and suggestions that improved the quality of the paper. G.N.\ acknowledges the support from the Wenner-Gren Foundations (UPD2020-0191 and UPD2022-0059) and the Royal Physiographic Society in Lund through the Stiftelsen Walter Gyllenbergs fond. N.R.\ and R.F.\ acknowledge support from the Royal Physiographic Society in Lund through the Stiftelsen Walter Gyllenbergs fond and Märta och Erik Holmbergs donation. R.F.\ also acknowledges support from the Göran Gustafsson Foundation for Research in Natural Sciences and Medicine.
M.M. acknowledges funding through VIDI grant "Pushing Galactic Archaeology to its limits" VI.Vidi.193.093, which is funded by the Dutch Research Council (NWO).
G.M.\...
H.J.\...
B.T.\ acknowledges the financial support from the Wenner-Gren Foundation (WGF2022-0041).
This work used The Immersion Grating Infrared Spectrometer (IGRINS) was developed under a collaboration between the University of Texas at Austin and the Korea Astronomy and Space Science Institute (KASI) with the financial support of the US National Science Foundation under grants AST-1229522, AST-1702267 and AST-1908892, McDonald Observatory of the University of Texas at Austin, the Korean GMT Project of KASI, the Mt. Cuba Astronomical Foundation and Gemini Observatory.
This work is based on observations obtained at the international Gemini Observatory, a program of NSF’s NOIRLab, which is managed by the Association of Universities for Research in Astronomy (AURA) under a cooperative agreement with the National Science Foundation on behalf of the Gemini Observatory partnership: the National Science Foundation (United States), National Research Council (Canada), Agencia Nacional de Investigaci\'{o}n y Desarrollo (Chile), Ministerio de Ciencia, Tecnolog\'{i}a e Innovaci\'{o}n (Argentina), Minist\'{e}rio da Ci\^{e}ncia, Tecnologia, Inova\c{c}\~{o}es e Comunica\c{c}\~{o}es (Brazil), and Korea Astronomy and Space Science Institute (Republic of Korea).
The following software and programming languages made this
research possible: TOPCAT (version 4.6; \citealt{topcat}); Python (version 3.8) and its packages ASTROPY (version 5.0; \citealt{astropy}), SCIPY \citep{scipy}, MATPLOTLIB \citep{matplotlib} and NUMPY \citep{numpy}.
\end{acknowledgements}

%
%


\bibliographystyle{aa}
\bibliography{references} 

\begin{thebibliography}{138}
\expandafter\ifx\csname natexlab\endcsname\relax\def\natexlab#1{#1}\fi

\bibitem[{{Abdurro'uf} {et~al.}(2022){Abdurro'uf}, {Accetta}, {Aerts}, {Silva Aguirre}, {Ahumada}, {Ajgaonkar}, {Filiz Ak}, {Alam}, {Allende Prieto}, {Almeida}, {Anders}, {Anderson}, {Andrews}, {Anguiano}, {Aquino-Ort{\'\i}z}, {Arag{\'o}n-Salamanca}, {Argudo-Fern{\'a}ndez}, {Ata}, {Aubert}, {Avila-Reese}, {Badenes}, {Barb{\'a}}, {Barger}, {Barrera-Ballesteros}, {Beaton}, {Beers}, {Belfiore}, {Bender}, {Bernardi}, {Bershady}, {Beutler}, {Bidin}, {Bird}, {Bizyaev}, {Blanc}, {Blanton}, {Boardman}, {Bolton}, {Boquien}, {Borissova}, {Bovy}, {Brandt}, {Brown}, {Brownstein}, {Brusa}, {Buchner}, {Bundy}, {Burchett}, {Bureau}, {Burgasser}, {Cabang}, {Campbell}, {Cappellari}, {Carlberg}, {Wanderley}, {Carrera}, {Cash}, {Chen}, {Chen}, {Cherinka}, {Chiappini}, {Choi}, {Chojnowski}, {Chung}, {Clerc}, {Cohen}, {Comerford}, {Comparat}, {da Costa}, {Covey}, {Crane}, {Cruz-Gonzalez}, {Culhane}, {Cunha}, {Dai}, {Damke}, {Darling}, {Davidson}, {Davies}, {Dawson}, {De Lee}, {Diamond-Stanic}, {Cano-D{\'\i}az}, {S{\'a}nchez},
  {Donor}, {Duckworth}, {Dwelly}, {Eisenstein}, {Elsworth}, {Emsellem}, {Eracleous}, {Escoffier}, {Fan}, {Farr}, {Feng}, {Fern{\'a}ndez-Trincado}, {Feuillet}, {Filipp}, {Fillingham}, {Frinchaboy}, {Fromenteau}, {Galbany}, {Garc{\'\i}a}, {Garc{\'\i}a-Hern{\'a}ndez}, {Ge}, {Geisler}, {Gelfand}, {G{\'e}ron}, {Gibson}, {Goddy}, {Godoy-Rivera}, {Grabowski}, {Green}, {Greener}, {Grier}, {Griffith}, {Guo}, {Guy}, {Hadjara}, {Harding}, {Hasselquist}, {Hayes}, {Hearty}, {Hern{\'a}ndez}, {Hill}, {Hogg}, {Holtzman}, {Horta}, {Hsieh}, {Hsu}, {Hsu}, {Huber}, {Huertas-Company}, {Hutchinson}, {Hwang}, {Ibarra-Medel}, {Chitham}, {Ilha}, {Imig}, {Jaekle}, {Jayasinghe}, {Ji}, {Johnson}, {Jones}, {J{\"o}nsson}, {Katkov}, {Khalatyan}, {Kinemuchi}, {Kisku}, {Knapen}, {Kneib}, {Kollmeier}, {Kong}, {Kounkel}, {Kreckel}, {Krishnarao}, {Lacerna}, {Lane}, {Langgin}, {Lavender}, {Law}, {Lazarz}, {Leung}, {Leung}, {Lewis}, {Li}, {Li}, {Lian}, {Liang}, {Lin}, {Lin}, {Lin}, {Lintott}, {Long}, {Longa-Pe{\~n}a}, {L{\'o}pez-Cob{\'a}}, {Lu},
  {Lundgren}, {Luo}, {Mackereth}, {de la Macorra}, {Mahadevan}, {Majewski}, {Manchado}, {Mandeville}, {Maraston}, {Margalef-Bentabol}, {Masseron}, {Masters}, {Mathur}, {McDermid}, {Mckay}, {Merloni}, {Merrifield}, {Meszaros}, {Miglio}, {Di Mille}, {Minniti}, {Minsley}, {Monachesi}, {Moon}, {Mosser}, {Mulchaey}, {Muna}, {Mu{\~n}oz}, {Myers}, {Myers}, {Nadathur}, {Nair}, {Nandra}, {Neumann}, {Newman}, {Nidever}, {Nikakhtar}, {Nitschelm}, {O'Connell}, {Garma-Oehmichen}, {Luan Souza de Oliveira}, {Olney}, {Oravetz}, {Ortigoza-Urdaneta}, {Osorio}, {Otter}, {Pace}, {Padilla}, {Pan}, {Pan}, {Parikh}, {Parker}, {Peirani}, {Pe{\~n}a Ram{\'\i}rez}, {Penny}, {Percival}, {Perez-Fournon}, {Pinsonneault}, {Poidevin}, {Poovelil}, {Price-Whelan}, {B{\'a}rbara de Andrade Queiroz}, {Raddick}, {Ray}, {Rembold}, {Riddle}, {Riffel}, {Riffel}, {Rix}, {Robin}, {Rodr{\'\i}guez-Puebla}, {Roman-Lopes}, {Rom{\'a}n-Z{\'u}{\~n}iga}, {Rose}, {Ross}, {Rossi}, {Rubin}, {Salvato}, {S{\'a}nchez}, {S{\'a}nchez-Gallego}, {Sanderson}, {Santana
  Rojas}, {Sarceno}, {Sarmiento}, {Sayres}, {Sazonova}, {Schaefer}, {Schiavon}, {Schlegel}, {Schneider}, {Schultheis}, {Schwope}, {Serenelli}, {Serna}, {Shao}, {Shapiro}, {Sharma}, {Shen}, {Shetrone}, {Shu}, {Simon}, {Skrutskie}, {Smethurst}, {Smith}, {Sobeck}, {Spoo}, {Sprague}, {Stark}, {Stassun}, {Steinmetz}, {Stello}, {Stone-Martinez}, {Storchi-Bergmann}, {Stringfellow}, {Stutz}, {Su}, {Taghizadeh-Popp}, {Talbot}, {Tayar}, {Telles}, {Teske}, {Thakar}, {Theissen}, {Tkachenko}, {Thomas}, {Tojeiro}, {Hernandez Toledo}, {Troup}, {Trump}, {Trussler}, {Turner}, {Tuttle}, {Unda-Sanzana}, {V{\'a}zquez-Mata}, {Valentini}, {Valenzuela}, {Vargas-Gonz{\'a}lez}, {Vargas-Maga{\~n}a}, {Alfaro}, {Villanova}, {Vincenzo}, {Wake}, {Warfield}, {Washington}, {Weaver}, {Weijmans}, {Weinberg}, {Weiss}, {Westfall}, {Wild}, {Wilde}, {Wilson}, {Wilson}, {Wilson}, {Wolf}, {Wood-Vasey}, {Yan}, {Zamora}, {Zasowski}, {Zhang}, {Zhao}, {Zheng}, {Zheng}, \& {Zhu}}]{ApogeeDr17}
{Abdurro'uf}, {Accetta}, K., {Aerts}, C., {et~al.} 2022, \apjs, 259, 35

\bibitem[{{Aboussa{\"i}d} {et~al.}(1996){Aboussa{\"i}d}, {Carleer}, {Hurtmans}, {Bi{\'e}mont}, \& {Godefroid}}]{ACHBG}
{Aboussa{\"i}d}, A., {Carleer}, M., {Hurtmans}, D., {Bi{\'e}mont}, E., \& {Godefroid}, M.~R. 1996, Physica Scripta, 53, 28

\bibitem[{{Af{\c{s}}ar} {et~al.}(2023){Af{\c{s}}ar}, {Bozkurt}, {Topcu}, {{\"O}zdemir}, {Sneden}, {Mace}, {Jaffe}, \& {L{\'o}pez-Valdivia}}]{afsar:2023}
{Af{\c{s}}ar}, M., {Bozkurt}, Z., {Topcu}, G.~B., {et~al.} 2023, \apj, 949, 86

\bibitem[{{Af{\c{s}}ar} {et~al.}(2016){Af{\c{s}}ar}, {Sneden}, {Frebel}, {Kim}, {Mace}, {Kaplan}, {Lee}, {Oh}, {Sok Oh}, {Pak}, {Park}, {Pavel}, {Yuk}, \& {Jaffe}}]{Afsar:2016}
{Af{\c{s}}ar}, M., {Sneden}, C., {Frebel}, A., {et~al.} 2016, \apj, 819, 103

\bibitem[{{Af{\c{s}}ar} {et~al.}(2018){Af{\c{s}}ar}, {Sneden}, {Wood}, {Lawler}, {Bozkurt}, {B{\"o}cek Topcu}, {Mace}, {Kim}, \& {Jaffe}}]{afsar:2018}
{Af{\c{s}}ar}, M., {Sneden}, C., {Wood}, M.~P., {et~al.} 2018, \apj, 865, 44

\bibitem[{{Amarsi} {et~al.}(2016){Amarsi}, {Lind}, {Asplund}, {Barklem}, \& {Collet}}]{amarsi16}
{Amarsi}, A.~M., {Lind}, K., {Asplund}, M., {Barklem}, P.~S., \& {Collet}, R. 2016, \mnras, 463, 1518

\bibitem[{{Amarsi} {et~al.}(2020){Amarsi}, {Lind}, {Osorio}, {Nordlander}, {Bergemann}, {Reggiani}, {Wang}, {Buder}, {Asplund}, {Barklem}, {Wehrhahn}, {Sk{\'u}lad{\'o}ttir}, {Kobayashi}, {Karakas}, {Gao}, {Bland-Hawthorn}, {de Silva}, {Kos}, {Lewis}, {Martell}, {Sharma}, {Simpson}, {Zucker}, {{\v{C}}otar}, {Horner}, \& {GALAH Collaboration}}]{NLTE}
{Amarsi}, A.~M., {Lind}, K., {Osorio}, Y., {et~al.} 2020, \aap, 642, A62

\bibitem[{{Amarsi} {et~al.}(2019){Amarsi}, {Nissen}, \& {Sk{\'u}lad{\'o}ttir}}]{Amarsi:2019}
{Amarsi}, A.~M., {Nissen}, P.~E., \& {Sk{\'u}lad{\'o}ttir}, {\'A}. 2019, \aap, 630, A104

\bibitem[{{Amarsi} {et~al.}(2018){Amarsi}, {Nordlander}, {Barklem}, {Asplund}, {Collet}, \& {Lind}}]{Amarsi:2018}
{Amarsi}, A.~M., {Nordlander}, T., {Barklem}, P.~S., {et~al.} 2018, \aap, 615, A139

\bibitem[{Anstee \& O'Mara(1991)}]{anstee_investigation_1991}
Anstee, S.~D. \& O'Mara, B.~J. 1991, MNRAS, 253, 549

\bibitem[{Anstee \& O'Mara(1995)}]{Anstee1995}
Anstee, S.~D. \& O'Mara, B.~J. 1995, MNRAS, 276, 859

\bibitem[{Arqueros(1988)}]{arqueros:1988}
Arqueros, F. 1988, Optics Communications, 67, 341

\bibitem[{{Astropy Collaboration} {et~al.}(2022){Astropy Collaboration}, {Price-Whelan}, {Lim}, {Earl}, {Starkman}, {Bradley}, {Shupe}, {Patil}, {Corrales}, {Brasseur}, {N{\"o}the}, {Donath}, {Tollerud}, {Morris}, {Ginsburg}, {Vaher}, {Weaver}, {Tocknell}, {Jamieson}, {van Kerkwijk}, {Robitaille}, {Merry}, {Bachetti}, {G{\"u}nther}, {Aldcroft}, {Alvarado-Montes}, {Archibald}, {B{\'o}di}, {Bapat}, {Barentsen}, {Baz{\'a}n}, {Biswas}, {Boquien}, {Burke}, {Cara}, {Cara}, {Conroy}, {Conseil}, {Craig}, {Cross}, {Cruz}, {D'Eugenio}, {Dencheva}, {Devillepoix}, {Dietrich}, {Eigenbrot}, {Erben}, {Ferreira}, {Foreman-Mackey}, {Fox}, {Freij}, {Garg}, {Geda}, {Glattly}, {Gondhalekar}, {Gordon}, {Grant}, {Greenfield}, {Groener}, {Guest}, {Gurovich}, {Handberg}, {Hart}, {Hatfield-Dodds}, {Homeier}, {Hosseinzadeh}, {Jenness}, {Jones}, {Joseph}, {Kalmbach}, {Karamehmetoglu}, {Ka{\l}uszy{\'n}ski}, {Kelley}, {Kern}, {Kerzendorf}, {Koch}, {Kulumani}, {Lee}, {Ly}, {Ma}, {MacBride}, {Maljaars}, {Muna}, {Murphy}, {Norman},
  {O'Steen}, {Oman}, {Pacifici}, {Pascual}, {Pascual-Granado}, {Patil}, {Perren}, {Pickering}, {Rastogi}, {Roulston}, {Ryan}, {Rykoff}, {Sabater}, {Sakurikar}, {Salgado}, {Sanghi}, {Saunders}, {Savchenko}, {Schwardt}, {Seifert-Eckert}, {Shih}, {Jain}, {Shukla}, {Sick}, {Simpson}, {Singanamalla}, {Singer}, {Singhal}, {Sinha}, {Sip{\H{o}}cz}, {Spitler}, {Stansby}, {Streicher}, {{\v{S}}umak}, {Swinbank}, {Taranu}, {Tewary}, {Tremblay}, {Val-Borro}, {Van Kooten}, {Vasovi{\'c}}, {Verma}, {de Miranda Cardoso}, {Williams}, {Wilson}, {Winkel}, {Wood-Vasey}, {Xue}, {Yoachim}, {Zhang}, {Zonca}, \& {Astropy Project Contributors}}]{astropy}
{Astropy Collaboration}, {Price-Whelan}, A.~M., {Lim}, P.~L., {et~al.} 2022, \apj, 935, 167

\bibitem[{{Baratella} {et~al.}(2021){Baratella}, {D'Orazi}, {Sheminova}, {Spina}, {Carraro}, {Gratton}, {Magrini}, {Randich}, {Lugaro}, {Pignatari}, {Romano}, {Biazzo}, {Bragaglia}, {Casali}, {Desidera}, {Frasca}, {de Silva}, {Melo}, {Van der Swaelmen}, {Tautvai{\v{s}}ien{\.{e}}}, {Jim{\'e}nez-Esteban}, {Gilmore}, {Bensby}, {Smiljanic}, {Bayo}, {Franciosini}, {Gonneau}, {Hourihane}, {Jofr{\'e}}, {Monaco}, {Morbidelli}, {Sacco}, {Sbordone}, {Worley}, \& {Zaggia}}]{Cu_ges:21}
{Baratella}, M., {D'Orazi}, V., {Sheminova}, V., {et~al.} 2021, \aap, 653, A67

\bibitem[{Barklem \& O'Mara(1997)}]{Barklem1997a}
Barklem, P.~S. \& O'Mara, B.~J. 1997, MNRAS, 290, 102

\bibitem[{Barklem {et~al.}(1998)Barklem, O'Mara, \& Ross}]{Barklem1998b}
Barklem, P.~S., O'Mara, B.~J., \& Ross, J.~E. 1998, MNRAS, 296, 1057

\bibitem[{{Battistini} \& {Bensby}(2015)}]{battistini:2015}
{Battistini}, C. \& {Bensby}, T. 2015, \aap, 577, A9

\bibitem[{{Battistini} \& {Bensby}(2016)}]{battistini:2016}
{Battistini}, C. \& {Bensby}, T. 2016, \aap, 586, A49

\bibitem[{{Bensby} {et~al.}(2014){Bensby}, {Feltzing}, \& {Oey}}]{bensby:2014}
{Bensby}, T., {Feltzing}, S., \& {Oey}, M.~S. 2014, \aap, 562, A71

\bibitem[{{Bergemann} {et~al.}(2019){Bergemann}, {Gallagher}, {Eitner}, {Bautista}, {Collet}, {Yakovleva}, {Mayriedl}, {Plez}, {Carlsson}, {Leenaarts}, {Belyaev}, \& {Hansen}}]{bergemann:19}
{Bergemann}, M., {Gallagher}, A.~J., {Eitner}, P., {et~al.} 2019, \aap, 631, A80

\bibitem[{{Bergemann} \& {Gehren}(2008)}]{bergemann:08}
{Bergemann}, M. \& {Gehren}, T. 2008, \aap, 492, 823

\bibitem[{{Biemont} {et~al.}(1993){Biemont}, {Quinet}, \& {Zeippen}}]{BQZ}
{Biemont}, E., {Quinet}, P., \& {Zeippen}, C.~J. 1993, \aaps, 102, 435, (BQZ)

\bibitem[{{Bisterzo} {et~al.}(2014){Bisterzo}, {Travaglio}, {Gallino}, {Wiescher}, \& {K{\"a}ppeler}}]{bisterzo:14}
{Bisterzo}, S., {Travaglio}, C., {Gallino}, R., {Wiescher}, M., \& {K{\"a}ppeler}, F. 2014, \apj, 787, 10

\bibitem[{{B{\"o}cek Topcu} {et~al.}(2019){B{\"o}cek Topcu}, {Af{\c{s}}ar}, {Sneden}, {Pilachowski}, {Denissenkov}, {VandenBerg}, {Strickland}, {{\"O}zdemir}, {Mace}, {Kim}, \& {Jaffe}}]{afsar:2019}
{B{\"o}cek Topcu}, G., {Af{\c{s}}ar}, M., {Sneden}, C., {et~al.} 2019, \mnras, 485, 4625

\bibitem[{{B{\"o}cek Topcu} {et~al.}(2020){B{\"o}cek Topcu}, {Af{\c{s}}ar}, {Sneden}, {Pilachowski}, {Denissenkov}, {VandenBerg}, {Wright}, {Mace}, {Jaffe}, {Strickland}, {Kim}, \& {Sokal}}]{afsar:2020}
{B{\"o}cek Topcu}, G., {Af{\c{s}}ar}, M., {Sneden}, C., {et~al.} 2020, \mnras, 491, 544

\bibitem[{{Brady} {et~al.}(2023){Brady}, {Sneden}, {Pilachowski}, {Af{\c{s}}ar}, {Mace}, {Jaffe}, {Gosnell}, \& {Seifert}}]{Brady:2023}
{Brady}, K.~E., {Sneden}, C., {Pilachowski}, C.~A., {et~al.} 2023, \aj, 166, 154

\bibitem[{{Brooke} {et~al.}(2016){Brooke}, {Bernath}, {Western}, {Sneden}, {Af{\c{s}}ar}, {Li}, \& {Gordon}}]{brooke:2016}
{Brooke}, J. S.~A., {Bernath}, P.~F., {Western}, C.~M., {et~al.} 2016, \jqsrt, 168, 142

\bibitem[{{Carr} {et~al.}(2000){Carr}, {Sellgren}, \& {Balachandran}}]{carr:00}
{Carr}, J.~S., {Sellgren}, K., \& {Balachandran}, S.~C. 2000, \apj, 530, 307

\bibitem[{{Chang}(1990)}]{Chang:1990}
{Chang}, E.~S. 1990, Journal of Physical and Chemical Reference Data, 19, 119

\bibitem[{{Cirasuolo} {et~al.}(2012){Cirasuolo}, {Afonso}, {Bender}, {Bonifacio}, {Evans}, {Kaper}, {Oliva}, {Vanzi}, {Abreu}, {Atad-Ettedgui}, {Babusiaux}, {Bauer}, {Best}, {Bezawada}, {Bryson}, {Cabral}, {Caputi}, {Centrone}, {Chemla}, {Cimatti}, {Cioni}, {Clementini}, {Coelho}, {Daddi}, {Dunlop}, {Feltzing}, {Ferguson}, {Flores}, {Fontana}, {Fynbo}, {Garilli}, {Glauser}, {Guinouard}, {Hammer}, {Hastings}, {Hess}, {Ivison}, {Jagourel}, {Jarvis}, {Kauffman}, {Lawrence}, {Lee}, {Li Causi}, {Lilly}, {Lorenzetti}, {Maiolino}, {Mannucci}, {McLure}, {Minniti}, {Montgomery}, {Muschielok}, {Nandra}, {Navarro}, {Norberg}, {Origlia}, {Padilla}, {Peacock}, {Pedicini}, {Pentericci}, {Pragt}, {Puech}, {Randich}, {Renzini}, {Ryde}, {Rodrigues}, {Royer}, {Saglia}, {S{\'a}nchez}, {Schnetler}, {Sobral}, {Speziali}, {Todd}, {Tolstoy}, {Torres}, {Venema}, {Vitali}, {Wegner}, {Wells}, {Wild}, \& {Wright}}]{moons:12}
{Cirasuolo}, M., {Afonso}, J., {Bender}, R., {et~al.} 2012, in SPIE, Vol. 8446, Ground-based and Airborne Instrumentation for Astronomy IV, 84460S

\bibitem[{{Clayton}(2003)}]{clayton:2003}
{Clayton}, D. 2003, {Handbook of Isotopes in the Cosmos}

\bibitem[{{Costa Silva} {et~al.}(2020){Costa Silva}, {Delgado Mena}, \& {Tsantaki}}]{costa:20}
{Costa Silva}, A.~R., {Delgado Mena}, E., \& {Tsantaki}, M. 2020, \aap, 634, A136

\bibitem[{{Cowan}(1981)}]{Cowan:1981}
{Cowan}, R.~D. 1981, \physscr, 24, 615

\bibitem[{{Cunha} {et~al.}(2007){Cunha}, {Sellgren}, {Smith}, {Ramirez}, {Blum}, \& {Terndrup}}]{cunha:07}
{Cunha}, K., {Sellgren}, K., {Smith}, V.~V., {et~al.} 2007, \apj, 669, 1011

\bibitem[{{Cunha} {et~al.}(2017){Cunha}, {Smith}, {Hasselquist}, {Souto}, {Shetrone}, {Allende Prieto}, {Bizyaev}, {Frinchaboy}, {Garc{\'\i}a-Hern{\'a}ndez}, {Holtzman}, {Johnson}, {J{\H{o}}nsson}, {Majewski}, {M{\'e}sz{\'a}ros}, {Nidever}, {Pinsonneault}, {Schiavon}, {Sobeck}, {Skrutskie}, {Zamora}, {Zasowski}, \& {Fern{\'a}ndez-Trincado}}]{Cunha:2017}
{Cunha}, K., {Smith}, V.~V., {Hasselquist}, S., {et~al.} 2017, \apj, 844, 145

\bibitem[{{Curtis} {et~al.}(2019){Curtis}, {Ebinger}, {Fr{\"o}hlich}, {Hempel}, {Perego}, {Liebend{\"o}rfer}, \& {Thielemann}}]{curtis:2019}
{Curtis}, S., {Ebinger}, K., {Fr{\"o}hlich}, C., {et~al.} 2019, \apj, 870, 2

\bibitem[{{da Silveira} {et~al.}(2018){da Silveira}, {Barbuy}, {Fria{\c{c}}a}, {Hill}, {Zoccali}, {Rafelski}, {Gonzalez}, {Minniti}, {Renzini}, \& {Ortolani}}]{daSilveira:18}
{da Silveira}, C.~R., {Barbuy}, B., {Fria{\c{c}}a}, A.~C.~S., {et~al.} 2018, \aap, 614, A149

\bibitem[{{de los Reyes} {et~al.}(2020){de los Reyes}, {Kirby}, {Seitenzahl}, \& {Shen}}]{reyes:20}
{de los Reyes}, M. A.~C., {Kirby}, E.~N., {Seitenzahl}, I.~R., \& {Shen}, K.~J. 2020, \apj, 891, 85

\bibitem[{{Delgado Mena} {et~al.}(2017){Delgado Mena}, {Tsantaki}, {Adibekyan}, {Sousa}, {Santos}, {Gonz{\'a}lez Hern{\'a}ndez}, \& {Israelian}}]{delgado:17}
{Delgado Mena}, E., {Tsantaki}, M., {Adibekyan}, V.~Z., {et~al.} 2017, \aap, 606, A94

\bibitem[{{Demarque} {et~al.}(2004){Demarque}, {Woo}, {Kim}, \& {Yi}}]{Demarque:2004}
{Demarque}, P., {Woo}, J.-H., {Kim}, Y.-C., \& {Yi}, S.~K. 2004, \apjs, 155, 667

\bibitem[{{Ernandes} {et~al.}(2020){Ernandes}, {Barbuy}, {Fria{\c{c}}a}, {Hill}, {Zoccali}, {Minniti}, {Renzini}, \& {Ortolani}}]{heitor:20}
{Ernandes}, H., {Barbuy}, B., {Fria{\c{c}}a}, A.~C.~S., {et~al.} 2020, \aap, 640, A89

\bibitem[{{Ertmer} \& {Hofer}(1976)}]{EH}
{Ertmer}, W. \& {Hofer}, B. 1976, Zeitschrift fur Physik A Hadrons and Nuclei, 276, 9

\bibitem[{{Follert} {et~al.}(2014){Follert}, {Dorn}, {Oliva}, {Lizon}, {Hatzes}, {Piskunov}, {Reiners}, {Seemann}, {Stempels}, {Heiter}, {Marquart}, {Lockhart}, {Anglada-Escude}, {L{\"o}winger}, {Baade}, {Grunhut}, {Bristow}, {Klein}, {Jung}, {Ives}, {Kerber}, {Pozna}, {Paufique}, {Kaeufl}, {Origlia}, {Valenti}, {Gojak}, {Hilker}, {Pasquini}, {Smette}, \& {Smoker}}]{crires}
{Follert}, R., {Dorn}, R.~J., {Oliva}, E., {et~al.} 2014, in Society of Photo-Optical Instrumentation Engineers (SPIE) Conference Series, Vol. 9147, Ground-based and Airborne Instrumentation for Astronomy V, ed. S.~K. {Ramsay}, I.~S. {McLean}, \& H.~{Takami}, 914719

\bibitem[{{Forsberg}(2023)}]{Rebecca_phd}
{Forsberg}, R. 2023, PhD thesis, Lund University, Sweden

\bibitem[{{Forsberg} {et~al.}(2019){Forsberg}, {J{\"o}nsson}, {Ryde}, \& {Matteucci}}]{forsberg:19}
{Forsberg}, R., {J{\"o}nsson}, H., {Ryde}, N., \& {Matteucci}, F. 2019, \aap, 631, A113

\bibitem[{{Frid\'en}({2023})}]{friden:23}
{Frid\'en}, E. {2023}, {The chemical evolution of the Milky Way: pushing APOGEE to higher precision and accuracy}, {Student Paper}

\bibitem[{{Frogel} {et~al.}(1999){Frogel}, {Tiede}, \& {Kuchinski}}]{frogel:99}
{Frogel}, J.~A., {Tiede}, G.~P., \& {Kuchinski}, L.~E. 1999, \aj, 117, 2296

\bibitem[{{Garc{\'\i}a-Hern{\'a}ndez} {et~al.}(2023){Garc{\'\i}a-Hern{\'a}ndez}, {Rao}, {Lambert}, {Eriksson}, {Reddy}, \& {Masseron}}]{carbonstar:2023}
{Garc{\'\i}a-Hern{\'a}ndez}, D.~A., {Rao}, N.~K., {Lambert}, D.~L., {et~al.} 2023, \apj, 948, 15

\bibitem[{{Garro} {et~al.}(2023){Garro}, {Fern{\'a}ndez-Trincado}, {Minniti}, {Moya}, {Palma}, {Beers}, {Placco}, {Barbuy}, {Sneden}, {Alves-Brito}, {Dias}, {Af{\c{s}}ar}, {Frelijj}, \& {Lane}}]{YSOsi:2023}
{Garro}, E.~R., {Fern{\'a}ndez-Trincado}, J.~G., {Minniti}, D., {et~al.} 2023, \aap, 669, A136

\bibitem[{{Genoni} {et~al.}(2016){Genoni}, {Riva}, {Pariani}, {Aliverti}, \& {Moschetti}}]{hires:16}
{Genoni}, M., {Riva}, M., {Pariani}, G., {Aliverti}, M., \& {Moschetti}, M. 2016, in Society of Photo-Optical Instrumentation Engineers (SPIE) Conference Series, Vol. 9911, Modeling, Systems Engineering, and Project Management for Astronomy VI, ed. G.~Z. {Angeli} \& P.~{Dierickx}, 99112L

\bibitem[{{Gilmozzi} \& {Spyromilio}(2008)}]{ELT2008}
{Gilmozzi}, R. \& {Spyromilio}, J. 2008, in Society of Photo-Optical Instrumentation Engineers (SPIE) Conference Series, Vol. 7012, Ground-based and Airborne Telescopes II, ed. L.~M. {Stepp} \& R.~{Gilmozzi}, 701219

\bibitem[{{Grevesse} {et~al.}(2007){Grevesse}, {Asplund}, \& {Sauval}}]{solar:sme}
{Grevesse}, N., {Asplund}, M., \& {Sauval}, A.~J. 2007, \ssr, 130, 105

\bibitem[{{Grevesse} {et~al.}(2015){Grevesse}, {Scott}, {Asplund}, \& {Sauval}}]{Grevesse:2015}
{Grevesse}, N., {Scott}, P., {Asplund}, M., \& {Sauval}, A.~J. 2015, \aap, 573, A27

\bibitem[{{Grisoni} {et~al.}(2020){Grisoni}, {Romano}, {Spitoni}, {Matteucci}, {Ryde}, \& {J{\"o}nsson}}]{Grisoni:2020}
{Grisoni}, V., {Romano}, D., {Spitoni}, E., {et~al.} 2020, \mnras, 498, 1252

\bibitem[{{Guer{\c{c}}o} {et~al.}(2022){Guer{\c{c}}o}, {Ram{\'\i}rez}, {Cunha}, {Smith}, {Prantzos}, {Sellgren}, \& {Daflon}}]{Guerco:2022}
{Guer{\c{c}}o}, R., {Ram{\'\i}rez}, S., {Cunha}, K., {et~al.} 2022, \apj, 929, 24

\bibitem[{{Gully-Santiago} {et~al.}(2012){Gully-Santiago}, {Wang}, {Deen}, \& {Jaffe}}]{Gully:2012}
{Gully-Santiago}, M., {Wang}, W., {Deen}, C., \& {Jaffe}, D. 2012, in Society of Photo-Optical Instrumentation Engineers (SPIE) Conference Series, Vol. 8450, Modern Technologies in Space- and Ground-based Telescopes and Instrumentation II, ed. R.~{Navarro}, C.~R. {Cunningham}, \& E.~{Prieto}, 84502S

\bibitem[{{Gustafsson} {et~al.}(2008){Gustafsson}, {Edvardsson}, {Eriksson}, {et~al.}}]{marcs:08}
{Gustafsson}, B., {Edvardsson}, B., {Eriksson}, K., {et~al.} 2008, \aap, 486, 951

\bibitem[{{Han} {et~al.}(2012){Han}, {Yuk}, {Ko}, {Oh}, {Nah}, {Oh}, {Park}, {Lee}, {Kim}, {Chun}, {Jaffe}, {Pak}, \& {Gully-Santiago}}]{Han:2012}
{Han}, J.-Y., {Yuk}, I.-S., {Ko}, K., {et~al.} 2012, in Society of Photo-Optical Instrumentation Engineers (SPIE) Conference Series, Vol. 8550, Optical Systems Design 2012, ed. D.~G. {Smith}, J.-L.~M. {Tissot}, L.~{Mazuray}, T.~E. {Kidger}, F.~{Wyrowski}, J.~M. {Raynor}, R.~{Wartmann}, S.~{David}, A.~{Erdmann}, A.~P. {Wood}, P.~{Ben{\'\i}tez}, \& M.~C. {de la Fuente}, 85501B

\bibitem[{{Hasselquist} {et~al.}(2016){Hasselquist}, {Shetrone}, {Cunha}, {Smith}, {Holtzman}, {Lawler}, {Allende Prieto}, {Beers}, {Chojnowski}, {Fern{\'a}ndez-Trincado}, {Garc{\'\i}a-Hern{\'a}ndez}, {Hearty}, {Majewski}, {Pereira}, {Placco}, {Villanova}, \& {Zamora}}]{Hasselquist:2016}
{Hasselquist}, S., {Shetrone}, M., {Cunha}, K., {et~al.} 2016, \apj, 833, 81

\bibitem[{{Hayes} {et~al.}(2022){Hayes}, {Masseron}, {Sobeck}, {Garc{\'\i}a-Hern{\'a}ndez}, {Allende Prieto}, {Beaton}, {Cunha}, {Hasselquist}, {Holtzman}, {J{\"o}nsson}, {Majewski}, {Shetrone}, {Smith}, \& {Almeida}}]{Hayes:2022}
{Hayes}, C.~R., {Masseron}, T., {Sobeck}, J., {et~al.} 2022, \apjs, 262, 34

\bibitem[{{Hinkle} {et~al.}(1995){Hinkle}, {Wallace}, \& {Livingston}}]{Hinkle:1995}
{Hinkle}, K., {Wallace}, L., \& {Livingston}, W. 1995, \pasp, 107, 1042

\bibitem[{{Holanda} {et~al.}(2024){Holanda}, {Roriz}, {Drake}, {Junqueira}, {Daflon}, {da Silva}, \& {Pereira}}]{Holanda:2024}
{Holanda}, N., {Roriz}, M.~P., {Drake}, N.~A., {et~al.} 2024, \mnras, 527, 1389

\bibitem[{{Hunter}(2007)}]{matplotlib}
{Hunter}, J.~D. 2007, Computing in Science and Engineering, 9, 90

\bibitem[{{Ikeda} {et~al.}(2016){Ikeda}, {Kobayashi}, {Kondo}, {Otsubo}, {Hamano}, {Sameshima}, {Yoshikawa}, {Fukue}, {Nakanishi}, {Kawanishi}, {Nakaoka}, {Kinoshita}, {Kitano}, {Asano}, {Takenaka}, {Watase}, {Mito}, {Yasui}, {Minami}, {Izumu}, {Yamamoto}, {Mizumoto}, {Arasaki}, {Arai}, {Matsunaga}, \& {Kawakita}}]{winered}
{Ikeda}, Y., {Kobayashi}, N., {Kondo}, S., {et~al.} 2016, in Society of Photo-Optical Instrumentation Engineers (SPIE) Conference Series, Vol. 9908, Ground-based and Airborne Instrumentation for Astronomy VI, ed. C.~J. {Evans}, L.~{Simard}, \& H.~{Takami}, 99085Z

\bibitem[{{Jeong} {et~al.}(2014){Jeong}, {Chun}, {Oh}, {Park}, {Yuk}, {Oh}, {Kim}, {Ko}, {Pavel}, {Yu}, \& {Jaffe}}]{Jeong:2014}
{Jeong}, U., {Chun}, M.-Y., {Oh}, J.~S., {et~al.} 2014, in Society of Photo-Optical Instrumentation Engineers (SPIE) Conference Series, Vol. 9154, High Energy, Optical, and Infrared Detectors for Astronomy VI, ed. A.~D. {Holland} \& J.~{Beletic}, 91541X

\bibitem[{{J{\"o}nsson} {et~al.}(2020){J{\"o}nsson}, {Holtzman}, {Allende Prieto}, {Cunha}, {Garc{\'\i}a-Hern{\'a}ndez}, {Hasselquist}, {Masseron}, {Osorio}, {Shetrone}, {Smith}, {Stringfellow}, {Bizyaev}, {Edvardsson}, {Majewski}, {M{\'e}sz{\'a}ros}, {Souto}, {Zamora}, {Beaton}, {Bovy}, {Donor}, {Pinsonneault}, {Poovelil}, \& {Sobeck}}]{ApogeeDR16}
{J{\"o}nsson}, H., {Holtzman}, J.~A., {Allende Prieto}, C., {et~al.} 2020, \aj, 160, 120

\bibitem[{{J{\"o}nsson} {et~al.}(2017){J{\"o}nsson}, {Ryde}, {Nordlander}, {Pehlivan Rhodin}, {Hartman}, {J{\"o}nsson}, \& {Eriksson}}]{jonsson:17}
{J{\"o}nsson}, H., {Ryde}, N., {Nordlander}, T., {et~al.} 2017, \aap, 598, A100

\bibitem[{{Kaplan} {et~al.}(2021){Kaplan}, {Dinerstein}, {Kim}, \& {Jaffe}}]{Kaplan:2021}
{Kaplan}, K.~F., {Dinerstein}, H.~L., {Kim}, H., \& {Jaffe}, D.~T. 2021, \apj, 919, 27

\bibitem[{{Kaplan} {et~al.}(2017){Kaplan}, {Dinerstein}, {Oh}, {Mace}, {Kim}, {Sokal}, {Pavel}, {Lee}, {Pak}, {Park}, {Sok Oh}, \& {Jaffe}}]{Kaplan:2017}
{Kaplan}, K.~F., {Dinerstein}, H.~L., {Oh}, H., {et~al.} 2017, \apj, 838, 152

\bibitem[{{Kobayashi} {et~al.}(2020){Kobayashi}, {Karakas}, \& {Lugaro}}]{kobayashi:20}
{Kobayashi}, C., {Karakas}, A.~I., \& {Lugaro}, M. 2020, \apj, 900, 179

\bibitem[{{Kobayashi} \& {Nakasato}(2011)}]{kobayashi:11}
{Kobayashi}, C. \& {Nakasato}, N. 2011, \apj, 729, 16

\bibitem[{{Kupka} {et~al.}(2000){Kupka}, {Ryabchikova}, {Piskunov}, {Stempels}, \& {Weiss}}]{vald4}
{Kupka}, F.~G., {Ryabchikova}, T.~A., {Piskunov}, N.~E., {Stempels}, H.~C., \& {Weiss}, W.~W. 2000, Baltic Astronomy, 9, 590

\bibitem[{{Kurucz}(2006)}]{K06}
{Kurucz}, R.~L. 2006, Robert L. Kurucz on-line database of observed and predicted atomic transitions

\bibitem[{{Kurucz}(2008)}]{K08}
{Kurucz}, R.~L. 2008, Robert L. Kurucz on-line database of observed and predicted atomic transitions

\bibitem[{{Lee} {et~al.}(2017){Lee}, {Gullikson}, \& {Kaplan}}]{Lee:2017}
{Lee}, J.-J., {Gullikson}, K., \& {Kaplan}, K. 2017, {Igrins/Plp 2.2.0}, Zenodo

\bibitem[{{Lee} {et~al.}(2016){Lee}, {Lee}, {Park}, {Lee}, {Kidder}, {Mace}, \& {Jaffe}}]{Lee:2016}
{Lee}, S., {Lee}, J.-E., {Park}, S., {et~al.} 2016, \apj, 826, 179

\bibitem[{{Li} {et~al.}(2015){Li}, {Gordon}, {Rothman}, {Tan}, {Hu}, {Kassi}, {Campargue}, \& {Medvedev}}]{li:2015}
{Li}, G., {Gordon}, I.~E., {Rothman}, L.~S., {et~al.} 2015, \apjs, 216, 15

\bibitem[{{Lim} {et~al.}(2022){Lim}, {Koch-Hansen}, {Chun}, {Hong}, \& {Lee}}]{Lim:2022}
{Lim}, D., {Koch-Hansen}, A.~J., {Chun}, S.-H., {Hong}, S., \& {Lee}, Y.-W. 2022, \aap, 666, A62

\bibitem[{{Lind} {et~al.}(2017){Lind}, {Amarsi}, {Asplund}, {Barklem}, {Bautista}, {Bergemann}, {Collet}, {Kiselman}, {Leenaarts}, \& {Pereira}}]{lind17}
{Lind}, K., {Amarsi}, A.~M., {Asplund}, M., {et~al.} 2017, \mnras, 468, 4311

\bibitem[{{Lodders}(2003)}]{Lodders:2003}
{Lodders}, K. 2003, \apj, 591, 1220

\bibitem[{{Lomaeva} {et~al.}(2019){Lomaeva}, {J{\"o}nsson}, {Ryde}, {Schultheis}, \& {Thorsbro}}]{lomaeva:19}
{Lomaeva}, M., {J{\"o}nsson}, H., {Ryde}, N., {Schultheis}, M., \& {Thorsbro}, B. 2019, \aap, 625, A141

\bibitem[{{L{\'o}pez-Valdivia} {et~al.}(2023){L{\'o}pez-Valdivia}, {Mace}, {Han}, {Sawczynec}, {Hern{\'a}ndez}, {Prato}, {Johns-Krull}, {Oh}, {Lee}, {Kraus}, {Llama}, \& {Jaffe}}]{YSOsii:2023}
{L{\'o}pez-Valdivia}, R., {Mace}, G.~N., {Han}, E., {et~al.} 2023, \apj, 943, 49

\bibitem[{{Mace} {et~al.}(2016){Mace}, {Kim}, {Jaffe}, {Park}, {Lee}, {Kaplan}, {Yu}, {Yuk}, {Chun}, {Pak}, {Kim}, {Lee}, {Sneden}, {Afsar}, {Pavel}, {Lee}, {Oh}, {Jeong}, {Park}, {Kidder}, {Lee}, {Nguyen Le}, {McLane}, {Gully-Santiago}, {Oh}, {Lee}, {Hwang}, \& {Park}}]{Mcdonald}
{Mace}, G., {Kim}, H., {Jaffe}, D.~T., {et~al.} 2016, in Society of Photo-Optical Instrumentation Engineers (SPIE) Conference Series, Vol. 9908, Ground-based and Airborne Instrumentation for Astronomy VI, ed. C.~J. {Evans}, L.~{Simard}, \& H.~{Takami}, 99080C

\bibitem[{{Mace} {et~al.}(2018){Mace}, {Sokal}, {Lee}, {Oh}, {Park}, {Lee}, {Good}, {MacQueen}, {Oh}, {Kaplan}, {Kidder}, {Chun}, {Yuk}, {Jeong}, {Pak}, {Kim}, {Nah}, {Lee}, {Yu}, {Hwang}, {Park}, {Kim}, {Chinn}, {Peck}, {Diaz}, {Rutten}, {Prato}, {Jacoby}, {Cornelius}, {Hardesty}, {DeGroff}, {Dunham}, {Levine}, {Nofi}, {Lopez-Valdivia}, {Weinberger}, \& {Jaffe}}]{Mace:2018}
{Mace}, G., {Sokal}, K., {Lee}, J.-J., {et~al.} 2018, in Society of Photo-Optical Instrumentation Engineers (SPIE) Conference Series, Vol. 10702, Ground-based and Airborne Instrumentation for Astronomy VII, ed. C.~J. {Evans}, L.~{Simard}, \& H.~{Takami}, 107020Q

\bibitem[{{Madonna} {et~al.}(2018){Madonna}, {Bautista}, {Dinerstein}, {Sterling}, {Garc{\'\i}a-Rojas}, {Kaplan}, {Rubio-D{\'\i}ez}, {Castro-Rodr{\'\i}guez}, \& {Garz{\'o}n}}]{Madonna:2018}
{Madonna}, S., {Bautista}, M., {Dinerstein}, H.~L., {et~al.} 2018, \apjl, 861, L8

\bibitem[{{Maiolino} {et~al.}(2013){Maiolino}, {Haehnelt}, {Murphy}, {Queloz}, {Origlia}, {Alcala}, {Alibert}, {Amado}, {Allende Prieto}, {Ammler-von Eiff}, {Asplund}, {Barstow}, {Becker}, {Bonfils}, {Bouchy}, {Bragaglia}, {Burleigh}, {Chiavassa}, {Cimatti}, {Cirasuolo}, {Cristiani}, {D'Odorico}, {Dravins}, {Emsellem}, {Farihi}, {Figueira}, {Fynbo}, {Gansicke}, {Gillon}, {Gustafsson}, {Hill}, {Israelyan}, {Korn}, {Larsen}, {De Laverny}, {Liske}, {Lovis}, {Marconi}, {Martins}, {Molaro}, {Nisini}, {Oliva}, {Petitjean}, {Pettini}, {Recio Blanco}, {Rebolo}, {Reiners}, {Rodriguez-Lopez}, {Ryde}, {Santos}, {Savaglio}, {Snellen}, {Strassmeier}, {Tanvir}, {Testi}, {Tolstoy}, {Triaud}, {Vanzi}, {Viel}, \& {Volonteri}}]{hires:13}
{Maiolino}, R., {Haehnelt}, M., {Murphy}, M.~T., {et~al.} 2013, arXiv e-prints, arXiv:1310.3163

\bibitem[{{Manea} {et~al.}(2023){Manea}, {Hawkins}, {Ness}, {Buder}, {Martell}, \& {Zucker}}]{manea:23}
{Manea}, C., {Hawkins}, K., {Ness}, M.~K., {et~al.} 2023, arXiv e-prints, arXiv:2310.15257

\bibitem[{{Mawet} {et~al.}(2019){Mawet}, {Fitzgerald}, {Konopacky}, {Beichman}, {Jovanovic}, {Dekany}, {Hover}, {Chisholm}, {Ciardi}, {Artigau}, {Banyal}, {Beatty}, {Benneke}, {Blake}, {Burgasser}, {Canalizo}, {Chen}, {Do}, {Doppmann}, {Doyon}, {Dressing}, {Fang}, {Greene}, {Hillenbrand}, {Howard}, {Kane}, {Kataria}, {Kempton}, {Knutson}, {Kotani}, {Lafreni{\`e}re}, {Liu}, {Nishiyama}, {Pandey}, {Plavchan}, {Prato}, {Rajaguru}, {Robertson}, {Salyk}, {Sato}, {Schlawin}, {Sengupta}, {Sivarani}, {Skidmore}, {Tamura}, {Terada}, {Vasisht}, {Wang}, \& {Zhang}}]{tmt_nir:19}
{Mawet}, D., {Fitzgerald}, M., {Konopacky}, Q., {et~al.} 2019, in Bulletin of the American Astronomical Society, Vol.~51, 134

\bibitem[{{Mishenina} {et~al.}(2002){Mishenina}, {Kovtyukh}, {Soubiran}, {Travaglio}, \& {Busso}}]{mishenina:02}
{Mishenina}, T.~V., {Kovtyukh}, V.~V., {Soubiran}, C., {Travaglio}, C., \& {Busso}, M. 2002, \aap, 396, 189

\bibitem[{{Montelius}({2021})}]{montelius:21}
{Montelius}, M. {2021}, {Towards a high resolution view of infrared line formation}, {Student Paper}

\bibitem[{{Montelius} {et~al.}(2022){Montelius}, {Forsberg}, {Ryde}, {J{\"o}nsson}, {Af{\c{s}}ar}, {Johansen}, {Kaplan}, {Kim}, {Mace}, {Sneden}, \& {Thorsbro}}]{montelius:22}
{Montelius}, M., {Forsberg}, R., {Ryde}, N., {et~al.} 2022, \aap, 665, A135

\bibitem[{{Moon} {et~al.}(2012){Moon}, {Wang}, {Park}, {Yuk}, {Chun}, \& {Jaffe}}]{Moon:2012}
{Moon}, B., {Wang}, W., {Park}, C., {et~al.} 2012, in Society of Photo-Optical Instrumentation Engineers (SPIE) Conference Series, Vol. 8450, Modern Technologies in Space- and Ground-based Telescopes and Instrumentation II, ed. R.~{Navarro}, C.~R. {Cunningham}, \& E.~{Prieto}, 845048

\bibitem[{{Nandakumar} {et~al.}(2023{\natexlab{a}}){Nandakumar}, {Ryde}, {Casagrande}, \& {Mace}}]{Nandakumar:2023}
{Nandakumar}, G., {Ryde}, N., {Casagrande}, L., \& {Mace}, G. 2023{\natexlab{a}}, \aap, 675, A23

\bibitem[{{Nandakumar} {et~al.}(2023{\natexlab{b}}){Nandakumar}, {Ryde}, \& {Mace}}]{Nandakumar:2023b}
{Nandakumar}, G., {Ryde}, N., \& {Mace}, G. 2023{\natexlab{b}}, \aap, 676, A79

\bibitem[{{Nandakumar} {et~al.}(2018){Nandakumar}, {Ryde}, {Schultheis}, {Thorsbro}, {J{\"o}nsson}, {Barklem}, {Rich}, \& {Fragkoudi}}]{Nandakumar:18}
{Nandakumar}, G., {Ryde}, N., {Schultheis}, M., {et~al.} 2018, \mnras, 478, 4374

\bibitem[{{Nieuwmunster} {et~al.}(2023){Nieuwmunster}, {Nandakumar}, {Spitoni}, {Ryde}, {Schultheis}, {Rich}, {Barklem}, {Agertz}, {Renaud}, \& {Matteucci}}]{Nieuwmunster:2023}
{Nieuwmunster}, N., {Nandakumar}, G., {Spitoni}, E., {et~al.} 2023, \aap, 671, A94

\bibitem[{{Nordlander} \& {Lind}(2017)}]{Nordlander:2017}
{Nordlander}, T. \& {Lind}, K. 2017, \aap, 607, A75

\bibitem[{{Oh} {et~al.}(2014){Oh}, {Park}, {Cha}, {Yuk}, {Kim}, {Chun}, {Ko}, {Oh}, {Jeong}, {Nah}, {Lee}, {Pavel}, \& {Jaffe}}]{Oh:2014}
{Oh}, J.~S., {Park}, C., {Cha}, S.-M., {et~al.} 2014, in Society of Photo-Optical Instrumentation Engineers (SPIE) Conference Series, Vol. 9147, Ground-based and Airborne Instrumentation for Astronomy V, ed. S.~K. {Ramsay}, I.~S. {McLean}, \& H.~{Takami}, 914739

\bibitem[{{Origlia} {et~al.}(2014){Origlia}, {Oliva}, {Baffa}, {Falcini}, {Giani}, {Massi}, {Montegriffo}, {Sanna}, {Scuderi}, {Sozzi}, {Tozzi}, {Carleo}, {Gratton}, {Ghinassi}, \& {Lodi}}]{Origlia:2014}
{Origlia}, L., {Oliva}, E., {Baffa}, C., {et~al.} 2014, in Society of Photo-Optical Instrumentation Engineers (SPIE) Conference Series, Vol. 9147, Ground-based and Airborne Instrumentation for Astronomy V, ed. S.~K. {Ramsay}, I.~S. {McLean}, \& H.~{Takami}, 91471E

\bibitem[{{Park} {et~al.}(2014){Park}, {Jaffe}, {Yuk}, {Chun}, {Pak}, {Kim}, {Pavel}, {Lee}, {Oh}, {Jeong}, {Sim}, {Lee}, {Nguyen Le}, {Strubhar}, {Gully-Santiago}, {Oh}, {Cha}, {Moon}, {Park}, {Brooks}, {Ko}, {Han}, {Nah}, {Hill}, {Lee}, {Barnes}, {Yu}, {Kaplan}, {Mace}, {Kim}, {Lee}, {Hwang}, \& {Park}}]{Park:2014}
{Park}, C., {Jaffe}, D.~T., {Yuk}, I.-S., {et~al.} 2014, in Society of Photo-Optical Instrumentation Engineers (SPIE) Conference Series, Vol. 9147, Ground-based and Airborne Instrumentation for Astronomy V, ed. S.~K. {Ramsay}, I.~S. {McLean}, \& H.~{Takami}, 91471D

\bibitem[{{Park} {et~al.}(2018){Park}, {Lee}, {Kang}, {Lee}, {Chun}, {Kim}, {Yuk}, {Lee}, {Mace}, {Kim}, {Kaplan}, {Park}, {Sok Oh}, {Lee}, \& {Jaffe}}]{park:18}
{Park}, S., {Lee}, J.-E., {Kang}, W., {et~al.} 2018, \apjs, 238, 29

\bibitem[{{Pehlivan} {et~al.}(2015){Pehlivan}, {Nilsson}, \& {Hartman}}]{Pehlivan:2015}
{Pehlivan}, A., {Nilsson}, H., \& {Hartman}, H. 2015, \aap, 582, A98

\bibitem[{{Perdigon} {et~al.}(2021){Perdigon}, {de Laverny}, {Recio-Blanco}, {Fernandez-Alvar}, {Santos-Peral}, {Kordopatis}, \& {{\'A}lvarez}}]{Perdigon:2021}
{Perdigon}, J., {de Laverny}, P., {Recio-Blanco}, A., {et~al.} 2021, \aap, 647, A162

\bibitem[{{Pignatari} {et~al.}(2010){Pignatari}, {Gallino}, {Heil}, {Wiescher}, {K{\"a}ppeler}, {Herwig}, \& {Bisterzo}}]{pagin:10}
{Pignatari}, M., {Gallino}, R., {Heil}, M., {et~al.} 2010, \apj, 710, 1557

\bibitem[{{Piskunov} {et~al.}(1995){Piskunov}, {Kupka}, {Ryabchikova}, {Weiss}, \& {Jeffery}}]{vald}
{Piskunov}, N.~E., {Kupka}, F., {Ryabchikova}, T.~A., {Weiss}, W.~W., \& {Jeffery}, C.~S. 1995, A\&AS, 112, 525

\bibitem[{{Prantzos} {et~al.}(2020){Prantzos}, {Abia}, {Cristallo}, {Limongi}, \& {Chieffi}}]{prantzos:20}
{Prantzos}, N., {Abia}, C., {Cristallo}, S., {Limongi}, M., \& {Chieffi}, A. 2020, \mnras, 491, 1832

\bibitem[{{Prantzos} {et~al.}(2018){Prantzos}, {Abia}, {Limongi}, {Chieffi}, \& {Cristallo}}]{prantzos:18}
{Prantzos}, N., {Abia}, C., {Limongi}, M., {Chieffi}, A., \& {Cristallo}, S. 2018, \mnras, 476, 3432

\bibitem[{{Ram{\'\i}rez} \& {Allende Prieto}(2011)}]{Ramirez:2011}
{Ram{\'\i}rez}, I. \& {Allende Prieto}, C. 2011, \apj, 743, 135

\bibitem[{{Ram{\'{\i}}rez} {et~al.}(2000){Ram{\'{\i}}rez}, {Stephens}, {Frogel}, \& {DePoy}}]{ramirez:00}
{Ram{\'{\i}}rez}, S.~V., {Stephens}, A.~W., {Frogel}, J.~A., \& {DePoy}, D.~L. 2000, AJ, 120, 833

\bibitem[{{Rich} {et~al.}(2007){Rich}, {Origlia}, \& {Valenti}}]{rich:07}
{Rich}, R.~M., {Origlia}, L., \& {Valenti}, E. 2007, \apjl, 665, L119

\bibitem[{{Rich} {et~al.}(2012){Rich}, {Origlia}, \& {Valenti}}]{rich:12}
{Rich}, R.~M., {Origlia}, L., \& {Valenti}, E. 2012, \apj, 746, 59

\bibitem[{{Ryabchikova} {et~al.}(2015){Ryabchikova}, {Piskunov}, {Kurucz}, {Stempels}, {Heiter}, {Pakhomov}, \& {Barklem}}]{vald5}
{Ryabchikova}, T., {Piskunov}, N., {Kurucz}, R.~L., {et~al.} 2015, \physscr, 90, 054005

\bibitem[{{Ryde} {et~al.}(2016{\natexlab{a}}){Ryde}, {Fritz}, {Rich}, {Thorsbro}, {Schultheis}, {Origlia}, \& {Chatzopoulos}}]{ryde:2016_metalpoor}
{Ryde}, N., {Fritz}, T.~K., {Rich}, R.~M., {et~al.} 2016{\natexlab{a}}, \apj, 831, 40

\bibitem[{{Ryde} \& {Schultheis}(2015)}]{ryde:15}
{Ryde}, N. \& {Schultheis}, M. 2015, \aap, 573, A14

\bibitem[{{Ryde} {et~al.}(2016{\natexlab{b}}){Ryde}, {Schultheis}, {Grieco}, {Matteucci}, {Rich}, \& {Uttenthaler}}]{ryde:2016_bp2}
{Ryde}, N., {Schultheis}, M., {Grieco}, V., {et~al.} 2016{\natexlab{b}}, \aj, 151, 1

\bibitem[{{Sales-Silva} {et~al.}(2022){Sales-Silva}, {Daflon}, {Cunha}, {Souto}, {Smith}, {Chiappini}, {Donor}, {Frinchaboy}, {Garc{\'\i}a-Hern{\'a}ndez}, {Hayes}, {Majewski}, {Masseron}, {Schiavon}, {Weinberg}, {Beaton}, {Fern{\'a}ndez-Trincado}, {J{\"o}nsson}, {Lane}, {Minniti}, {Manchado}, {Moni Bidin}, {Nitschelm}, {O'Connell}, \& {Villanova}}]{sales:22}
{Sales-Silva}, J.~V., {Daflon}, S., {Cunha}, K., {et~al.} 2022, \apj, 926, 154

\bibitem[{{Sawczynec} {et~al.}(2022){Sawczynec}, {Mace}, {Gully-Santiago}, \& {Jaffe}}]{rrisa}
{Sawczynec}, E., {Mace}, G., {Gully-Santiago}, M., \& {Jaffe}, D. 2022, in American Astronomical Society Meeting Abstracts, Vol.~54, American Astronomical Society Meeting Abstracts, 203.06

\bibitem[{{Skidmore} {et~al.}(2022){Skidmore}, {Bernstein}, {Dumas}, {Goodrich}, {Millan-Gabet}, {Ramsay}, {Travouillon}, \& {Vernet}}]{TMT}
{Skidmore}, W., {Bernstein}, R., {Dumas}, C., {et~al.} 2022, Journal of Astronomical Telescopes, Instruments, and Systems, 8, 021510

\bibitem[{{Sneden} {et~al.}(2016){Sneden}, {Cowan}, {Kobayashi}, {Pignatari}, {Lawler}, {Den Hartog}, \& {Wood}}]{sneden:16}
{Sneden}, C., {Cowan}, J.~J., {Kobayashi}, C., {et~al.} 2016, \apj, 817, 53

\bibitem[{{Sneden} {et~al.}(2014){Sneden}, {Lucatello}, {Ram}, {Brooke}, \& {Bernath}}]{sneden:2014}
{Sneden}, C., {Lucatello}, S., {Ram}, R.~S., {Brooke}, J. S.~A., \& {Bernath}, P. 2014, \apjs, 214, 26

\bibitem[{{Sterling} {et~al.}(2016){Sterling}, {Dinerstein}, {Kaplan}, \& {Bautista}}]{Sterling:2016}
{Sterling}, N.~C., {Dinerstein}, H.~L., {Kaplan}, K.~F., \& {Bautista}, M.~A. 2016, \apjl, 819, L9

\bibitem[{{Takeda}(2020)}]{takeda:20}
{Takeda}, Y. 2020, arXiv e-prints, arXiv:2001.04588

\bibitem[{{Takeda} {et~al.}(2002){Takeda}, {Zhao}, {Chen}, {Qiu}, \& {Takada-Hidai}}]{takeda:02}
{Takeda}, Y., {Zhao}, G., {Chen}, Y.-Q., {Qiu}, H.-M., \& {Takada-Hidai}, M. 2002, \pasj, 54, 275

\bibitem[{{Tamai} \& {Spyromilio}(2014)}]{ELT2014}
{Tamai}, R. \& {Spyromilio}, J. 2014, in Society of Photo-Optical Instrumentation Engineers (SPIE) Conference Series, Vol. 9145, Ground-based and Airborne Telescopes V, ed. L.~M. {Stepp}, R.~{Gilmozzi}, \& H.~J. {Hall}, 91451E

\bibitem[{{Tautvai{\v{s}}ien{\.{e}}} {et~al.}(2021){Tautvai{\v{s}}ien{\.{e}}}, {Viscasillas V{\'a}zquez}, {Mikolaitis}, {Stonkut{\.{e}}}, {Minkevi{\v{c}}i{\={u}}t{\.{e}}}, {Drazdauskas}, \& {Bagdonas}}]{taut:21}
{Tautvai{\v{s}}ien{\.{e}}}, G., {Viscasillas V{\'a}zquez}, C., {Mikolaitis}, {\v{S}}., {et~al.} 2021, \aap, 649, A126

\bibitem[{{Taylor}(2005)}]{topcat}
{Taylor}, M.~B. 2005, in Astronomical Society of the Pacific Conference Series, Vol. 347, Astronomical Data Analysis Software and Systems XIV, ed. P.~{Shopbell}, M.~{Britton}, \& R.~{Ebert}, 29

\bibitem[{{Thorsbro} {et~al.}(2018){Thorsbro}, {Ryde}, {Schultheis}, {Hartman}, {Rich}, {Lomaeva}, {Origlia}, \& {J{\"o}nsson}}]{thorsbro:2018}
{Thorsbro}, B., {Ryde}, N., {Schultheis}, M., {et~al.} 2018, \apj, 866, 52

\bibitem[{{Tody}(1993)}]{IRAF}
{Tody}, D. 1993, in ASP Conf. Ser. 52: Astronomical Data Analysis Software and Systems II, ed. R.~J. {Hanisch}, R.~J.~V. {Brissenden}, \& J.~{Barnes}, 173

\bibitem[{{Valenti} \& {Piskunov}(1996)}]{sme}
{Valenti}, J.~A. \& {Piskunov}, N. 1996, \aaps, 118, 595

\bibitem[{{Valenti} \& {Piskunov}(2012)}]{sme_code}
{Valenti}, J.~A. \& {Piskunov}, N. 2012, {SME: Spectroscopy Made Easy}, astrophysics Source Code Library

\bibitem[{van~der Walt {et~al.}(2011)van~der Walt, Colbert, \& Varoquaux}]{numpy}
van~der Walt, S., Colbert, S.~C., \& Varoquaux, G. 2011, Computing in Science and Engineering, 13, 22

\bibitem[{Vasini {et~al.}(2023)Vasini, Spitoni, \& Matteucci}]{vasini:2023}
Vasini, A., Spitoni, E., \& Matteucci, F. 2023, Galactic Archaeology with [Mg/Mn] versus [Al/Fe] abundance ratios -- Uncertainties and caveats

\bibitem[{{Virtanen} {et~al.}(2020){Virtanen}, {Gommers}, {Oliphant}, {Haberland}, {Reddy}, {Cournapeau}, {Burovski}, {Peterson}, {Weckesser}, {Bright}, {van der Walt}, {Brett}, {Wilson}, {Millman}, {Mayorov}, {Nelson}, {Jones}, {Kern}, {Larson}, {Carey}, {Polat}, {Feng}, {Moore}, {VanderPlas}, {Laxalde}, {Perktold}, {Cimrman}, {Henriksen}, {Quintero}, {Harris}, {Archibald}, {Ribeiro}, {Pedregosa}, {van Mulbregt}, \& {SciPy 1. 0 Contributors}}]{scipy}
{Virtanen}, P., {Gommers}, R., {Oliphant}, T.~E., {et~al.} 2020, Nature Methods, 17, 261

\bibitem[{{Wallace} \& {Livingston}(2003)}]{Wallace:2003}
{Wallace}, L. \& {Livingston}, W. 2003, {An atlas of the solar spectrum in the infrared from 1850 to 9000 cm-1 (1.1 to 5.4 micrometer)}

\bibitem[{{Wang} {et~al.}(2010){Wang}, {Gully-Santiago}, {Deen}, {Mar}, \& {Jaffe}}]{Wang:2010}
{Wang}, W., {Gully-Santiago}, M., {Deen}, C., {Mar}, D.~J., \& {Jaffe}, D.~T. 2010, in Society of Photo-Optical Instrumentation Engineers (SPIE) Conference Series, Vol. 7739, Modern Technologies in Space- and Ground-based Telescopes and Instrumentation, ed. E.~{Atad-Ettedgui} \& D.~{Lemke}, 77394L

\bibitem[{{Woosley} \& {Weaver}(1995)}]{Woosley:1995}
{Woosley}, S.~E. \& {Weaver}, T.~A. 1995, \apjs, 101, 181

\bibitem[{{Yuk} {et~al.}(2010){Yuk}, {Jaffe}, {Barnes}, {Chun}, {Park}, {Lee}, {Lee}, {Wang}, {Park}, {Pak}, {Strubhar}, {Deen}, {Oh}, {Seo}, {Pyo}, {Park}, {Lacy}, {Goertz}, {Rand}, \& {Gully-Santiago}}]{Yuk:2010}
{Yuk}, I.-S., {Jaffe}, D.~T., {Barnes}, S., {et~al.} 2010, in Society of Photo-Optical Instrumentation Engineers (SPIE) Conference Series, Vol. 7735, Ground-based and Airborne Instrumentation for Astronomy III, ed. I.~S. {McLean}, S.~K. {Ramsay}, \& H.~{Takami}, 77351M

\bibitem[{{Zasowski} {et~al.}(2019){Zasowski}, {Schultheis}, {Hasselquist}, {Cunha}, {Sobeck}, {Johnson}, {Rojas-Arriagada}, {Majewski}, {Andrews}, {J{\"o}nsson}, {Beers}, {Chojnowski}, {Frinchaboy}, {Holtzman}, {Minniti}, {Nidever}, \& {Nitschelm}}]{Zasowski:2019}
{Zasowski}, G., {Schultheis}, M., {Hasselquist}, S., {et~al.} 2019, \apj, 870, 138

\end{thebibliography}




\begin{appendix} 

\section{Additional figures}

\begin{figure*}
  \includegraphics[width=\textwidth]{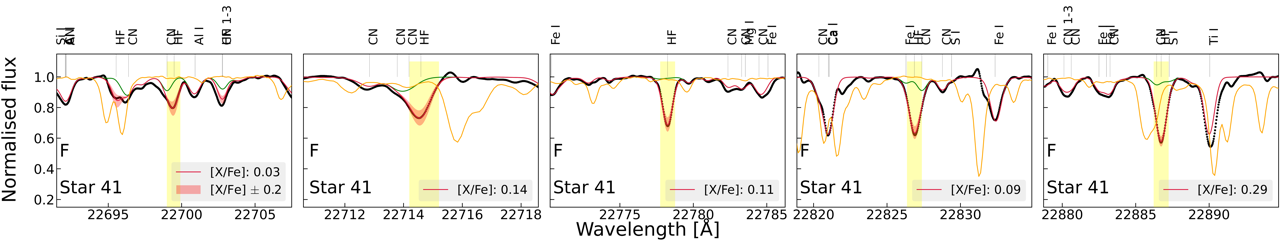}
  \caption{ Wavelength regions centered at the five selected fluorine lines for the star 2M18103303-1626220  (star 41). In each panel, the black circles denote the observed spectrum, the crimson line denotes the best-fit synthetic spectrum, and the red band denotes the variation in the synthetic spectrum for a difference of $\pm$0.2\,dex in the [F/Fe]. The yellow bands in each panel represent the line masks defined for the F lines, wherein SME fits observed spectra by varying the fluorine abundance and finds the best synthetic spectra fit by $\chi^{2}$ minimization. The green line shows the synthetic spectrum without F, also indicating any possible blends in the line, and the orange line shows the telluric spectrum of the standard star that is used to correct for telluric contaminations in the observed star spectrum. the The [F/Fe] values corresponding to the best-fit case for each F line are listed in each panel. All identified atomic and molecular lines are also denoted in the top of each panel. }
  \label{fig:fspectra}%
\end{figure*}

\begin{figure*}
  \includegraphics[width=\textwidth]{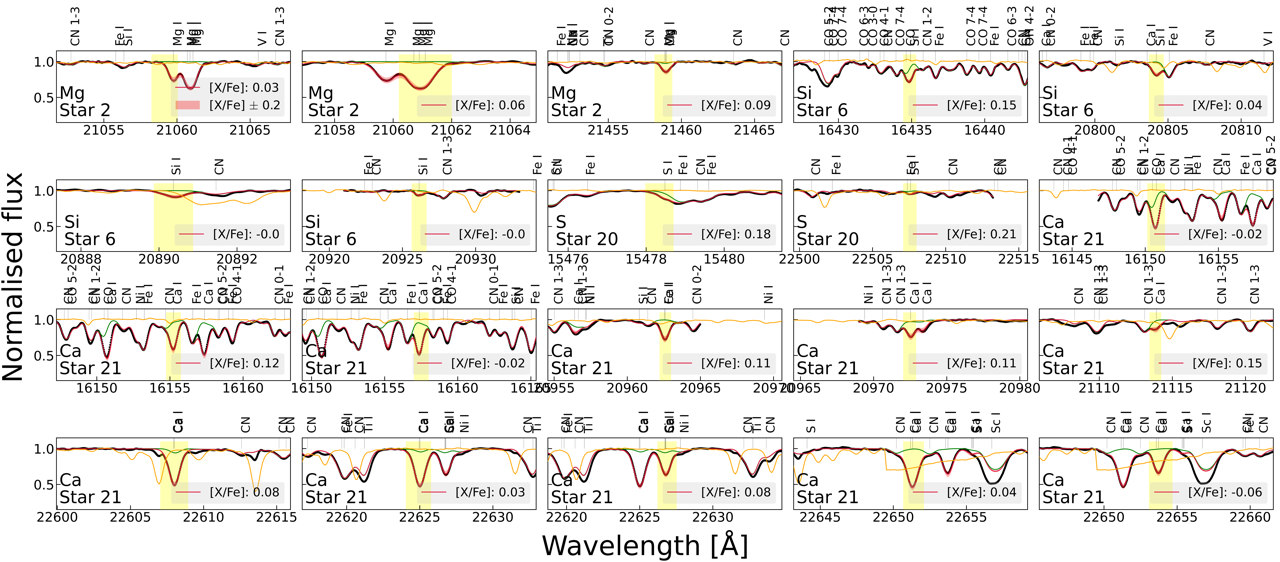}
  \caption{ Alpha elements: Wavelength regions centered at the three selected magnesium lines for the star 2M05594446-7212111 (star 2), four selected silicon lines for the star 2M06074096-0530332 (star 6), two selected sulphur lines for the star 2M14131192-4849280 (star 20), and 11 selected calcium lines for the  star 2M14240039-6252516 (star 21). Arrangement of figures and plot descriptions are similar to Figure~\ref{fig:fspectra}. }
  \label{fig:alphaspectra}%
\end{figure*}

\begin{figure*}
  \includegraphics[width=\textwidth]{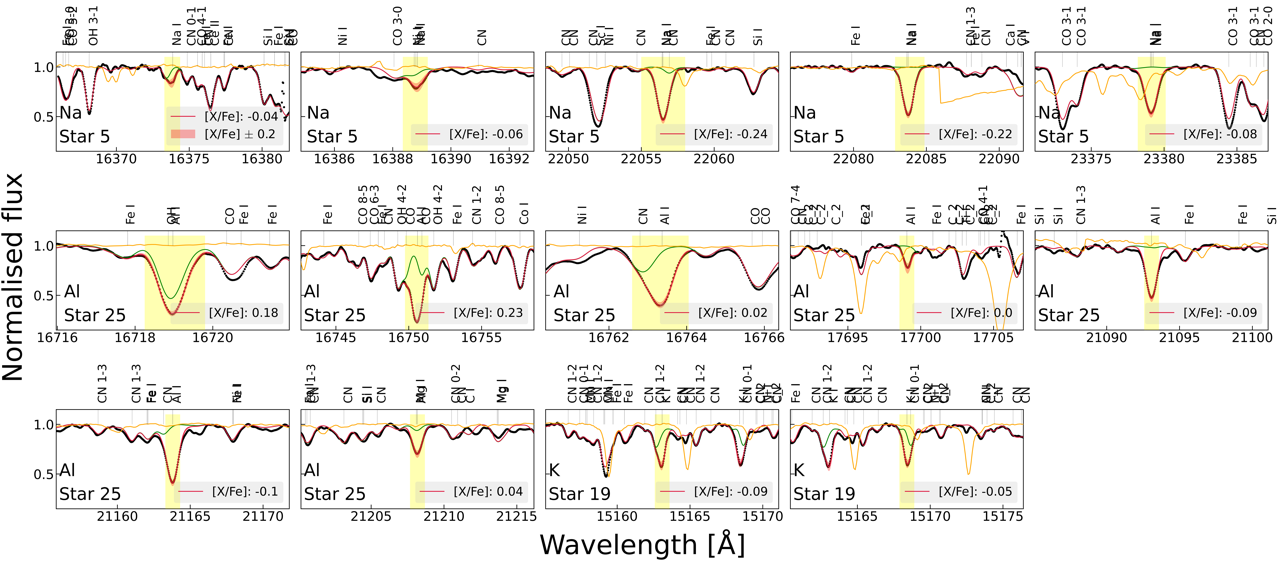}
  \caption{ Odd-Z elements: Wavelength regions centered at the five selected sodium lines (note that only the two H-band lines are used to estimate mean [Na/Fe] in Figure~\ref{fig:na_trend}) for the star 2M06052796-0553384 (star 5), seven selected aluminium lines for the star 2M14275833-6147534 (star 25), and two selected potassium lines for the  star 2M13403516-5040261 (star 19). Arrangement of figures and plot descriptions are similar to Figure~\ref{fig:fspectra}. }
  \label{fig:oddzspectra}%
\end{figure*}

\begin{figure*}
  \includegraphics[width=\textwidth]{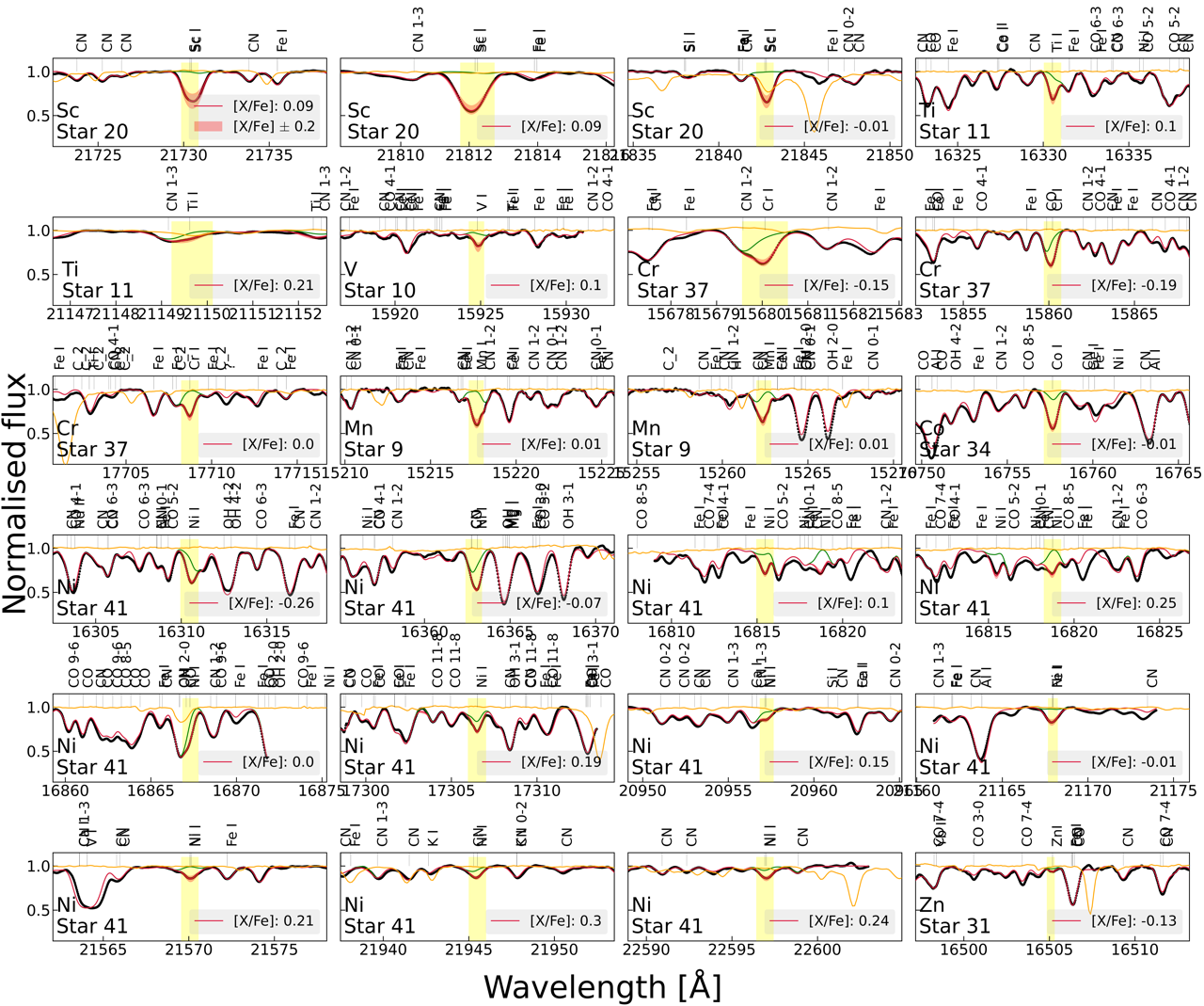}
  \caption{ Iron-peak elements and zinc: Wavelength regions centered at the three selected scandium lines for the star 2M14131192-4849280 (star 20), two selected titanium lines for the star 2M06223443-0443153 (star 11), one selected vanadium line for the star 2M06171159-7259319 (star 10), three selected chromium lines for the star 2M14371958-6251344 (star 37), two selected manganese lines for the star 2M06143705-0551064 (star 9), one selected chromium line for the star 2M14345114-6225509 (star 34), 11 selected nickel lines for the  star 2M18103303-1626220 (star 41), and one selected zinc line for the star 2M14332869-6211255 (star 31). Arrangement of figures and plot descriptions are similar to Figure~\ref{fig:fspectra}.  }
  \label{fig:ironpeakspectra}%
\end{figure*}

\begin{figure*}
  \includegraphics[width=\textwidth]{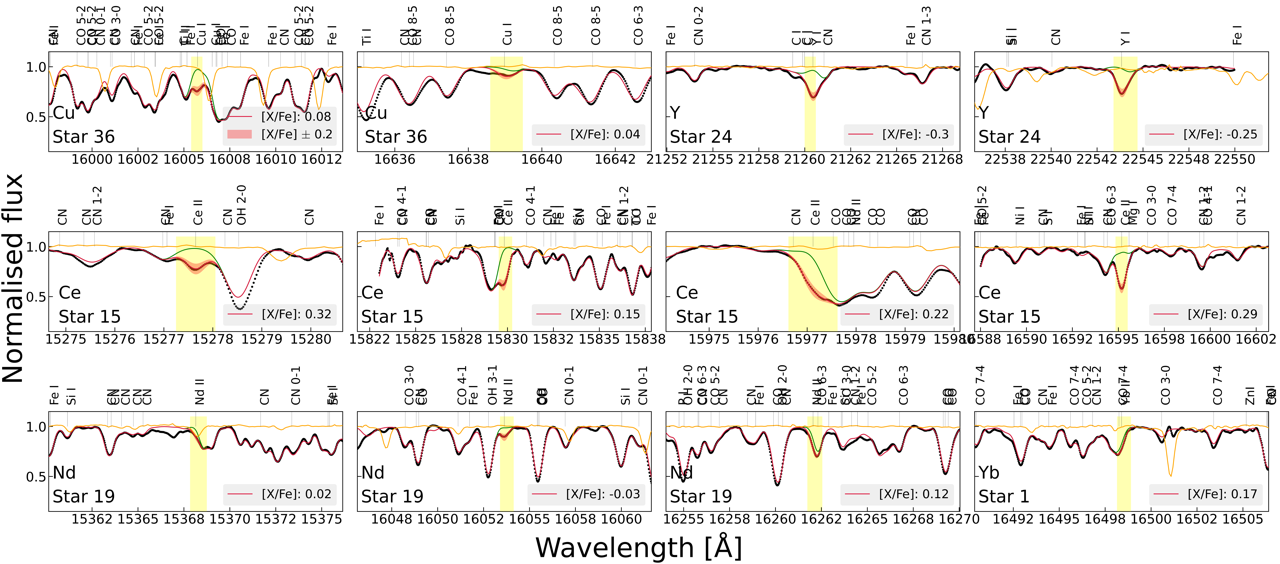}
  \caption{ Neutron-capture elements: Wavelength regions centered at the two selected copper lines for the star 2M14360935-6309399 (star 36), two selected yttrium lines for the star 2M14261117-6240220 (star 24), four selected cerium lines for the star 2M06574070-1231239 (star 15), three selected neodymium lines for the star 2M13403516-5040261 (star 19), and one selected ytterbium line for the  star 2M05484106-0602007 (star 1). Arrangement of figures and plot descriptions are similar to Figure~\ref{fig:fspectra}. }
  \label{fig:neutroncapturespectra}%
\end{figure*}

\end{appendix}

\end{document}